\documentclass[aps,preprint,floatfix,nofootinbib,showpacs]{revtex4-1}
\pdfoutput=1
\usepackage{dcolumn}
\usepackage{bm}
\usepackage{graphicx}

\usepackage{amssymb,amsmath}
\usepackage{multirow}
\usepackage{units,changes}
\usepackage{color,url}
\usepackage{tabu}
\usepackage{array}
\usepackage[colorlinks=true,urlcolor=blue,anchorcolor=blue
,citecolor=blue,filecolor=blue,linkcolor=blue,menucolor=blue
,linktocpage=true,pdfproducer=medialab,pdfa=true]{hyperref}
\usepackage{colordvi}
\usepackage{subcaption}
\def\beqa{\begin{eqnarray}}
\def\eeqa{\end{eqnarray}}

\usepackage{soul}

\begin{document}
\preprint{CPTNP-2025-044}

\title{Exploring muonphilic dark matter with the $Z_2$-even mediator at muon colliders}
\def\slash#1{#1\!\!\!/}

\author{Wanyun Chen$^{1,2}$, Haoqi Li$^1$, Chih-Ting Lu$^{1,3}$, Qiulei Wang$^1$}
\affiliation{
$^1$ Department of Physics and Institute of Theoretical Physics, Nanjing Normal University, Nanjing, 210023, P. R. China}  
\affiliation{
$^2$ Department of Physics, Konkuk University, 120 Neungdong-ro, Gwangjin-gu, Seoul 05029, Republic of Korea}
\affiliation{$^3$ Nanjing Key Laboratory of Particle Physics and Astrophysics, Nanjing, 210023, China} 

\begin{abstract}

The Galactic Center GeV Excess (GCE) remains a compelling but enigmatic signal from the inner region of our galaxy. Muonphilic dark matter (DM), which couples exclusively to muons via a new mediator, provides a viable explanation for the GCE and relic density while naturally evading constraints from direct detection, collider searches and other multi-messenger observations. Based on the viable non-resonant parameter space identified in previous global fits, we perform a comprehensive study exploring the prospects for discovering such muonphilic DM in the context of a future $3$ TeV muon collider, focusing on simplified models with a \(Z_2\)-even mediator. Four distinct search strategies are investigated: visible on-shell mediator decays (\(\mu^{+}\mu^{-}\gamma\) final state), invisible on-shell mediator decays (mono-photon plus missing energy), mono-photon production via off-shell mediators, and vector boson fusion production.Through a detailed signal-background analysis using cut-and-count methods, we project the exclusion limits at $95\%$ confidence level for seven representative models across a wide range of mediator masses. Our results demonstrate that the projected limits cover a significant portion of the viable parameter space that explains the GCE, establishing a muon collider as a decisive machine for testing the muonphilic DM hypothesis.  
\end{abstract}

\maketitle

\section{Introduction} 

Dark matter (DM) is a non-luminous form of matter that does not interact electromagnetically, yet its gravitational influence profoundly shapes the dynamics of galaxies and large-scale structures in the universe~\cite{Trimble:1987ee,Barack:2018yly}. As one of the most compelling pieces of evidence for physics beyond the Standard Model (SM), the existence of DM remains a major challenge to modern physics~\cite{Arbey:2021gdg,Cirelli:2024ssz}. Although DM constitutes approximately $27\%$ of the total mass-energy content of the universe, compared to merely $5\%$ from ordinary baryonic matter, its particle nature remains entirely unknown~\cite{Battaglieri:2017aum,Lin:2019uvt,Ferreira:2020fam,Belenchia:2021rfb,PerezdelosHeros:2020qyt}. Long-standing anomalies in indirect DM detection, such as the galactic center GeV excess (GCE)~\cite{Slatyer:2021qgc,Cholis:2021rpp,DiMauro:2021qcf}, the antiproton excess~\cite{Cui:2016ppb,Cuoco:2016eej,Calore:2022stf}, and the 511~keV~\cite{Kierans:2019aqz,Cai:2020fnq,Keith:2021guq} and 3.5~keV~\cite{Slatyer:2021qgc,Silich:2021sra} lines, have motivated extensive efforts to uncover potential connections to DM physics.

Among these indirect detection signals, the GCE, first identified by the Fermi-LAT satellite in 2009~\cite{Hooper:2013rwa,Goodenough:2009gk,Hooper:2010mq,Fermi-LAT:2015sau}, remains one of the most persistent and challenging puzzles in high-energy astrophysics and particle physics. This signal exhibits a significant gamma-ray excess in the 1–3~GeV range from the galactic center, with a spatial distribution consistent with the predicted NFW density profile of DM~\cite{Navarro:1995iw}. Its spectral shape also closely resembles expectations from weakly interacting massive particle (WIMP) annihilation channels, such as $b\bar{b}$~\cite{Goodenough:2009gk,Hooper:2010mq}. However, secondary particles (e.g., hadrons or electrons) produced in such annihilations would generate observable signals at other wavelengths (e.g., radio or X-rays), though current multi-wavelength observations have not yet confirmed such counterparts~\cite{McDaniel:2017ppt,Jeltema:2011bd}.

Various DM models have been proposed to explain the GCE~\cite{Dolan:2014ska,Ipek:2014gua,Berlin:2014tja,Alves:2014yha,Agrawal:2014una,Izaguirre:2014vva,Ko:2014gha,Abdullah:2014lla,Martin:2014sxa,Berlin:2014pya,Cheung:2014lqa,Agrawal:2014oha,Karwin:2016tsw,Abdughani:2021oit,Fan:2022dck,Fan:2024wvo}. Among these, muonphilic DM has emerged in recent years as a promising candidate due to its consistency with multi-messenger observational constraints~\cite{DiMauro:2021qcf,Abdughani:2021oit}. Conventional WIMP models, which remain leading explanations, typically involve annihilation into SM particles (e.g., $b\bar{b}$, $\tau^+\tau^-$, or $W^+ W^-$), producing gamma rays via inverse Compton scattering or hadronization processes. However, many such models face tensions with antiproton and positron spectra from AMS-02~\cite{AMS:2016oqu,AMS:2019rhg} and direct detection limits from experiments such as PandaX~\cite{PandaX:2024qfu}.

To resolve these tensions, muonphilic DM models have been proposed. These posit that DM couples exclusively to muon pairs ($\mu^+\mu^-$) via a mediator, with negligible interactions with other SM particles~\cite{Krnjaic:2019rsv,Garani:2019fpa,Saez:2021qta,Drees:2021rsg,Hapitas:2021ilr,Abdughani:2021oit,Borah:2021khc,Medina:2021ram,Heeck:2022znj,Baek:2022ozm,Manzari:2023gkt,Figueroa:2024tmn,Li:2025yzb,Wang:2025kit,Tang:2025vqf,Bell:2025acg,De:2025hay}. This feature avoids hadronic or electronic secondary emission, offering a consistent explanation for the GCE without conflicting with other observational data. When DM particles annihilate into $\mu^+\mu^-$ or produce four muons via the decays from a pair of mediators, final-state radiation (FSR) photons are generated predominantly in the 1–5~GeV range. The resulting spectrum exhibits a pronounced peak-like feature that aligns well with the GCE observed by Fermi-LAT~\cite{DiMauro:2021qcf,Abdughani:2021oit}. Compared to other channels (e.g., $b\bar{b}$), the $\mu^+\mu^-$ final state yields a harder FSR spectrum with a steeper rise and sharper peak around 1–3~GeV, making it ideal for reproducing the GCE spectral morphology~\cite{Elor:2015bho,DiMauro:2021raz}. Furthermore, since the model couples only to muons, it avoids detectable synchrotron radiation from electrons/positrons or hadronic emission, thereby evading constraints from radio (e.g., WMAP/Planck synchrotron limits) and X-ray (e.g., Chandra inverse Compton bounds) observations~\cite{Crocker:2010gy,Chan:2017aup,Hooper:2012jc}. This enables muonphilic DM models to resolve persistent multi-messenger tensions that challenge conventional DM scenarios.

In muonphilic DM models, DM couples exclusively to muons via a mediator, such as a dark photon or scalar boson, making it challenging to probe at conventional colliders like the Large Hadron Collider (LHC) and Large Electron-Positron Collider (LEP) as well as direct detection experiments, which consequently impose weak constraints. However, owing to the muon’s mass (approximately $200$ times that of the electron) and its absence from strong interactions, a high-energy muon collider (at the TeV scale) offers superior energy resolution and cleaner backgrounds compared to hadron colliders (e.g., LHC) or electron-positron colliders (e.g., CEPC/FCC-ee)~\cite{MuonCollider:2022xlm,Black:2022cth,Accettura:2023ked}. Thus, high-energy muon colliders represent a promising avenue for directly probing muonphilic DM\footnote{While several proposals exist to test DM models relevant to the GCE at colliders (see Ref.~\cite{Alves:2014yha,Berlin:2014tja,Agrawal:2014una,Izaguirre:2014vva,Ipek:2014gua,Cline:2014dwa,Cheung:2014lqa,Carena:2019pwq,Koechler:2025ryv,Hu:2025thq}), to the best of our knowledge, none have explored muonphilic DM models motivated by the GCE in the context of a muon collider, as investigated in this work.}.

In this work, we investigate the following signal processes at a muon collider: 
(1) Visible on-shell mediator decays: $\mu^+\mu^-\to\gamma +\text{MED}$, with $\text{MED}\to\mu^+\mu^-$, where MED denotes the mediator; 
(2) Mono-photon with on-shell mediators: $\mu^+\mu^-\to\gamma +\text{MED}$, with $\text{MED}\to \chi\overline{\chi}$, where the DM particles ($\chi$) yield missing energy and the photon energy distribution exhibits a peak-like structure; 
(3) Mono-photon with off-shell mediators: $\mu^+\mu^-\to\gamma +\chi\overline{\chi}$, producing photon and missing energy signatures similar to SM backgrounds, thereby complicating signal extraction; (4) Vector boson fusion production: $\mu^+\mu^-\to\nu_{\mu}\bar{\nu_{\mu}}\mu^+\mu^-\text{MED}$, with $\text{MED}\to\mu^+\mu^-/\chi\overline{\chi}$ . 
Using these search strategies, a muon collider can directly test whether muonphilic DM models resolve the longstanding GCE, precisely measure the DM-muon coupling strength and mediator mass, and complement the limitations of LHC searches and direct detection experiments. Combined with astrophysical observations from Fermi-LAT, AMS-02, and others, muon collider searches enable cross-validation of potential DM-induced astrophysical signals.

The rest of this paper is organized as follows: Sec.~\ref{sec:model} briefly reviews the simplified muonphilic DM models considered and their key parameter space. Based on the allowed parameter space of muonphilic DM models, Sec.~\ref{sec:search} proposes four search strategies and provides a state-of-the-art signal-background analysis. The resulting projected exclusion limits for muonphilic DM models at muon colliders are presented and discussed in Sec.~\ref{sec:exclusion}. Finally, we conclude in Sec.~\ref{sec:conclusion}. Collider constraints from LEP and LHC on muonphilic DM models are discussed in the Appendix~\ref{app:collider_constraints}.

\section{Simplified muonphilic DM models with the $Z_2$-even mediator}
\label{sec:model}

\subsection{The models} 

For simplified muonphilic DM models consisting of a Standard Model (SM) singlet DM particle and a mediator (MED), our previous work~\cite{Abdughani:2021oit} systematically introduced all renormalizable interaction types: 16 in the $Z_2$-even mediator framework and 7 in the $Z_2$-odd mediator framework. These models are constructed from different spin and interaction combinations of the DM and MED particles, with DM stability ensured by the $Z_2$ symmetry. Using a likelihood function incorporating experimental observations, including the Fermi-LAT GCE~\cite{DiMauro:2021raz}, Plank DM relic density~\cite{Planck:2015fie}, PandaX-4T direct detection limits~\cite{PandaX-4T:2021bab}, LEP collider bounds~\cite{ALEPH:2002gap}, and the muon $g-2$ anomaly $\delta a_\mu$~\cite{Muong-2:2021ojo}, we performed a global analysis of all 23 interaction types to exclude parameter spaces inconsistent with observational data.

In $Z_2$-even mediator models, interactions between DM pairs and SM muon pairs are mediated by a $Z_2$-even SM singlet mediator with spin $0$ or $1$. Such muonphilic DM models naturally evade LEP and LHC constraints from mono-photon and mono-jet searches~\cite{Fox:2011fx,Fox:2011pm}, thus allowing for the existence of DM at the electroweak scale. Based on the spin and interaction types of DM and MED, the 16 $Z_2$-even interactions can be categorized into three major classes: (1) fermionic DM ($\chi$) models; (2) scalar DM ($S$) models; (3) vector DM ($X_\mu$) models. When the DM mass ($m_D$) is less than the mediator mass ($M$), $m_D < M$, the observed relic density can only be achieved through the $2\mu$ final state. For $m_D > M$, the process DM+DM $\to$ MED+MED becomes kinematically accessible, producing a $4\mu$ final state. However, explaining the GCE in this case requires a significantly larger annihilation cross-section that typically conflicts with relic density constraints.

$Z_2$-even mediator models benefit from their ability to modulate early-time and present-day DM annihilation rates through $s$-channel resonance enhancement mechanism ($M \approx 2 m_D$), preserving the relic density while enhancing the GCE signal. For example, scalar mediator models ($\mathcal{L}_3$, $\mathcal{L}_9$, $\mathcal{L}_{13}$) can simultaneously satisfy relic density ($\langle\sigma v\rangle \approx 3\times10^{-26}~\text{cm}^3/\text{s}$)~\cite{Planck:2015fie} and the GCE signal ($\langle\sigma v\rangle \approx 4\times10^{-26}~\text{cm}^3/\text{s}$)~\cite{DiMauro:2021raz} requirements by tuning the mediator decay width $\Gamma_{\text{MED}}$, while also explaining the muon $g-2$ anomaly ($\delta a_\mu \approx 2.51\times10^{-9}$)~\cite{Muong-2:2021ojo}. Moreover, the DM-nucleon scattering cross-sections are suppressed to $\sigma_{\text{SI}} \sim 10^{-48}~\text{cm}^2$ due to two-loop processes, evading current direct detection limits.

In contrast, $Z_2$-odd mediator models feature $t$-channel dominated DM annihilation. Since the mediator mass always exceeds the DM mass, the only accessible final state is $2\mu$, precluding the resonant enhancement mechanism available $Z_2$-even mediator models. This makes it challenging to simultaneously explain both the GCE and DM relic abundance. Additionally, the electrically charged nature of the mediator subjects it to LEP mass limits ($M > 103.5~\text{GeV}$)~\cite{ALEPH:2002gap}, significantly constraining the viable parameter space~\cite{Abdughani:2021oit}. Satisfying the relic density requirement necessitates strong couplings ($g_D > 1$), which typically cause DM-nucleon scattering cross-sections to exceed current limits ($\sigma_{\text{SI}} > 10^{-46}~\text{cm}^2$)~\cite{PandaX-4T:2021bab}. In the end, the only $Z_2$-odd mediator model that marginally survives under thermal DM production is the vector DM model $\mathcal{L}_{23}$ ($g_D \bar{\psi} \gamma^\mu P_R \mu X_\mu^\dagger$), but it requires $M > 300~\text{GeV}$ and $g_D > 1.5$ to achieve the observed DM relic density. The parameter space is readily testable by LHC Run-3 data.  Consequently, $Z_2$-odd mediator models possess essentially no viable parameter space under current GCE, relic density, and direct detection constraints, while $Z_2$-even mediator models still retain viable regions in both resonance and non-resonance regimes. 

\begin{table}[h]
\centering
\begin{tabular}{|c|c|c|c|}
\hline
 & Scalar & Fermion & Vector \\
\hline \hline
Dark Matter & $S$ & $\chi$ & $X^{\mu}$ \\
\hline
Mediator & $\phi$ & $\psi$ & $V^{\mu}$ \\
\hline
\end{tabular}
\caption{The dark matter and mediator notations used in this work.}
\label{tab:particle_notataion}
\end{table}

\begin{table}[t]
    \centering
    \begin{tabular}{>{\centering\arraybackslash}p{3cm}|>{\centering\arraybackslash}p{8cm}}
\hline \hline
 Types  & Lagrangian   \\
\hline \hline
\multirow{2}{*}{$\chi$ and $\phi$}  
    & $\mathcal{L}_3=(g_D\bar \chi i \gamma^5 \chi + g_f \bar f f)\phi$   \\

    & $\mathcal{L}_4=(g_D\bar \chi i \gamma^5 \chi + g_f \bar f i \gamma^5 f)\phi$  \\
\hline
 \multirow{1}{*}{$\chi$ and $V_\mu$} 
 & $\mathcal{L}_8=(g_D \bar \chi \gamma^\mu\chi + g_f \bar f \gamma^\mu \gamma^5 f) V_\mu$ \\ 
\hline
\multirow{2}{*}{$S$ and $\phi$} & $\mathcal{L}_9=(M_{D\phi} S^\dagger S + g_f \bar f f) \phi$ \\ 
 & $\mathcal{L}_{10}= (M_{D\phi} S^\dagger S + g_f \bar f i \gamma^5 f) \phi$  \\ 
\hline 
 \multirow{2}{*}{$X_\mu$ and $\phi$} & $\mathcal{L}_{13}=
 (M_{D\phi} X^\mu X_\mu^\dagger + g_f \bar f f) \phi$  \\ 
  & $\mathcal{L}_{14}=(M_{D\phi} X^\mu X_\mu^\dagger + g_f \bar f i \gamma^5 f) \phi$  \\
\hline \hline
\end{tabular}
\caption{The Lagrangian of relevant dark matter simplified models in the $Z_2$-even mediator scenario. Here $g_D$ and $M_{D\phi}$ are relevant couplings between the mediator and a pair of DM particles with dimensionless and dimension-1, respectively. Similarly, $g_f$ is the dimensionless coupling between the mediator and a pair of muons.}
\label{tab:Z2even}
\end{table}

This work focuses on $Z_2$-even mediator models. The notations for relevant DM and MED candidates with various spin assignments are summarized in Table~\ref{tab:particle_notataion}. We specifically explore non-resonance regions of these muonphilic DM models at $3$ TeV muon colliders~\cite{MuonCollider:2022xlm}. The resonance regions are highly fine-tuned and sensitive to slight parameter variations within narrow parameter spaces, requiring specialized analysis that we defer to future work. Based on Table~II and Figure~3 of Ref.~\cite{Abdughani:2021oit}, we select simplified DM models ($\mathcal{L}_3$, $\mathcal{L}_4$, $\mathcal{L}_8$, $\mathcal{L}_9$, $\mathcal{L}_{10}$, $\mathcal{L}_{13}$, $\mathcal{L}_{14}$) with non-resonant parameter regions, collecting them in Table~\ref{tab:Z2even}. Here $g_D$ and $M_{D\phi}$ are relevant couplings between the mediator and a pair of DM particles with dimensionless and dimension-1, respectively. Similarly, $g_f$ is the dimensionless coupling between the mediator and a pair of muons. Detailed expressions for DM annihilation cross-sections and DM-nucleon scattering cross-sections are provided in our previous work~\cite{Abdughani:2021oit}.

\subsection{The predicted parameter space}
\label{sec:parameter}

\begin{figure}[h]
\centering 
\includegraphics[width=0.48\textwidth]{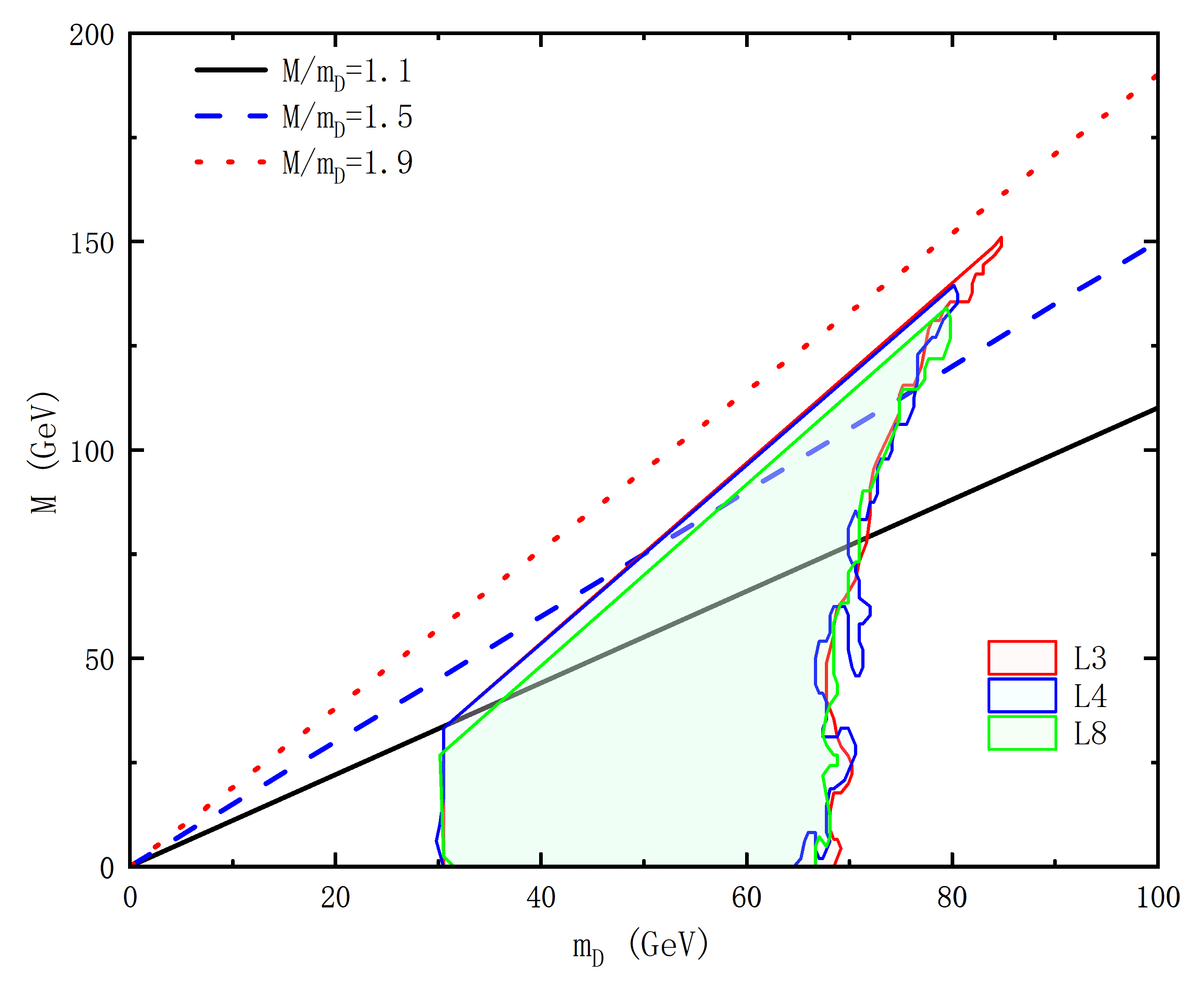}
\includegraphics[width=0.48\textwidth]{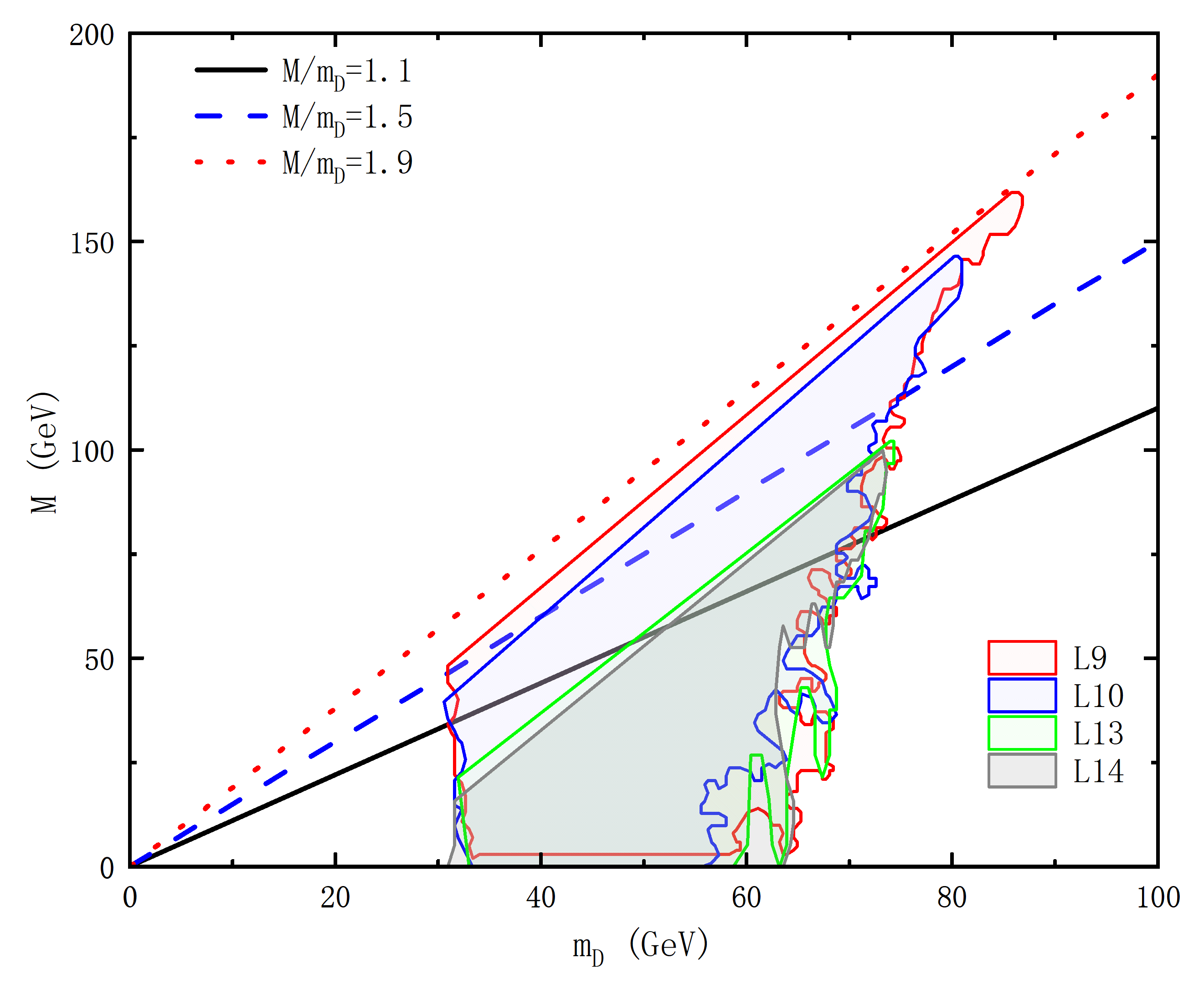}
\caption{The parameter space for relevant simplified muonphilic DM models with $M < 2m_D$, based on Ref.~\cite{Abdughani:2021oit}, is shown here. The left panel depicts $\mathcal{L}_3$, $\mathcal{L}_4$ and $\mathcal{L}_8$, while the right panel covers $\mathcal{L}_9$, $\mathcal{L}_{10}$, $\mathcal{L}_{13}$ and $\mathcal{L}_{14}$. For comparison, three benchmark ratios of mediator to DM mass are displayed: $M/m_D = 1.1$ (black-solid), $M/m_D = 1.5$ (blue-dashed) and $M/m_D = 1.9$ (red-dotted). } 
\label{fig:parameter_space}
\end{figure}

According to global fits in Ref.~\cite{Abdughani:2021oit}, only the simplified muonphilic DM models, $\mathcal{L}_3$, $\mathcal{L}_4$, $\mathcal{L}_8$, $\mathcal{L}_9$, $\mathcal{L}_{10}$, $\mathcal{L}_{13}$, and $\mathcal{L}_{14}$, exhibit allowed parameter space outside resonance regions\footnote{Here we do not consider the $\mathcal{L}_6$ case, as it only permits very light mediators outside the resonance region.}. The parameter space can be further divided into two scenarios: $M > 2m_D$ and $M < 2m_D$. 
For $M > 2m_D$, on-shell mediator production is possible at muon colliders if kinematically accessible. The mediator may subsequently decay into either a DM particle pair or a muon pair, depending on the branching ratios. In contrast, for $M < 2m_D$, the mediator decays exclusively to muon pairs. DM particle pairs can only be produced via off-shell mediators, making their detection more challenging and strongly dependent on the ratio $M/m_D$. We therefore present this ratio for the relevant DM models in Fig.~\ref{fig:parameter_space}, based on results from Ref.~\cite{Abdughani:2021oit}. 
We observe that only models $\mathcal{L}_9$ and $\mathcal{L}_{10}$ feature parameter space near $M/m_D = 1.9$. The parameter spaces of $\mathcal{L}_3$, $\mathcal{L}_4$, and $\mathcal{L}_8$ approach the ratio $M/m_D = 1.5$, while $\mathcal{L}_{13}$ and $\mathcal{L}_{14}$ exhibit relatively small regions near $M/m_D = 1.1$.

Note that for a very light mediator (\(M/m_D \ll 1\)), fixed-target experiments are generally more efficient than muon colliders in searching for its visible on-shell decays. In this regime, aside from \(\text{MED} \to \mu^+\mu^-\) at tree level, other decay channels such as \(\text{MED} \to e^+e^-, \gamma\gamma, \nu\bar{\nu}\) arise at loop level and depend on UV-completion details that lie beyond the scope of this work. We therefore restrict our discussion to mediator masses \(M > 2m_\mu\), ensuring that the decay into a muon pair is kinematically allowed. Such light mediators can be searched for at BaBar and Belle II experiments~\cite{BaBar:2016sci,Belle-II:2024wtd}. Additionally, the NA64\(\mu\) experiment is also sensitive to light muonphilic mediators~\cite{NA64:2024klw,NA64:2024nwj}; however, it primarily targets missing-momentum signatures from invisible decays, treating muon final states as background to be suppressed by veto systems~\cite{Chen:2018vkr}. Consequently, although NA64\(\mu\) has set constraints on muonphilic DM in the \(L_\mu - L_\tau\) model framework, retaining unique sensitivity to invisible decays where the mediator decays predominantly to DM pairs or neutrinos, the scenario studied here does not fall into that category. Instead, our work provides a complementary probe at future high-energy muon colliders.

We focus on three distinct types of searches for muonphilic DM models at muon colliders in this work: 
\begin{itemize}
\item \textbf{Visible on-shell mediator decays} \\ 
The mediator is produced on-shell at muon colliders and subsequently decays to a muon pair: $\mu^+\mu^- \to \gamma , \text{MED} \to \gamma (\mu^+\mu^-)$. The production cross-section depends only on $g_f$:
\begin{equation}
\sigma(\mu^+\mu^- \to \gamma , \text{MED}) \propto g_f^2,
\end{equation}
while the branching ratio $\mathcal{B}(\text{MED} \to \mu^+\mu^-)$ is a function of $g_f$, $g_D$, and $m_D$.
\item \textbf{Invisible on-shell mediator decays} \\
We consider the mono-photon signature in the regime where $M > 2m_D$ with $m_D < 100\;\text{GeV}$, and where the couplings satisfy $g_D \cdot g_f \gtrsim 10^{-2}$ for models $\mathcal{L}_3$, $\mathcal{L}_4$, and $\mathcal{L}_8$, or $M_{D\phi} \cdot g_f \gtrsim 10^{-4}\;\text{TeV}$ for models $\mathcal{L}_9$, $\mathcal{L}_{10}$, $\mathcal{L}_{13}$, and $\mathcal{L}_{14}$.
\item \textbf{Mono-photon with off-shell mediators} \\
We study the parameter region where $1.1 m_D \lesssim M \lesssim 1.9 m_D$ with $m_D < 100\;\text{GeV}$, and with coupling constraints $g_D \cdot g_f \gtrsim 10^{-2}$ for $\mathcal{L}_3$, $\mathcal{L}_4$, $\mathcal{L}_8$, and $M_{D\phi} \cdot g_f \gtrsim 10^{-4}\;\text{TeV}$ for $\mathcal{L}_9$, $\mathcal{L}_{10}$, $\mathcal{L}_{13}$, and $\mathcal{L}_{14}$. 
\end{itemize}

\section{Muonphilic DM search strategies at muon colliders}
\label{sec:search}

Based on the structure of parameter space, we develop four distinct search strategies for muonphilic DM models at muon colliders: Visible on-shell mediator decays, Sec.~\ref{sec:visible_onsell}; Invisible on-shell mediator decays, Sec.~\ref{sec:invisible_onsell}; Mono-photon signatures with off-shell mediators, Sec.~\ref{sec:invisible_offsell}; Vector boson fusion production, Sec.~\ref{sec:VBF}. In the following subsections, we employ the following computational pipeline: \texttt{FeynRules}~\cite{Alloul:2013bka} generates UFO model files for the Lagrangians listed in Table~\ref{tab:Z2even}; \texttt{MadGraph5\_aMC@NLO}~\cite{Alwall:2014hca} produces Monte Carlo events for signal and background hard processes; \texttt{Pythia8}~\cite{Sjostrand:2014zea} handles parton showering and hadronization; and \texttt{Delphes3}~\cite{deFavereau:2013fsa} with a muon collider template performs fast detector simulation. We have also performed a simplified analysis of collider constraints from LEP and the LHC, with the corresponding details provided in Appendix~\ref{app:collider_constraints}. These constraints, however, are considerably weaker than those obtained at future muon colliders, which are the primary focus of this work. 

\subsection{Visible on-shell mediator decays} 
\label{sec:visible_onsell}

Figure~3 in Ref.~\cite{Abdughani:2021oit} indicates allowed parameter spaces exist outside resonance regions for models \( \mathcal{L}_3, \mathcal{L}_4, \mathcal{L}_8, \mathcal{L}_9, \mathcal{L}_{10}, \mathcal{L}_{13} \), and \( \mathcal{L}_{14} \). However, when considering on-shell mediator production through the visible decay channel \( \mu^+\mu^- \to\gamma \text{MED} \) followed by \( \text{MED} \to \mu^+\mu^- \), only three interaction types for the mediators and a pair of muons emerge: scalar (\( \mathcal{L}_3, \mathcal{L}_9, \mathcal{L}_{13} \)), pseudoscalar (\( \mathcal{L}_4, \mathcal{L}_{10}, \mathcal{L}_{14} \)), and axial-vector (\( \mathcal{L}_8 \)). For the proposed 3 TeV muon collider, we focus on mediator masses \( M < 1.5 \ \text{TeV} \), covering all upper bounds of mediator masses in these models except \( \mathcal{L}_9 \) and \( \mathcal{L}_{10} \), where \( M \) may reach $3$ TeV. Our search strategies remain extensible to heavier mediators for higher-energy muon colliders. 

\begin{table}[htbp]
\centering
\footnotesize  
\begin{tabular}{|>{\centering\arraybackslash}p{1.2cm}|>{\centering\arraybackslash}p{4.8cm}|>{\centering\arraybackslash}p{2.1cm}|*{4}{>{\centering\arraybackslash}p{1.8cm}|}}
\hline
\multicolumn{2}{|c|}{\textbf{Cut description}} & \textbf{background} & \textbf{signal1} & \textbf{signal2} & \textbf{signal3} & \textbf{signal4}  \\
\hline
\multicolumn{2}{|c|}{\textbf{Cross-section [fb]}} & 179 & $3.628$ & $4.935$ & $5.471$ & $6.366$  \\
\hline
\textbf{Cut-1} & $N(\gamma),~N(\mu^+),~N(\mu^-) > 0$, \newline
$E(\gamma),~{p_T(\mu^{\pm})} > 20\,\mathrm{GeV}$,  \newline
$|\eta(\gamma)|,~|\eta(\mu^{\pm})| < 2.5$ & {\footnotesize 0.82} & {\footnotesize 0.78} & {\footnotesize 0.85} & {\footnotesize 0.84} & {\footnotesize 0.83}  \\
\hline
\textbf{Cut-2} & $E(\gamma) > 150\,\mathrm{GeV}$, \newline
$|\eta(\gamma)| < 1.8$ & {\footnotesize 0.19} & {\footnotesize 0.54} & {\footnotesize 0.65} & {\footnotesize 0.62} & {\footnotesize 0.61}  \\
\hline
\textbf{Cut-3} & $p_T(\mu^{\pm}) > 150\,\mathrm{GeV}$, \newline
$|\eta(\mu^{\pm})| < 1.6$ & {\footnotesize 0.08} & {\footnotesize 0.41} & {\footnotesize 0.53} & {\footnotesize 0.51} & {\footnotesize 0.51}  \\
\hline
\textbf{Cut-4} & $\Delta R (\mu^+\mu^-) < 3.0$ & {\footnotesize 0.01} & {\footnotesize 0.24} & {\footnotesize 0.40} & {\footnotesize 0.46} & {\footnotesize 0.45}  \\
\hline
\multirow{4}{*}{\textbf{Cut-5}} & \multirow{4}{*}{$|M(\mu^+\mu^-) - M| < 0.15M$} & {\footnotesize $2 \times 10^{-5}$} & {\footnotesize 0.23} & {\footnotesize /} & {\footnotesize /} & {\footnotesize /}  \\
\cline{3-7}
&& {\footnotesize $3.7 \times 10^{-4}$} & {\footnotesize /} & {\footnotesize 0.38} & {\footnotesize /} & {\footnotesize /}  \\
\cline{3-7}
&& {\footnotesize $1.34 \times 10^{-3}$} & {\footnotesize /} & {\footnotesize /} & {\footnotesize 0.44} & {\footnotesize /}  \\
\cline{3-7}
&& {\footnotesize $2.15 \times 10^{-3}$} & {\footnotesize /} & {\footnotesize /} & {\footnotesize /} & {\footnotesize 0.45}  \\
\hline
\end{tabular}
\caption{The cut-flow table for visible on-shell mediator decays of the $\mathcal{L}_3$ model and its relevant background with cumulative efficiencies from \textbf{Cut-1} to \textbf{Cut-5}. The four benchmark points are signal-1 ($M = 50$ GeV), signal-2 ($M = 500$ GeV), signal-3 ($M = 900$ GeV), signal-4 ($M = 1300$ GeV). }

\label{tab:L3_decay}
\end{table}

\begin{table}[htbp]
\centering
\footnotesize  
\begin{tabular}{|>{\centering\arraybackslash}p{1.2cm}|>{\centering\arraybackslash}p{4.8cm}|>{\centering\arraybackslash}p{2.1cm}|*{4}{>{\centering\arraybackslash}p{1.8cm}|}}
\hline
\multicolumn{2}{|c|}{\textbf{Cut description}} & \textbf{background} & \textbf{signal1} & \textbf{signal2} & \textbf{signal3} & \textbf{signal4}  \\
\hline
\multicolumn{2}{|c|}{\textbf{Cross-section [fb]}} & 179 & 7.99 & 10.08 & 11.43 & 13.89  \\
\hline
\textbf{Cut-1} & $N(\gamma),~N(\mu^+),~N(\mu^-) > 0$, \newline
$E(\gamma),~{p_T(\mu^{\pm})} > 20\,\mathrm{GeV}$,  \newline
$|\eta(\gamma)|,~|\eta(\mu^{\pm})| < 2.5$ & {\footnotesize 0.82} & {\footnotesize 0.77} & {\footnotesize 0.85} & {\footnotesize 0.84} & {\footnotesize 0.83}  \\
\hline
\textbf{Cut-2} & $E(\gamma) > 150\,\mathrm{GeV}$, \newline
$|\eta(\gamma)| < 1.8$ & {\footnotesize 0.19} & {\footnotesize 0.53} & {\footnotesize 0.60} & {\footnotesize 0.59} & {\footnotesize 0.58}  \\
\hline
\textbf{Cut-3} & $p_T(\mu^{\pm}) > 150\,\mathrm{GeV}$, \newline
$|\eta(\mu^{\pm})| < 1.6$ & {\footnotesize 0.08} & {\footnotesize 0.40} & {\footnotesize 0.47} & {\footnotesize 0.46} & {\footnotesize 0.48}  \\
\hline
\textbf{Cut-4} & $\Delta R (\mu^+\mu^-) < 3.0$ & {\footnotesize 0.01} & {\footnotesize 0.20} & {\footnotesize 0.32} & {\footnotesize 0.40} & {\footnotesize 0.41}  \\
\hline
\multirow{4}{*}{\textbf{Cut-5}} & \multirow{4}{*}{$|M(\mu^+\mu^-) - M| < 0.15M$} & {\footnotesize $2 \times 10^{-5}$} & {\footnotesize 0.19} & {\footnotesize /} & {\footnotesize /} & {\footnotesize /}  \\
\cline{3-7}
&& {\footnotesize $3.7 \times 10^{-4}$} & {\footnotesize /} & {\footnotesize 0.31} & {\footnotesize /} & {\footnotesize /}  \\
\cline{3-7}
&& {\footnotesize $1.34 \times 10^{-3}$} & {\footnotesize /} & {\footnotesize /} & {\footnotesize 0.39} & {\footnotesize /}  \\
\cline{3-7}
&& {\footnotesize $2.15 \times 10^{-3}$} & {\footnotesize /} & {\footnotesize /} & {\footnotesize /} & {\footnotesize 0.40}  \\
\hline
\end{tabular}
\caption{Similar to Table~\ref{tab:L3_decay}, but for the $\mathcal{L}_8$ model. }
\label{tab:L8_decay}
\end{table}

\begin{figure}[h]
\centering 
\includegraphics[width=0.45\textwidth]{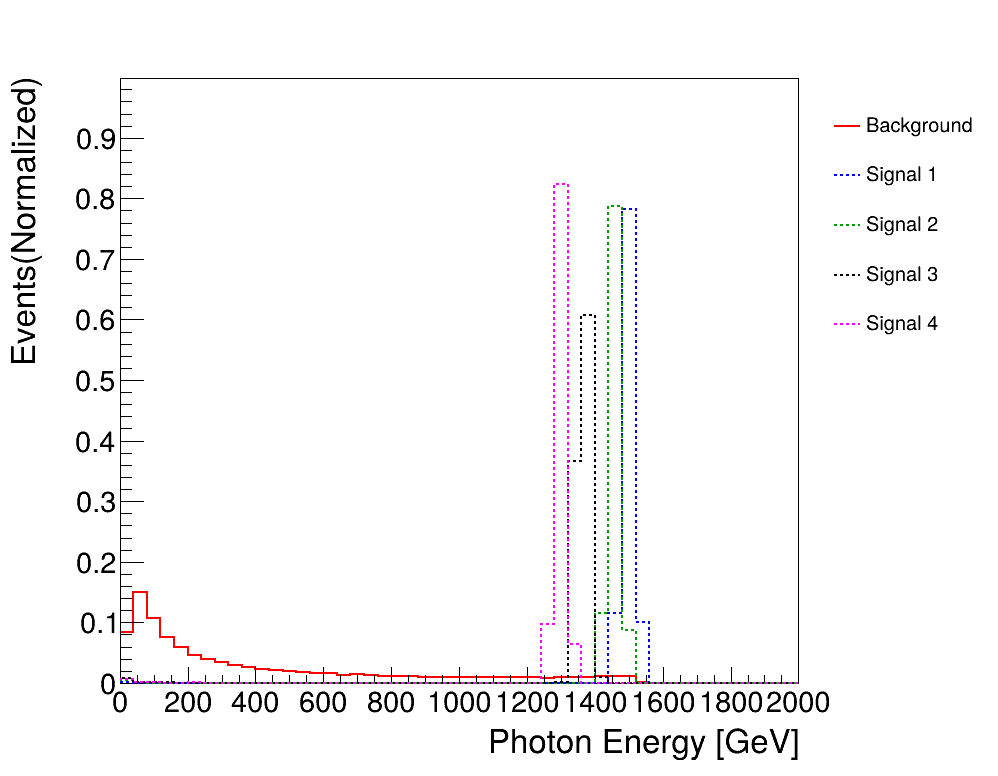} 
\includegraphics[width=0.45\textwidth]{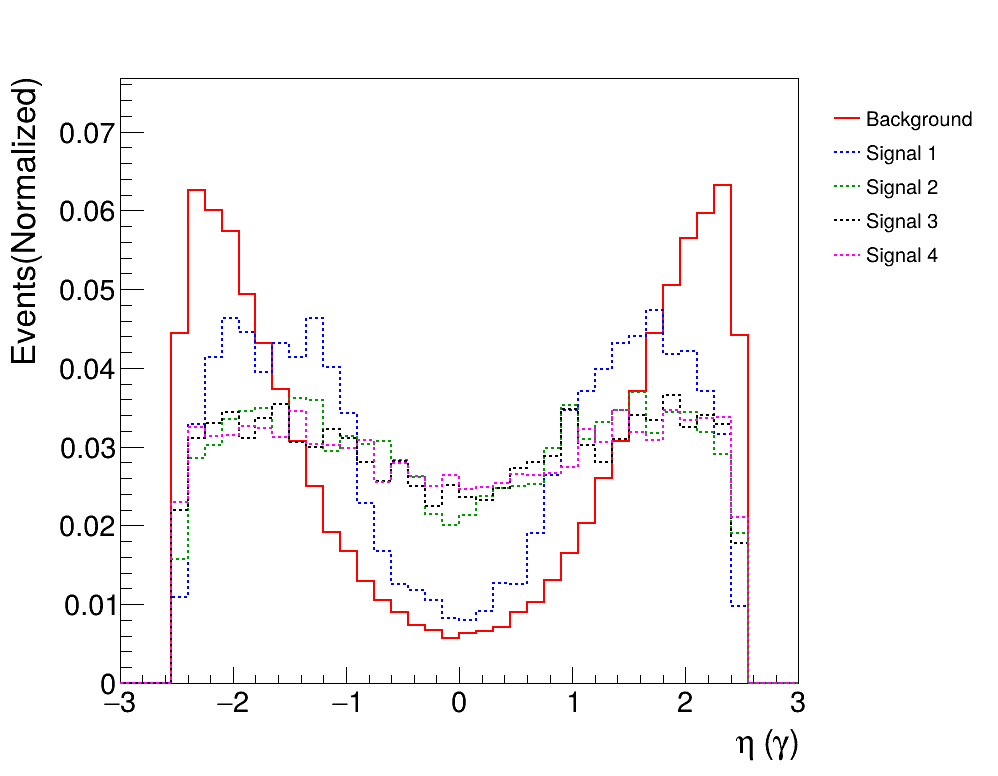}
\includegraphics[width=0.45\textwidth]{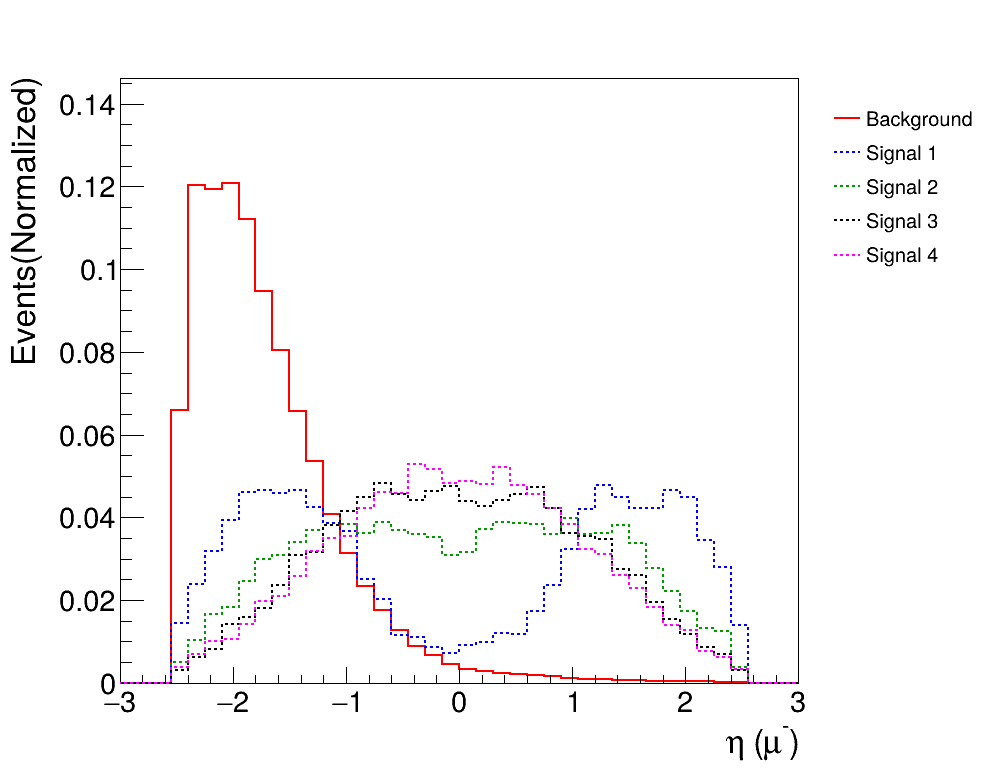}
\includegraphics[width=0.45\textwidth]{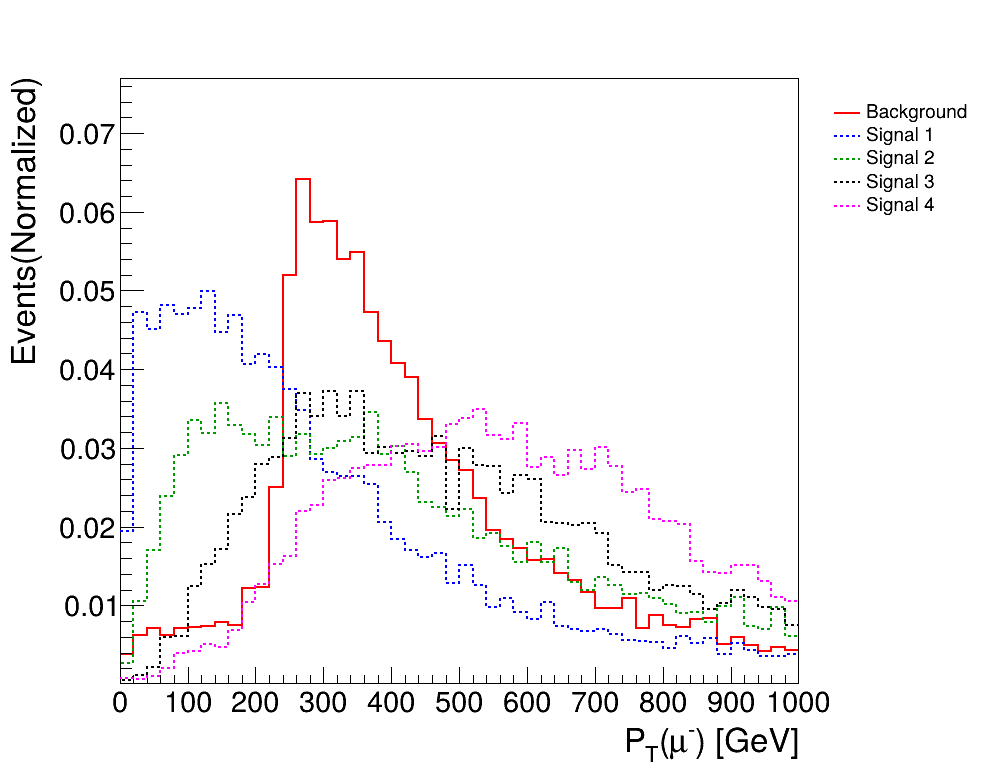}
\includegraphics[width=0.45\textwidth]{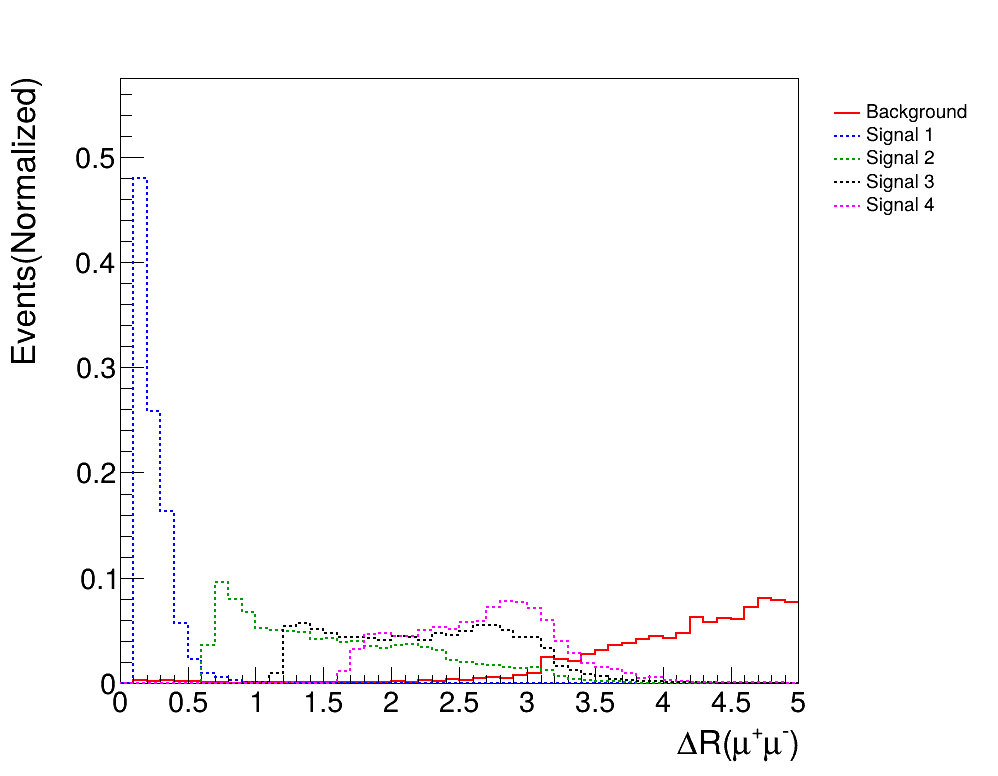}
\includegraphics[width=0.45\textwidth]{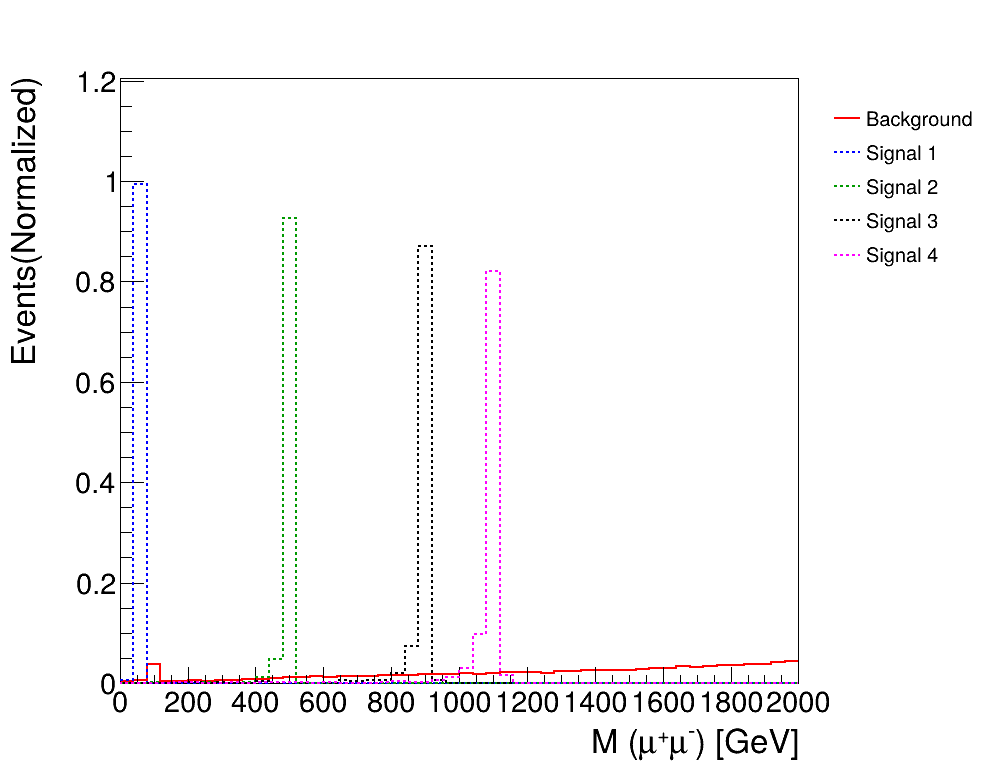}
\caption{
Visible on-shell mediator decays of the $\mathcal{L}_3$ model, representative kinematic distributions include photon energy $E({\gamma})$ (top-left), photon pseudorapidity $\eta({\gamma})$ (top-right), $\mu^-$ pseudorapidity $\eta(\mu^-)$ (middle-left), $\mu^-$ transverse momentum $p_T(\mu^-)$ (middle-right), $\Delta R$ of a muon pair (bottom-left), and invariant mass $M(\mu^+\mu^-)$ of a muon pair (bottom-right) for signal-1 ($M = 50$ GeV, blue-dotted line), signal-2 ($M = 500$ GeV, green-dotted line), signal-3 ($M = 900$ GeV, black-dotted line), signal-4 ($M = 1300$ GeV, purple-dotted line) and background (red-solid line). } 
\label{fig:L3_decay}
\end{figure}

In this subsection, we demonstrate our methodology using the \( \mathcal{L}_3 \) and \( \mathcal{L}_8 \) models, presenting benchmark points (BPs), kinematic distributions, and event selection criteria. The analysis for other models is presented in Sec.~\ref{sec:exclusion}. Note that for \( M > 2m_D \), kinematically allowed invisible mediator decays lead to model-dependent variations in the \( \text{MED} \to \mu^+\mu^- \) branching ratios. Therefore, we set the mass ratio \(M/m_D = 2.5\) and fix the coupling constant ratio for each model, thereby ensuring that the branching ratio of \(\text{MED} \to \mu^+\mu^-\) is no less than $99\%$. Four benchmark mediator masses are considered in this study: (1) Signal-1: $M = 50$ GeV; (2) Signal-2: $M = 500$ GeV; (3) Signal-3: $M = 900$ GeV; (4) Signal-4: $M = 1300$ GeV. For the \(\mathcal{L}_3\) model, the benchmark coupling parameters are set to \(g_f = 0.08\) and \(g_D = 0.01\), corresponding to a coupling constant ratio of \(g_D/g_f = 1/8\). For the \(\mathcal{L}_8\) model, the benchmark coupling parameters are set to \(g_f = 0.09\) and \(g_D = 0.01\), with the corresponding coupling constant ratio being \(g_D/g_f = 1/9\). 
For the signal process $\mu^+\mu^-\to\gamma\text{MED}$ with $\text{MED}\to\mu^+\mu^-$, the irreducible SM background process is $\mu^+\mu^-\to\mu^+\mu^-\gamma$\footnote{The relevant VBF background processes including $\mu^+ \mu^- \to \mu^+ \nu_{\mu}\gamma W^-$, with $W^- \to \mu^- \bar{\nu_{\mu}}$,
and $\mu^+ \mu^- \to \mu^- \bar{\nu}_{\mu}\gamma W^+$, with $W^+ \to \mu^+ \nu_{\mu}$ are also considered. However, these backgrounds have negligible impact on the final exclusion results, and their contributions can be safely neglected.}. The corresponding production cross-sections for the four BPs and the SM background are listed in the second row of Tables~\ref{tab:L3_decay} and~\ref{tab:L8_decay}. Representative kinematic distributions for signal and background processes are shown in Fig.~\ref{fig:L3_decay}.

We find that the peak of the background photon energy \(E(\gamma)\) appears at $50$ GeV, whereas the peaks for the four signal BPs occur at higher energies and decrease with increasing  mediator mass, as shown in the top-left panel of Fig.~\ref{fig:L3_decay}. The photon pseudorapidity \(\eta(\gamma)\) for the background and the light mediator case (signal-1) displays a pronounced forward-backward distribution. However, this behavior does not appear for heavier mediators, resulting in a flatter \(\eta(\gamma)\) distribution (top-right panel). A similar forward-backward distribution is also observed in the \(\eta(\mu^{\pm})\) variable for both the background and signal-1 (middle-left panel). The transverse momentum distribution \(p_T(\mu^-)\) peaks around $350$ GeV for the background. For the signal processes, a heavier mediator leads to a harder \(p_T(\mu^-)\) spectrum (middle-right panel). The opening angle distribution \(\Delta R (\mu^+\mu^-)\) provides a key distinction: a light mediator (signal-1) is produced with a large boost, leading to a smaller \(\Delta R (\mu^+\mu^-)\) compared to the background. The background exhibits a significantly larger \(\Delta R (\mu^+\mu^-)\) than all four signal BPs (bottom-left panel). Finally, the dimuon invariant mass distributions \(M(\mu^+\mu^-)\) for the signals peak at their respective mediator masses, while the background distribution is broad and featureless (bottom-right panel).

Based on the above analysis of kinematic distributions, the following event selection criteria for signal and background events are adopted:
\begin{itemize} 
\item Cut-1 (basic cuts): $N(\gamma),~N(\mu^+),~N(\mu^-) > 0$, with leading photon, a pair of muons satisfying $E(\gamma),~{p_T(\mu^{\pm})} > 20\,\mathrm{GeV}$, and $|\eta(\gamma)|,~|\eta(\mu^{\pm})| < 2.5$;   

\item Cut-2: $E(\gamma) > 150\,\mathrm{GeV}$, and
$|\eta(\gamma)| < 1.8$; 

\item Cut-3: $p_T(\mu^{\pm}) > 150\,\mathrm{GeV}$, and $|\eta(\mu^{\pm})| < 1.6$; 

\item Cut-4: $\Delta R (\mu^+\mu^-) < 3.0$; 

\item Cut-5: $|M(\mu^+\mu^-) - M| < 0.15M$. 
\end{itemize}
After applying these selection criteria, the signal and background efficiencies are calculated. Tables~\ref{tab:L3_decay} and~\ref{tab:L8_decay} summarize the selection efficiencies for the four BPs in the \(\mathcal{L}_3\) and \(\mathcal{L}_8\) models, respectively. The final background efficiency is suppressed to $\mathcal{O}(10^{-3})$, while the signal efficiencies remain above $19\%$ and can exceed $40\%$ for larger mediator masses.


\subsection{Invisible on-shell mediator decays} 
\label{sec:invisible_onsell}

In this context, global fitting results from Fig.~3 of Ref.~\cite{Abdughani:2021oit} show \( M > 2m_D \) with \( m_D < 100 \ \text{GeV} \). With the exception of models \( \mathcal{L}_9 \) and \( \mathcal{L}_{10} \), where the mediator masses \( M \) can reach upper bounds of \( 3 \ \text{TeV} \), all other models (\( \mathcal{L}_3 \), \( \mathcal{L}_4 \), \( \mathcal{L}_8 \), \( \mathcal{L}_{13} \), \( \mathcal{L}_{14} \)) exhibit upper bounds of $M$ below \( 1.5 \ \text{TeV} \). Given our focus on the \( 3 \ \text{TeV} \) muon collider proposal, we restrict our study to mediator masses \( M\lesssim 1.5 \ \text{TeV} \). Notably, on-shell mediator production via muon pairs occurs through scalar-type couplings (\( \mathcal{L}_3 \), \( \mathcal{L}_9 \), \( \mathcal{L}_{13} \)), pseudoscalar-type couplings (\( \mathcal{L}_4 \), \( \mathcal{L}_{10} \), \( \mathcal{L}_{14} \)), or axial-vector-type couplings (\( \mathcal{L}_8 \)). However, branching ratios for mediator decays to DM particle pairs vary across models due to their distinct interaction types.

\begin{table}[htbp]
\centering
\footnotesize  
\begin{tabular}{|>{\centering\arraybackslash}p{6cm}|>{\centering\arraybackslash}p{5cm}|>{\centering\arraybackslash}p{2.2cm}|*{4}{>{\centering\arraybackslash}p{1.5cm}|}}
\hline
\multicolumn{2}{|c|}{\textbf{Cut description}} & \textbf{background} & \textbf{signal1} & \textbf{signal2} & \textbf{signal3} & \textbf{signal4}  \\
\hline
\multicolumn{2}{|c|}{\textbf{Cross-section [fb]}} & 2980 & $8.21\times 10^{-2}$ & $8.46\times 10^{-2}$ & $9.10\times 10^{-2}$ & 0.10 \\
\hline
\multicolumn{1}{|c|}{\multirow{3}{*}{\textbf{Cut-1}}} & $N(\gamma)>0$ & {\footnotesize 0.92} & {\footnotesize 0.91} & {\footnotesize 0.91} & {\footnotesize 0.91} & {\footnotesize 0.91}  \\
\cline{2-7}
\multicolumn{1}{|c|}{} & $E(\gamma) > 100\,\mathrm{GeV} \ \&\ |\eta(\gamma)| < 2.5$ & {\footnotesize 0.35} & {\footnotesize 0.91} & {\footnotesize 0.91} & {\footnotesize 0.91} & {\footnotesize 0.91}  \\
\cline{2-7}
\multicolumn{1}{|c|}{} & $ {\:/\!\!\!\! E}_T > 40\,\mathrm{GeV} $ & {\footnotesize 0.30} & {\footnotesize 0.91} & {\footnotesize 0.91} & {\footnotesize 0.91} & {\footnotesize 0.91}  \\
\hline
\multicolumn{1}{|c|}{\multirow{1}{*}{\textbf{Cut-2}}} & $E(\gamma) > 1200\,\mathrm{GeV} \ $ & {\footnotesize $7.4\times 10^{-3}$} & {\footnotesize 0.91} & {\footnotesize 0.91} & {\footnotesize 0.91} & {\footnotesize 0.83}  \\
\hline
\multicolumn{1}{|c|}{\multirow{1}{*}{\textbf{Cut-3}}} & $P_T(\gamma)/{\:/\!\!\!\! E}>0.4$ & {\footnotesize $1.7\times 10^{-3}$} & {\footnotesize 0.57} & {\footnotesize 0.56} & {\footnotesize 0.51} & {\footnotesize 0.38}  \\
\hline
\multicolumn{1}{|c|}{\multirow{1}{*}{\textbf{Cut-4}}} & $M_T > 1600\,\mathrm{GeV}$ & {\footnotesize $1.0\times 10^{-3}$} & {\footnotesize 0.45} & {\footnotesize 0.45} & {\footnotesize 0.42} & {\footnotesize 0.32}  \\
\hline
\end{tabular}
\caption{The cut-flow table for the invisible  on-shell mediator decay scenario ($M/m_D=2.5$) of the $\mathcal{L}_3$ model and its relevant background with cumulative efficiencies from \textbf{Cut-1} to \textbf{Cut-4}. The four benchmark points are signal-1 ($M = 50$ GeV), signal-2 ($M = 500$ GeV), signal-3 ($M = 900$ GeV), signal-4 ($M = 1300$ GeV). }
\label{tab:L3_onshell}
\end{table}

\begin{figure}[h]
\centering 
\includegraphics[width=0.48\textwidth]{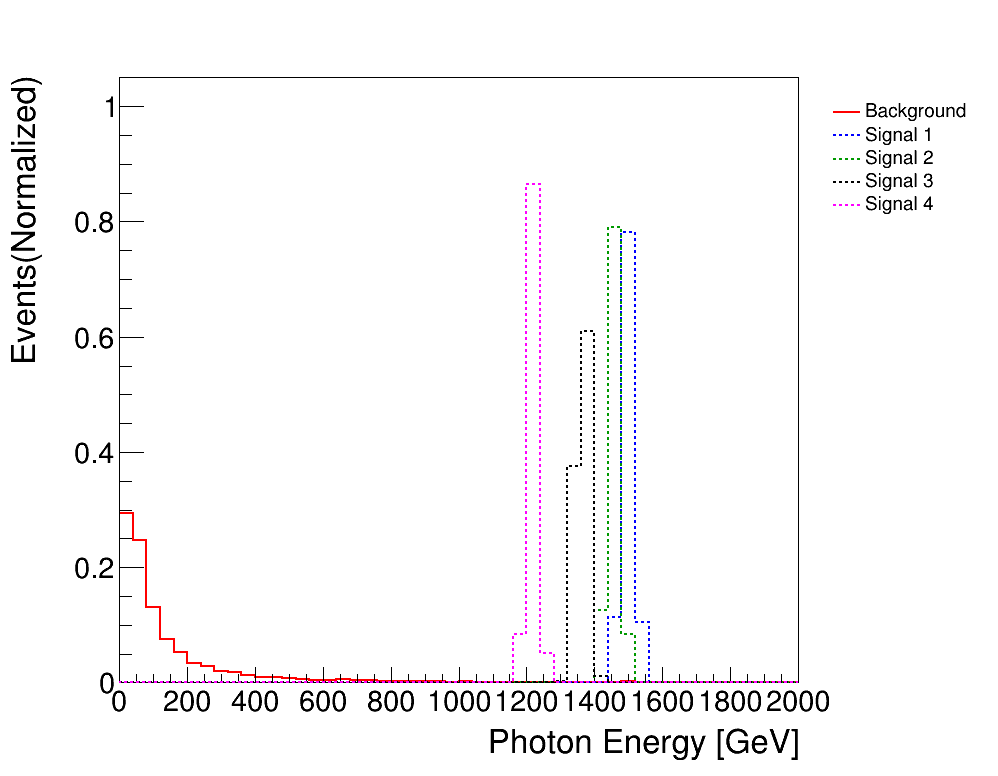} 
\includegraphics[width=0.48\textwidth]{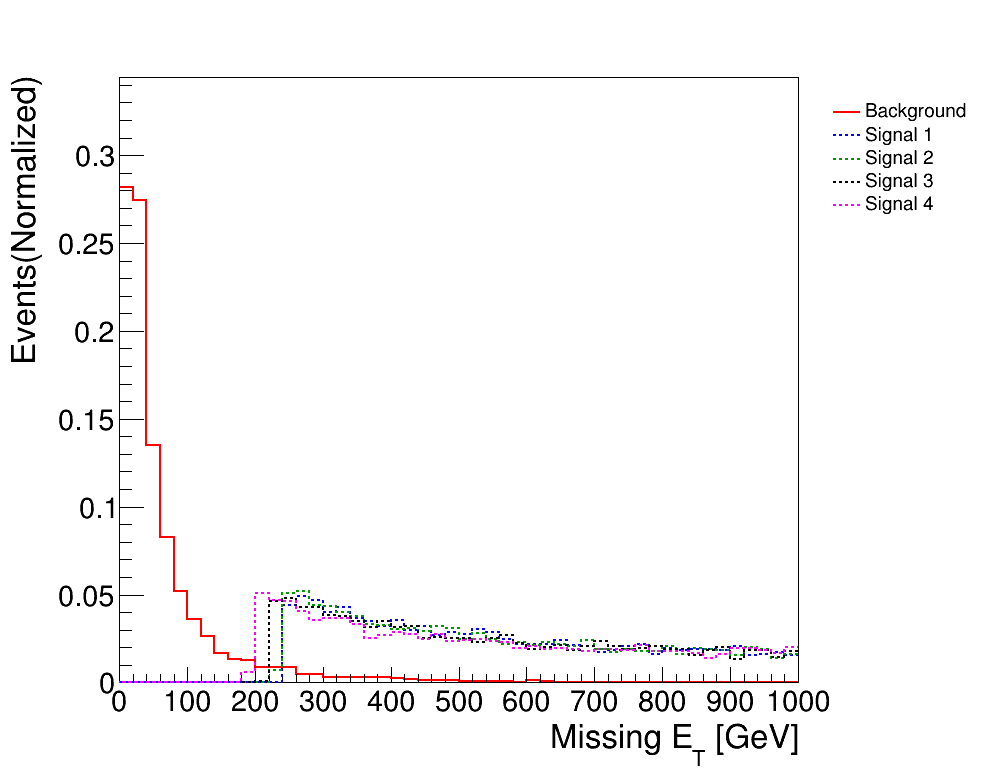}
\includegraphics[width=0.48\textwidth]{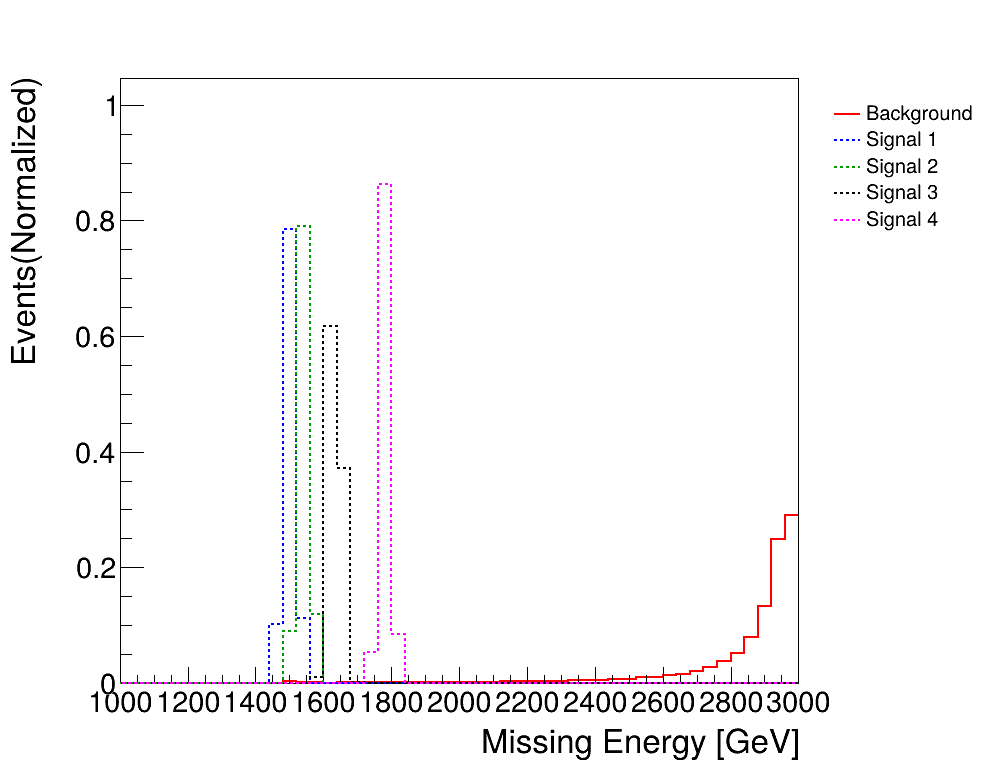}
\includegraphics[width=0.48\textwidth]{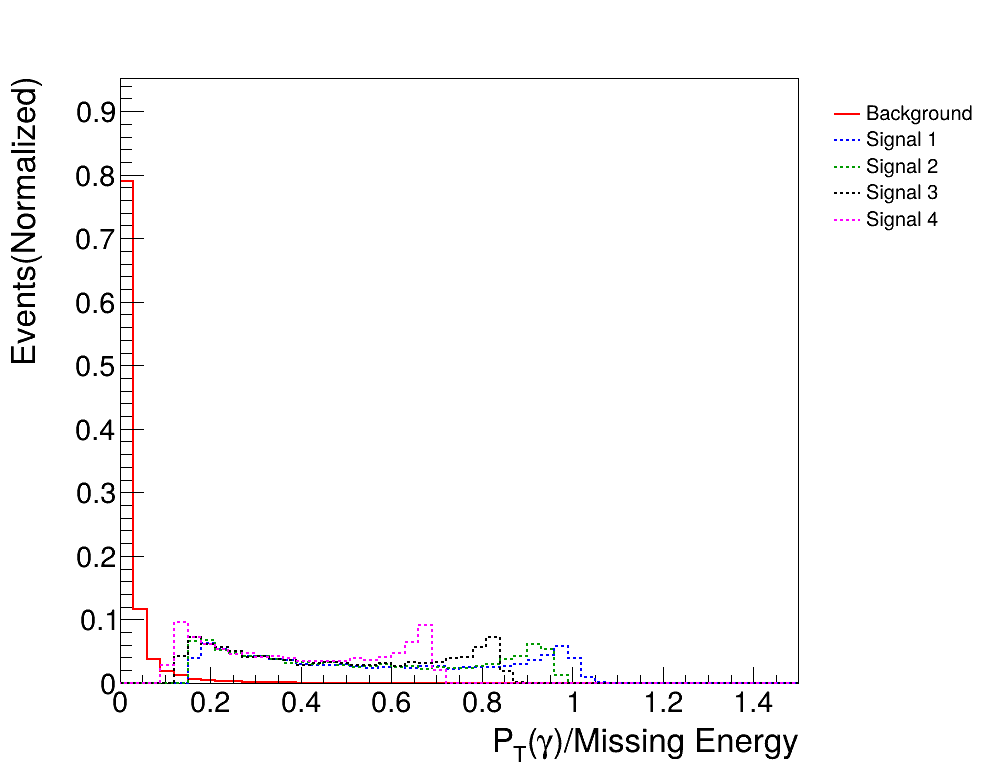}
\caption{In the invisible on-shell mediator decay scenario ($M/m_D=2.5$) of the $\mathcal{L}_3$ model, representative kinematic distributions include photon energy $E({\gamma})$ (top-left), missing transverse energy $ {\:/\!\!\!\! E}_T$ (top-right), missing energy ${\:/\!\!\!\! E}$ (bottom-left), and the ratio of $P_T(\gamma)/{\:/\!\!\!\! E}$ (bottom-right) for signal-1 ($M = 50$ GeV, blue-dotted line), signal-2 ($M = 500$ GeV, green-dotted line), signal-3 ($M = 900$ GeV, black-dotted line), signal-4 ($M = 1300$ GeV, purple-dotted line) and background (red-solid line). } 
\label{fig:L3_onshell}
\end{figure}

To demonstrate our search methodology, we use the \( \mathcal{L}_3 \) model as a representative example, presenting BPs, kinematic distributions, and event selection criteria. The analysis is subsequently extended to other models. We select four mediator masses: Signal-1 ($M = 50$ GeV), signal-2 ($M = 500$ GeV), signal-3 ($M = 900$ GeV), and signal-4 ($M = 1300$ GeV), with a fixed mediator-to-DM mass ratio $M/m_D=2.5$, and choose $g_D = 0.15$ and $g_f = 0.01$ as our benchmark couplings. The signal process is $\mu^+\mu^-\to\phi\gamma$, followed by the on-shell mediator decay to a pair of DM particles ($\phi\to\chi\bar{\chi}$). The irreducible SM background process is $\mu^+\mu^-\to\nu\bar{\nu}\gamma$\footnote{Similarly, we have considered the relevant VBF process $\mu^+ \mu^- \to \nu_{\mu}\bar{\nu_{\mu}}\gamma Z$, with $Z \to \nu_{\ell}\bar{\nu_{\ell}}$, but its contribution is highly suppressed and can be neglected in this analysis.}. The corresponding production cross-sections for the four BPs and the SM background are listed in the second row of Table~\ref{tab:L3_onshell}. The representative kinematic distributions for signal and background processes are shown in Fig.~\ref{fig:L3_onshell}.

For on-shell mediator production with an initial state radiation (ISR) photon, the photon energy $E(\gamma)$ exhibits a monochromatic peak, while the background $E(\gamma)$ distribution decreases smoothly, as shown in the top-left panel of Fig.~\ref{fig:L3_onshell}. The transverse missing energy ${\:/\!\!\!\! E}_T$ is more energetic for signals than for background (top-right panel). Conversely, the missing energy $ {\:/\!\!\!\! E} $ distribution shows the opposite trend to $E(\gamma)$ distribution (bottom-left panel). Finally, since the photon transverse momentum and missing energy distributions differ significantly between signals and background, we examine the ratio $P_T(\gamma)/{\:/\!\!\!\! E}$ distribution (bottom-right panel) to enhance discrimination. 

\begin{table}[htbp]
\footnotesize  
\begin{tabular}{|>{\centering\arraybackslash}p{6cm}|>{\centering\arraybackslash}p{5cm}|>{\centering\arraybackslash}p{2.2cm}|*{4}{>{\centering\arraybackslash}p{1.5cm}|}}
\hline
\multicolumn{2}{|c|}{\textbf{Cut description}} & \textbf{background} & \textbf{signal1} & \textbf{signal2} & \textbf{signal3} & \textbf{signal4}  \\
\hline
\multicolumn{2}{|c|}{\textbf{Cross-section [fb]}} & 2980 & 0.132 & 0.137 & 0.152 & 0.183 \\
\hline
\multicolumn{1}{|c|}{\multirow{3}{*}{\textbf{Cut-1}}} & $N(\gamma)>0$ & {\footnotesize 0.92} & {\footnotesize 0.91} & {\footnotesize 0.91} & {\footnotesize 0.91} & {\footnotesize 0.91}  \\
\cline{2-7}
\multicolumn{1}{|c|}{} & $E(\gamma) > 100\,\mathrm{GeV} \ \&\ |\eta(\gamma)| < 2.5$ & {\footnotesize 0.35} & {\footnotesize 0.91} & {\footnotesize 0.91} & {\footnotesize 0.91} & {\footnotesize 0.91}  \\
\cline{2-7}
\multicolumn{1}{|c|}{} & $ {\:/\!\!\!\! E}_T > 40\,\mathrm{GeV} $ & {\footnotesize 0.30} & {\footnotesize 0.91} & {\footnotesize 0.91} & {\footnotesize 0.91} & {\footnotesize 0.91}  \\
\hline
\multicolumn{1}{|c|}{\multirow{1}{*}{\textbf{Cut-2}}} & $E(\gamma) > 1200\,\mathrm{GeV} \ $ & {\footnotesize $7.4\times 10^{-3}$} & {\footnotesize 0.91} & {\footnotesize 0.91} & {\footnotesize 0.91} & {\footnotesize 0.83}  \\
\hline
\multicolumn{1}{|c|}{\multirow{1}{*}{\textbf{Cut-3}}} & $P_T(\gamma)/{\:/\!\!\!\! E}>0.4$ & {\footnotesize $1.7\times 10^{-3}$} & {\footnotesize 0.50} & {\footnotesize 0.49} & {\footnotesize 0.47} & {\footnotesize 0.34}  \\
\hline
\multicolumn{1}{|c|}{\multirow{1}{*}{\textbf{Cut-4}}} & $M_T > 1600\,\mathrm{GeV}$ & {\footnotesize $1.0\times 10^{-3}$} & {\footnotesize 0.37} & {\footnotesize 0.37} & {\footnotesize 0.36} & {\footnotesize 0.29}  \\
\hline
\end{tabular}
\caption{Similar to Table~\ref{tab:L3_onshell}, but for the $\mathcal{L}_8$ model. }
\label{tab:L8_onshell}
\end{table}

Based on these kinematic distributions, we implement the following event selections to signal and background events: 
\begin{itemize} 
\item Cut-1 (basic cuts): $N(\gamma)>0$, with the leading photon satisfying $E(\gamma) > 100\,\mathrm{GeV}$ and $|\eta(\gamma)| < 2.5$, plus ${\:/\!\!\!\! E}_T > 40\,\mathrm{GeV}$; 
\item Cut-2: $E(\gamma) > 1200\,\mathrm{GeV}$;  
\item Cut-3: $P_T(\gamma)/{\:/\!\!\!\! E} > 0.4$;  
\item Cut-4: $M_T > 1600$ GeV, where the transverse mass for the mono-photon plus ${\:/\!\!\!\! E}_T$ is defined as 
\begin{equation}
M_T = \sqrt{2 P_T(\gamma) \cdot \not{E}_T \cdot (1 - \cos\Delta\phi)},
\end{equation}
with $\Delta\phi$ being the azimuthal angle between the leading photon and ${\:/\!\!\!\! E}_T$. 
\end{itemize} 
Here, the final $M_T$ selection is used to further distinguish signal and background events. Table~\ref{tab:L3_onshell} summarizes the selection efficiencies for four signal BPs in the $\mathcal{L}_3$ model and background. As efficiency table for the $\mathcal{L}_4$ model closely resemble to the $\mathcal{L}_3$ model's, we show only the $\mathcal{L}_8$ model in Table~\ref{tab:L8_onshell}. Here we adopt the benchmark couplings of \(g_D = 1.2\) and \(g_f = 0.1\) in the $\mathcal{L}_8$ model. 

\subsection{Mono-photon signature with off-shell mediator}
\label{sec:invisible_offsell}

In regions of parameter space where \( M < 2m_D \) for models \( \mathcal{L}_3, \mathcal{L}_4, \mathcal{L}_8, \mathcal{L}_9, \mathcal{L}_{10}, \mathcal{L}_{13}, \mathcal{L}_{14} \), visible on-shell mediator decays provide optimal search sensitivity. However, when couplings to DM particles significantly exceed those to muon pairs, mono-photon signatures with off-shell mediators offer complementary probes. We therefore investigate this situation through the off-shell mediator mono-photon channel. DM masses are constrainted to \( m_D < 100 \ \text{GeV} \) from global fitting results~\cite{Abdughani:2021oit}. We consider three mediator-to-DM mass ratios: \( M/m_D = \{1.1, 1.5, 1.9\} \) as mentioned in Sec.~\ref{sec:parameter}. To demonstrate our methodology, we select three representative models: \( \mathcal{L}_3 \), \( \mathcal{L}_9 \), and \( \mathcal{L}_{13} \), presenting their BPs, kinematic distributions and event selection criteria before generalizing to other models in Sec.~\ref{sec:exclusion}. 

\begin{table}[htbp]
\centering
\footnotesize  
\begin{tabular}{|>{\centering\arraybackslash}p{6cm}|>{\centering\arraybackslash}p{5cm}|>{\centering\arraybackslash}p{2.2cm}|*{4}{>{\centering\arraybackslash}p{1.5cm}|}}
\hline
\multicolumn{2}{|c|}{\textbf{Cut description}} & \textbf{background} & \textbf{signal1} & \textbf{signal2} & \textbf{signal3}  \\
\hline 
\multicolumn{2}{|c|}{\textbf{Cross-section [fb]}} & 2980 & 0.42 & 0.45 & 0.53  \\
\hline
\multicolumn{1}{|c|}{\multirow{3}{*}{\textbf{Cut-1}}} & $N(\gamma)>0$ & {\footnotesize 0.92} & {\footnotesize 0.91} & {\footnotesize 0.91} & {\footnotesize 0.91}   \\
\cline{2-6}
\multicolumn{1}{|c|}{} & $E(\gamma) > 100\,\mathrm{GeV} \ \&\ |\eta(\gamma)| < 2.5$ & {\footnotesize 0.35} & {\footnotesize 0.75} & {\footnotesize 0.75} & {\footnotesize 0.78}   \\
\cline{2-6}
\multicolumn{1}{|c|}{} & $ {\:/\!\!\!\! E}_T > 40\,\mathrm{GeV} $ & {\footnotesize 0.30} & {\footnotesize 0.73} & {\footnotesize 0.73} & {\footnotesize 0.76}  \\
\hline
\multicolumn{1}{|c|}{\multirow{1}{*}{\textbf{Cut-2}}} & $E(\gamma) > 1200\,\mathrm{GeV} \ $ & {\footnotesize $7.4\times 10^{-3}$} & {\footnotesize 0.41} & {\footnotesize 0.43} & {\footnotesize 0.51} \\
\hline
\multicolumn{1}{|c|}{\multirow{1}{*}{\textbf{Cut-3}}} & $P_T(\gamma)/{\:/\!\!\!\! E}>0.4$ & {\footnotesize $1.7\times 10^{-3}$} & {\footnotesize 0.24} & {\footnotesize 0.26} & {\footnotesize 0.32} \\
\hline
\multicolumn{1}{|c|}{\multirow{1}{*}{\textbf{Cut-4}}} & $M_T > 1600\,\mathrm{GeV}$ & {\footnotesize $1.0\times 10^{-3}$} & {\footnotesize 0.19} & {\footnotesize 0.21} & {\footnotesize 0.25} \\
\hline
\end{tabular}
\caption{The cut-flow table for the off-shell mediator scenario of the $\mathcal{L}_3$ model with $m_D=20$ GeV and its relevant background with cumulative efficiencies from \textbf{Cut-1} to \textbf{Cut-4}. The three benchmark points are signal-1 ($M/m_D=1.1$), signal-2 ($M/m_D=1.5$), signal-3 ($M/m_D=1.9$). }
\label{tab:L3_20_offshell}
\end{table}

\begin{table}[htbp]
\centering
\footnotesize  
\begin{tabular}{|>{\centering\arraybackslash}p{6cm}|>{\centering\arraybackslash}p{5cm}|>{\centering\arraybackslash}p{2.2cm}|*{4}{>{\centering\arraybackslash}p{1.5cm}|}}
\hline
\multicolumn{2}{|c|}{\textbf{Cut description}} & \textbf{background} & \textbf{signal1} & \textbf{signal2} & \textbf{signal3}  \\
\hline 
\multicolumn{2}{|c|}{\textbf{Cross-section [fb]}} & 2980 & 0.34 & 0.36 & 0.45  \\
\hline
\multicolumn{1}{|c|}{\multirow{3}{*}{\textbf{Cut-1}}} & $N(\gamma)>0$ & {\footnotesize 0.92} & {\footnotesize 0.91} & {\footnotesize 0.91} & {\footnotesize 0.91}   \\
\cline{2-6}
\multicolumn{1}{|c|}{} & $E(\gamma) > 100\,\mathrm{GeV} \ \&\ |\eta(\gamma)| < 2.5$ & {\footnotesize 0.35} & {\footnotesize 0.69} & {\footnotesize 0.70} & {\footnotesize 0.74}   \\
\cline{2-6}
\multicolumn{1}{|c|}{} & $ {\:/\!\!\!\! E}_T > 40\,\mathrm{GeV} $ & {\footnotesize 0.30} & {\footnotesize 0.67} & {\footnotesize 0.68} & {\footnotesize 0.72}  \\
\hline
\multicolumn{1}{|c|}{\multirow{1}{*}{\textbf{Cut-2}}} & $E(\gamma) > 1200\,\mathrm{GeV} \ $ & {\footnotesize $7.4\times 10^{-3}$} & {\footnotesize 0.28} & {\footnotesize 0.31} & {\footnotesize 0.43} \\
\hline
\multicolumn{1}{|c|}{\multirow{1}{*}{\textbf{Cut-3}}} & $P_T(\gamma)/{\:/\!\!\!\! E}>0.4$ & {\footnotesize $1.7\times 10^{-3}$} & {\footnotesize 0.16} & {\footnotesize 0.18} & {\footnotesize 0.26} \\
\hline
\multicolumn{1}{|c|}{\multirow{1}{*}{\textbf{Cut-4}}} & $M_T > 1600\,\mathrm{GeV}$ & {\footnotesize $1.0\times 10^{-3}$} & {\footnotesize 0.13} & {\footnotesize 0.15} & {\footnotesize 0.21} \\
\hline
\end{tabular}
\caption{Similar to Table~\ref{tab:L3_20_offshell}, but for the $\mathcal{L}_3$ model with $m_D=100$ GeV. } 
\label{tab:L3_100_offshell}
\end{table}

\begin{table}[htbp]
\centering
\footnotesize  
\begin{tabular}{|>{\centering\arraybackslash}p{6cm}|>{\centering\arraybackslash}p{5cm}|>{\centering\arraybackslash}p{2.2cm}|*{4}{>{\centering\arraybackslash}p{1.5cm}|}}
\hline
\multicolumn{2}{|c|}{\textbf{Cut description}} & \textbf{background} & \textbf{signal1} & \textbf{signal2} & \textbf{signal3}  \\
\hline 
\multicolumn{2}{|c|}{\textbf{Cross-section [fb]}} & 2980 & 0.30 & 0.42 & 1.20  \\
\hline 
\multicolumn{1}{|c|}{\multirow{3}{*}{\textbf{Cut-1}}} & $N(\gamma)>0$ & {\footnotesize 0.92} & {\footnotesize 0.91} & {\footnotesize 0.91} & {\footnotesize 0.91}   \\
\cline{2-6}
\multicolumn{1}{|c|}{} & $E(\gamma) > 100\,\mathrm{GeV} \ \&\ |\eta(\gamma)| < 2.5$ & {\footnotesize 0.35} & {\footnotesize 0.91} & {\footnotesize 0.91} & {\footnotesize 0.91}   \\
\cline{2-6}
\multicolumn{1}{|c|}{} & $ {\:/\!\!\!\! E}_T > 40\,\mathrm{GeV} $ & {\footnotesize 0.30} & {\footnotesize 0.91} & {\footnotesize 0.91} & {\footnotesize 0.91}  \\
\hline
\multicolumn{1}{|c|}{\multirow{1}{*}{\textbf{Cut-2}}} & $E(\gamma) > 1200\,\mathrm{GeV} \ $ & {\footnotesize $7.4\times 10^{-3}$} & {\footnotesize 0.91} & {\footnotesize 0.91} & {\footnotesize 0.91} \\
\hline
\multicolumn{1}{|c|}{\multirow{1}{*}{\textbf{Cut-3}}} & $P_T(\gamma)/{\:/\!\!\!\! E}>0.4$ & {\footnotesize $1.7\times 10^{-3}$} & {\footnotesize 0.57} & {\footnotesize 0.58} & {\footnotesize 0.58} \\
\hline
\multicolumn{1}{|c|}{\multirow{1}{*}{\textbf{Cut-4}}} & $M_T > 1600\,\mathrm{GeV}$ & {\footnotesize $1.0\times 10^{-3}$} & {\footnotesize 0.45} & {\footnotesize 0.46} & {\footnotesize 0.46} \\
\hline
\end{tabular}
\caption{Similar to Table~\ref{tab:L3_20_offshell}, but for the $\mathcal{L}_9$ model with $m_D=20$ GeV. } 
\label{tab:L9_20_offshell}
\end{table}

\begin{table}[htbp]
\centering
\footnotesize  
\begin{tabular}{|>{\centering\arraybackslash}p{6cm}|>{\centering\arraybackslash}p{5cm}|>{\centering\arraybackslash}p{2.2cm}|*{4}{>{\centering\arraybackslash}p{1.5cm}|}}
\hline
\multicolumn{2}{|c|}{\textbf{Cut description}} & \textbf{background} & \textbf{signal1} & \textbf{signal2} & \textbf{signal3}  \\
\hline 
\multicolumn{2}{|c|}{\textbf{Cross-section [fb]}} & 2980 & $1.3\times 10^{-2}$ & $1.8\times 10^{-2}$ & $4.9\times 10^{-2}$  \\
\hline 
\multicolumn{1}{|c|}{\multirow{3}{*}{\textbf{Cut-1}}} & $N(\gamma)>0$ & {\footnotesize 0.92} & {\footnotesize 0.91} & {\footnotesize 0.91} & {\footnotesize 0.91}   \\
\cline{2-6}
\multicolumn{1}{|c|}{} & $E(\gamma) > 100\,\mathrm{GeV} \ \&\ |\eta(\gamma)| < 2.5$ & {\footnotesize 0.35} & {\footnotesize 0.90} & {\footnotesize 0.90} & {\footnotesize 0.91}   \\
\cline{2-6}
\multicolumn{1}{|c|}{} & $ {\:/\!\!\!\! E}_T > 40\,\mathrm{GeV} $ & {\footnotesize 0.30} & {\footnotesize 0.90} & {\footnotesize 0.90} & {\footnotesize 0.91}  \\
\hline
\multicolumn{1}{|c|}{\multirow{1}{*}{\textbf{Cut-2}}} & $E(\gamma) > 1200\,\mathrm{GeV} \ $ & {\footnotesize $7.4\times 10^{-3}$} & {\footnotesize 0.85} & {\footnotesize 0.87} & {\footnotesize 0.90} \\
\hline
\multicolumn{1}{|c|}{\multirow{1}{*}{\textbf{Cut-3}}} & $P_T(\gamma)/{\:/\!\!\!\! E}>0.4$ & {\footnotesize $1.7\times 10^{-3}$} & {\footnotesize 0.53} & {\footnotesize 0.53} & {\footnotesize 0.56} \\
\hline
\multicolumn{1}{|c|}{\multirow{1}{*}{\textbf{Cut-4}}} & $M_T > 1600\,\mathrm{GeV}$ & {\footnotesize $1.0\times 10^{-3}$} & {\footnotesize 0.42} & {\footnotesize 0.43} & {\footnotesize 0.44} \\
\hline
\end{tabular}
\caption{Similar to Table~\ref{tab:L3_20_offshell}, but for the $\mathcal{L}_9$ model with $m_D=100$ GeV. } 
\label{tab:L9_100_offshell}
\end{table}

\begin{table}[htbp]
\centering
\footnotesize  
\begin{tabular}{|>{\centering\arraybackslash}p{6cm}|>{\centering\arraybackslash}p{5cm}|>{\centering\arraybackslash}p{2.2cm}|*{4}{>{\centering\arraybackslash}p{1.5cm}|}}
\hline
\multicolumn{2}{|c|}{\textbf{Cut description}} & \textbf{background} & \textbf{signal1} & \textbf{signal2} & \textbf{signal3}  \\
\hline 
\multicolumn{2}{|c|}{\textbf{Cross-section [fb]}} & 2980 & 518 & 533 & 521  \\
\hline
\multicolumn{1}{|c|}{\multirow{3}{*}{\textbf{Cut-1}}} & $N(\gamma)>0$ & {\footnotesize 0.92} & {\footnotesize 0.91} & {\footnotesize 0.91} & {\footnotesize 0.91}   \\
\cline{2-6}
\multicolumn{1}{|c|}{} & $E(\gamma) > 100\,\mathrm{GeV} \ \&\ |\eta(\gamma)| < 2.5$ & {\footnotesize 0.35} & {\footnotesize 0.54} & {\footnotesize 0.54} & {\footnotesize 0.53}   \\
\cline{2-6}
\multicolumn{1}{|c|}{} & $ {\:/\!\!\!\! E}_T > 40\,\mathrm{GeV} $ & {\footnotesize 0.30} & {\footnotesize 0.51} & {\footnotesize 0.50} & {\footnotesize 0.49}  \\
\hline
\multicolumn{1}{|c|}{\multirow{1}{*}{\textbf{Cut-2}}} & $E(\gamma) > 1200\,\mathrm{GeV} \ $ & {\footnotesize $7.4\times 10^{-3}$} & {\footnotesize $3.0\times 10^{-2}$} & {\footnotesize $2.8\times 10^{-2}$} & {\footnotesize $3.2\times 10^{-2}$} \\
\hline
\multicolumn{1}{|c|}{\multirow{1}{*}{\textbf{Cut-3}}} & $P_T(\gamma)/{\:/\!\!\!\! E}>0.4$ & {\footnotesize $1.7\times 10^{-3}$} & {\footnotesize $1.5\times 10^{-2}$} & {\footnotesize $1.6\times 10^{-2}$} & {\footnotesize $1.8\times 10^{-2}$} \\
\hline
\multicolumn{1}{|c|}{\multirow{1}{*}{\textbf{Cut-4}}} & $M_T > 1600\,\mathrm{GeV}$ & {\footnotesize $1.0\times 10^{-3}$} & {\footnotesize $1.3\times 10^{-2}$} & {\footnotesize $1.3\times 10^{-2}$} & {\footnotesize $1.6\times 10^{-2}$} \\
\hline
\end{tabular}
\caption{Similar to Table~\ref{tab:L3_20_offshell}, but for the $\mathcal{L}_{13}$ model with $m_D=20$ GeV. } 
\label{tab:L13_20_offshell}
\end{table}

\begin{table}[htbp]
\centering
\footnotesize  
\begin{tabular}{|>{\centering\arraybackslash}p{6cm}|>{\centering\arraybackslash}p{5cm}|>{\centering\arraybackslash}p{2.2cm}|*{4}{>{\centering\arraybackslash}p{1.5cm}|}}
\hline
\multicolumn{2}{|c|}{\textbf{Cut description}} & \textbf{background} & \textbf{signal1} & \textbf{signal2} & \textbf{signal3}  \\
\hline 
\multicolumn{2}{|c|}{\textbf{Cross-section [fb]}} & 2980 & 0.83 & 0.84 & 0.84  \\
\hline 
\multicolumn{1}{|c|}{\multirow{3}{*}{\textbf{Cut-1}}} & $N(\gamma)>0$ & {\footnotesize 0.92} & {\footnotesize 0.91} & {\footnotesize 0.91} & {\footnotesize 0.92}   \\
\cline{2-6}
\multicolumn{1}{|c|}{} & $E(\gamma) > 100\,\mathrm{GeV} \ \&\ |\eta(\gamma)| < 2.5$ & {\footnotesize 0.35} & {\footnotesize 0.53} & {\footnotesize 0.53} & {\footnotesize 0.52}   \\
\cline{2-6}
\multicolumn{1}{|c|}{} & $ {\:/\!\!\!\! E}_T > 40\,\mathrm{GeV} $ & {\footnotesize 0.30} & {\footnotesize 0.48} & {\footnotesize 0.49} & {\footnotesize 0.49}  \\
\hline
\multicolumn{1}{|c|}{\multirow{1}{*}{\textbf{Cut-2}}} & $E(\gamma) > 1200\,\mathrm{GeV} \ $ & {\footnotesize $7.4\times 10^{-3}$} & {\footnotesize $2.6\times 10^{-2}$} & {\footnotesize $2.7\times 10^{-2}$} & {\footnotesize $3.2\times 10^{-2}$} \\
\hline
\multicolumn{1}{|c|}{\multirow{1}{*}{\textbf{Cut-3}}} & $P_T(\gamma)/{\:/\!\!\!\! E}>0.4$ & {\footnotesize $1.7\times 10^{-3}$} & {\footnotesize $1.4\times 10^{-2}$} & {\footnotesize $1.4\times 10^{-2}$} & {\footnotesize $1.8\times 10^{-2}$} \\
\hline
\multicolumn{1}{|c|}{\multirow{1}{*}{\textbf{Cut-4}}} & $M_T > 1600\,\mathrm{GeV}$ & {\footnotesize $1.0\times 10^{-3}$} & {\footnotesize $1.3\times 10^{-2}$} & {\footnotesize $1.2\times 10^{-2}$} & {\footnotesize $1.5\times 10^{-2}$} \\
\hline
\end{tabular}
\caption{Similar to Table~\ref{tab:L3_20_offshell}, but for the $\mathcal{L}_{13}$ model with $m_D=100$ GeV. } 
\label{tab:L13_100_offshell}
\end{table}

\begin{figure}[h]
\centering 
\includegraphics[width=0.48\textwidth]{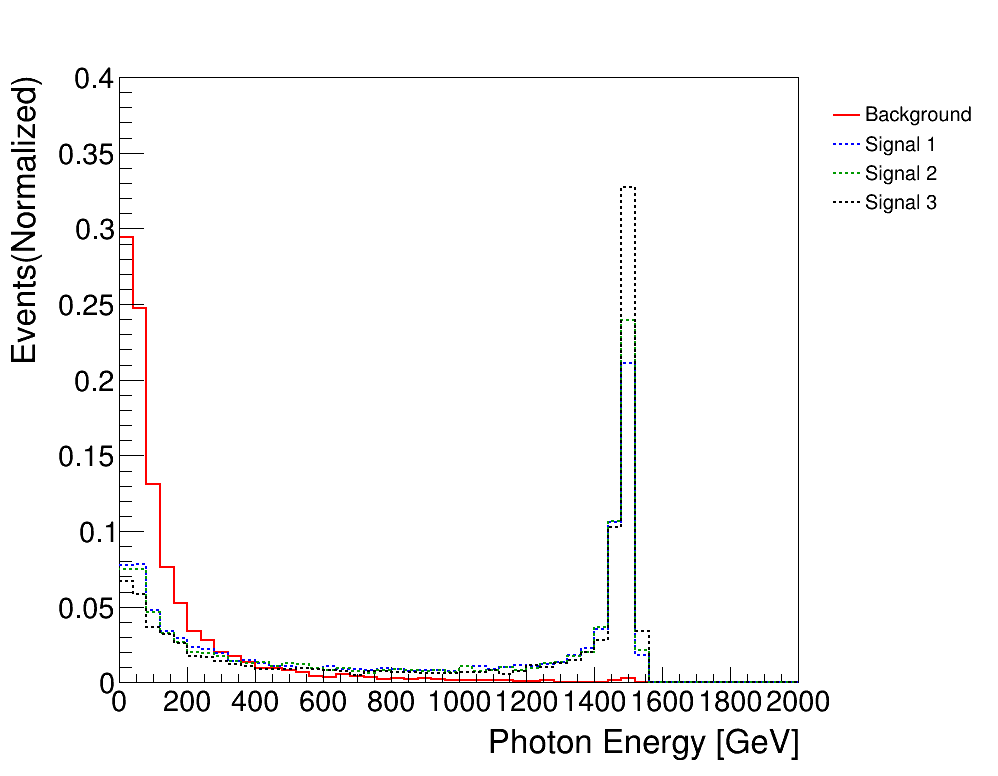} 
\includegraphics[width=0.48\textwidth]{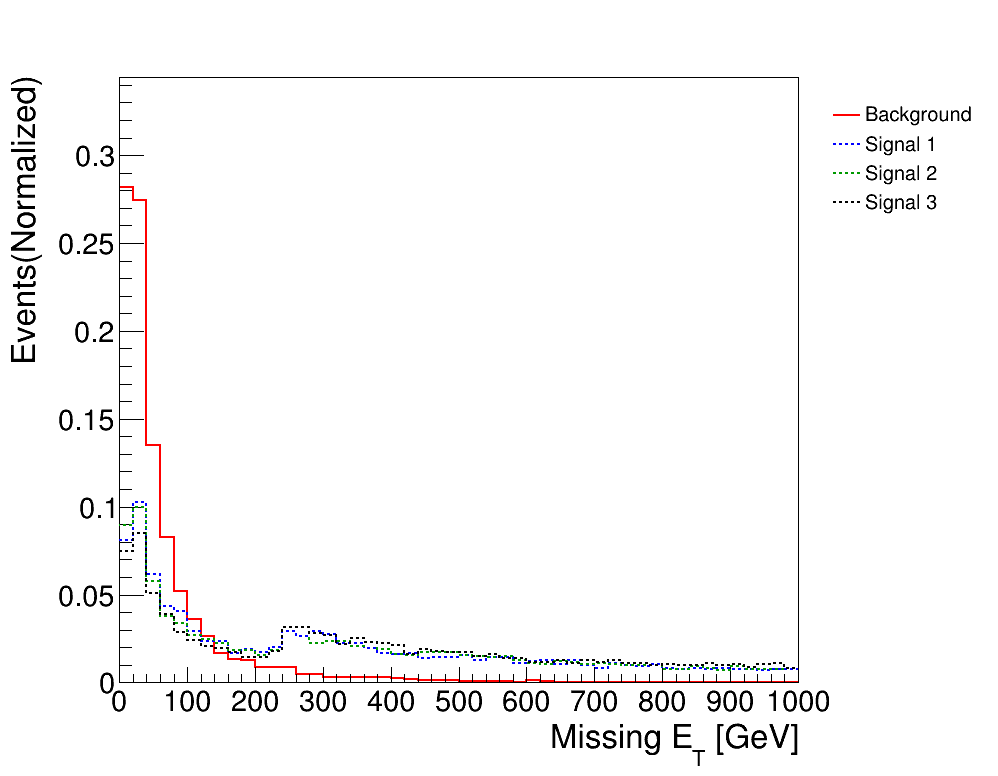}
\includegraphics[width=0.48\textwidth]{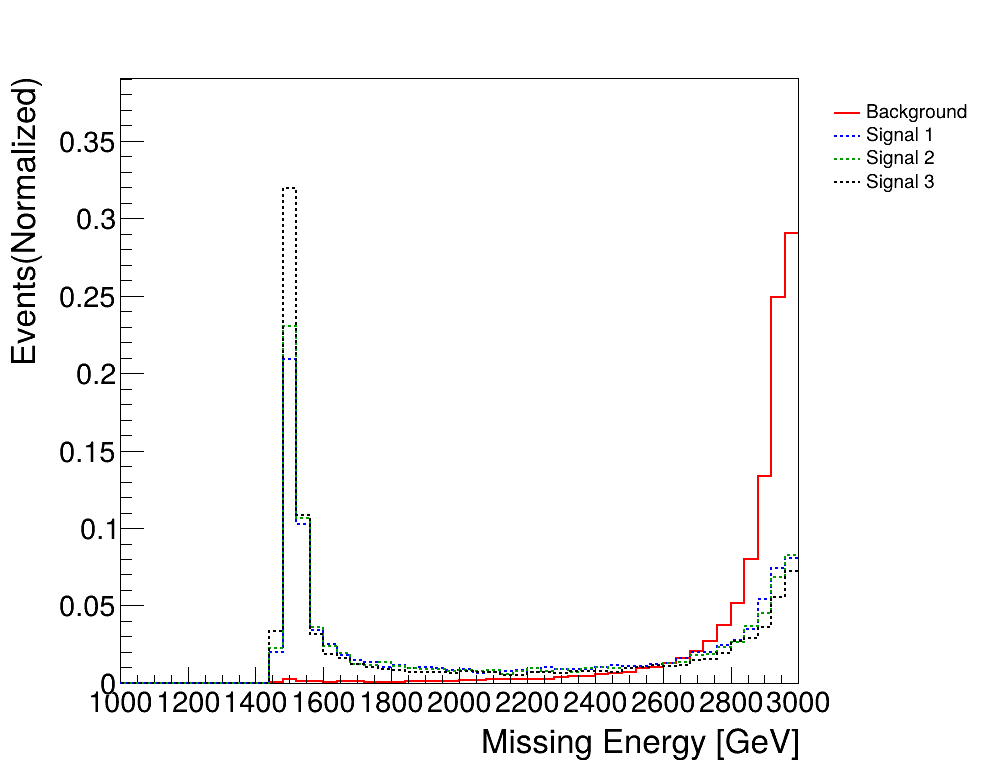}
\includegraphics[width=0.48\textwidth]{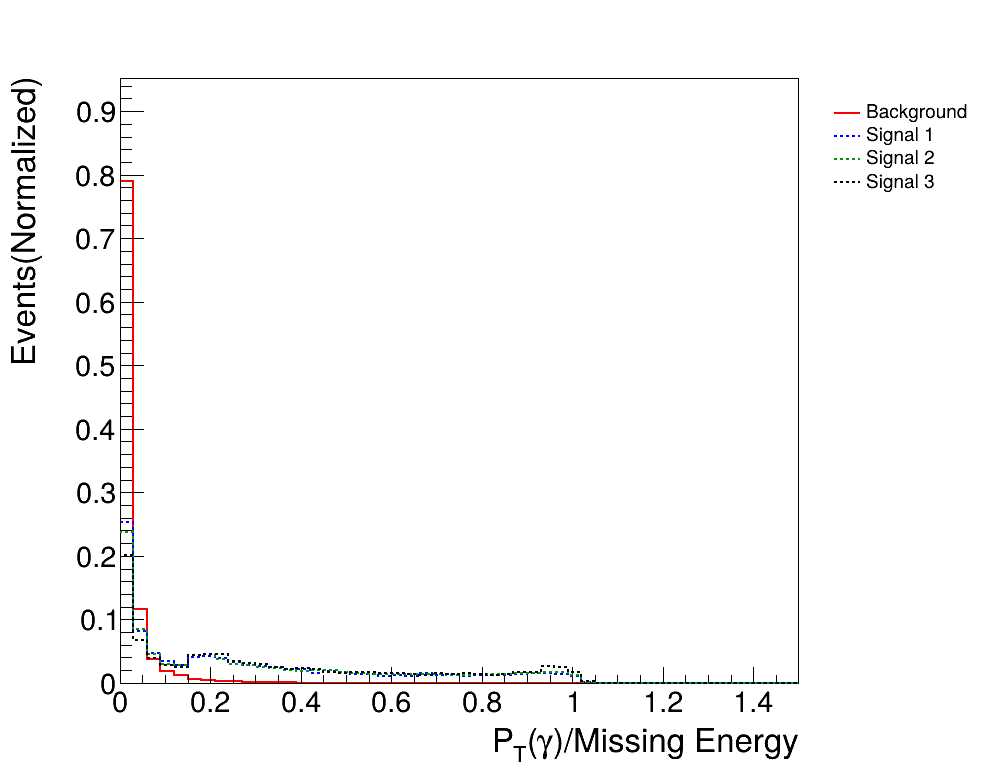}
\caption{In the off-shell mediator scenario of the $\mathcal{L}_3$ model with $m_D=20$ GeV, representative kinematic distributions include photon energy $E({\gamma})$ (top-left), missing transverse energy $ {\:/\!\!\!\! E}_T$ (top-right), missing energy ${\:/\!\!\!\! E}$ (bottom-left), and the ratio of $P_T(\gamma)/{\:/\!\!\!\! E}$ (bottom-right) for signal-1 ($M/m_D=1.1$, blue-dotted line), signal-2 ($M/m_D=1.5$, green-dotted line), signal-3 ($M/m_D=1.9$, black-dotted line), and background (red-solid line). } 
\label{fig:L3_20_offshell}
\end{figure}

\begin{figure}[h]
\centering 
\includegraphics[width=0.48\textwidth]{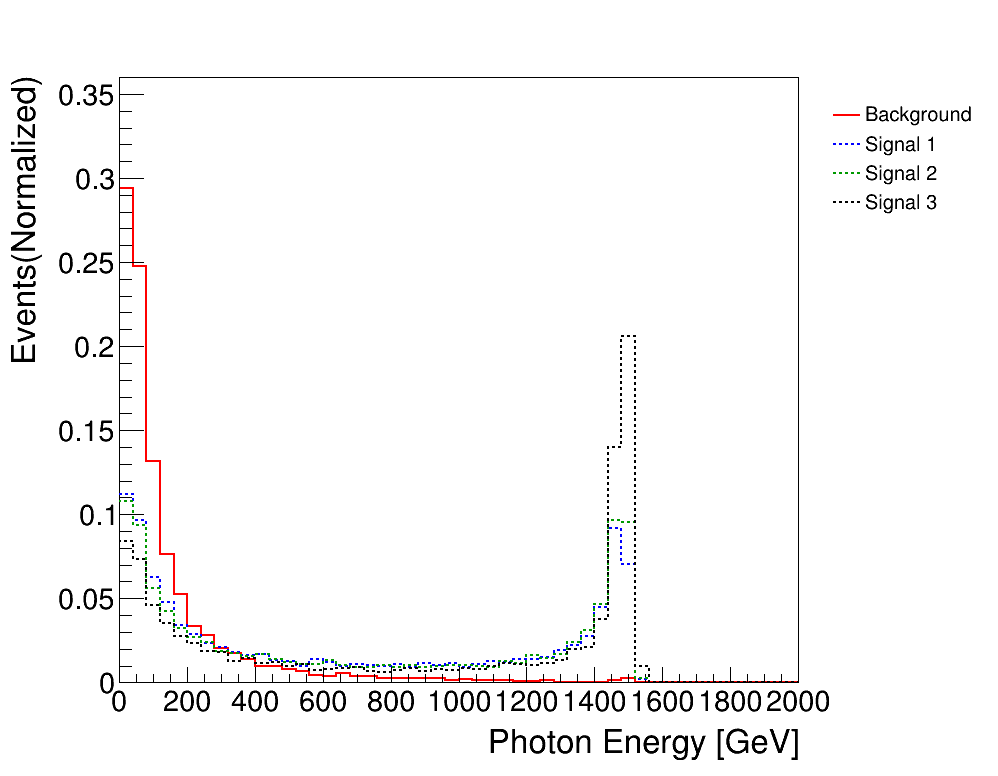} 
\includegraphics[width=0.48\textwidth]{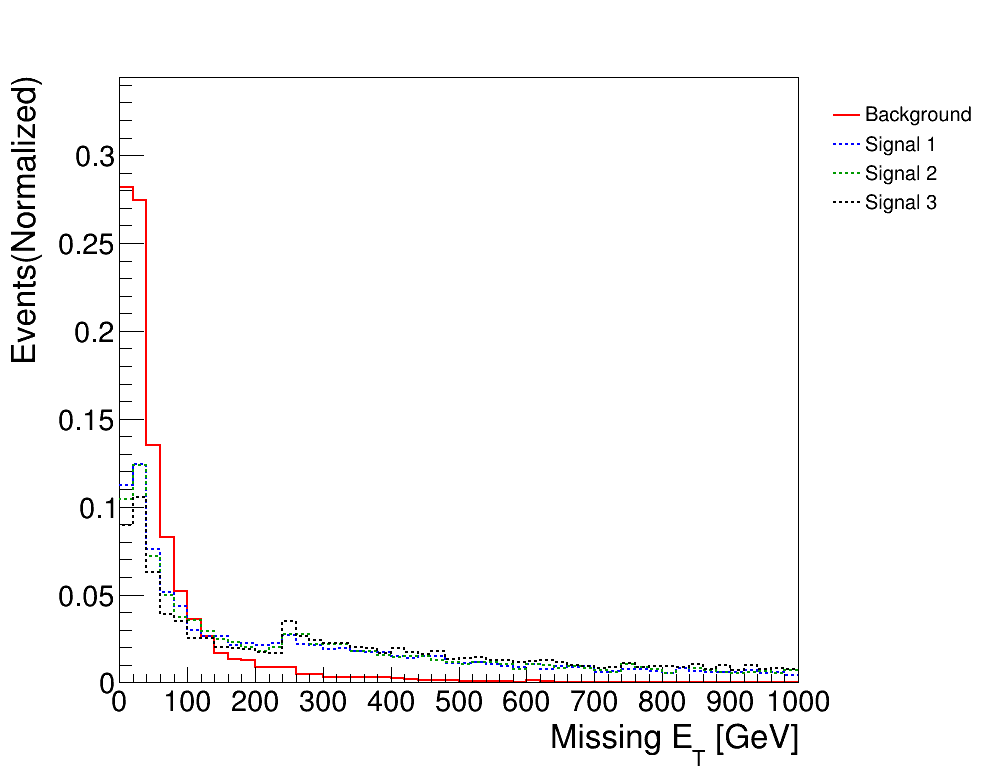}
\includegraphics[width=0.48\textwidth]{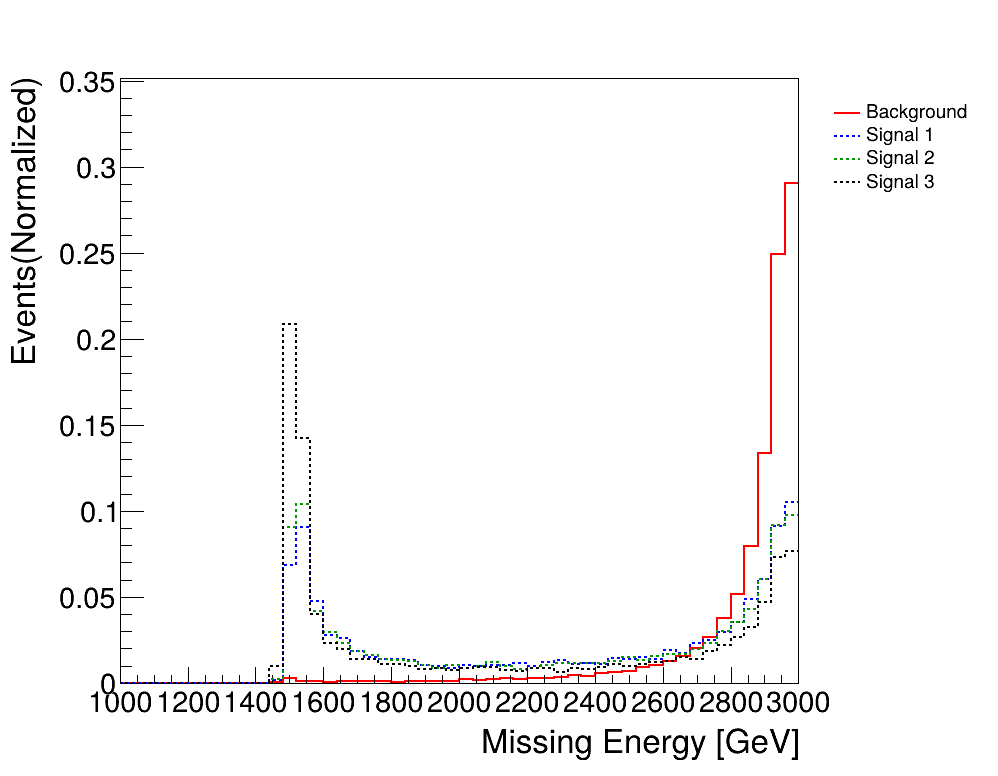}
\includegraphics[width=0.48\textwidth]{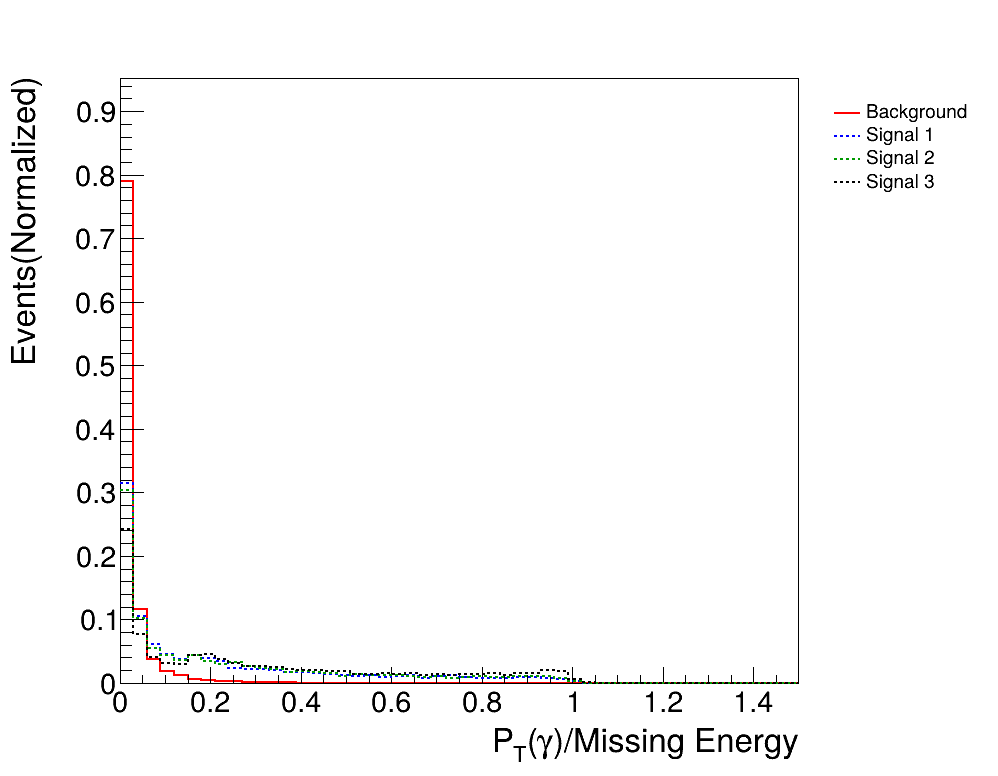}
\caption{Similar to Fig.~\ref{fig:L3_20_offshell}, but for the $\mathcal{L}_3$ model with $m_D = 100$ GeV.} 
\label{fig:L3_100_offshell}
\end{figure}

\begin{figure}[h]
\centering 
\includegraphics[width=0.48\textwidth]{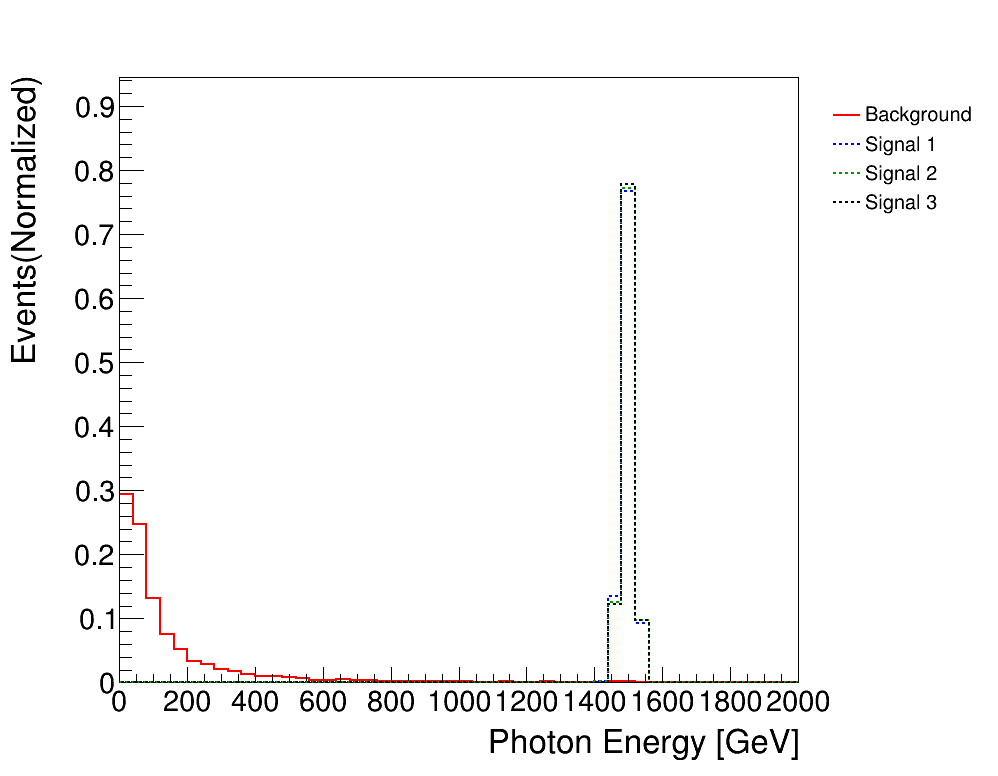} 
\includegraphics[width=0.48\textwidth]{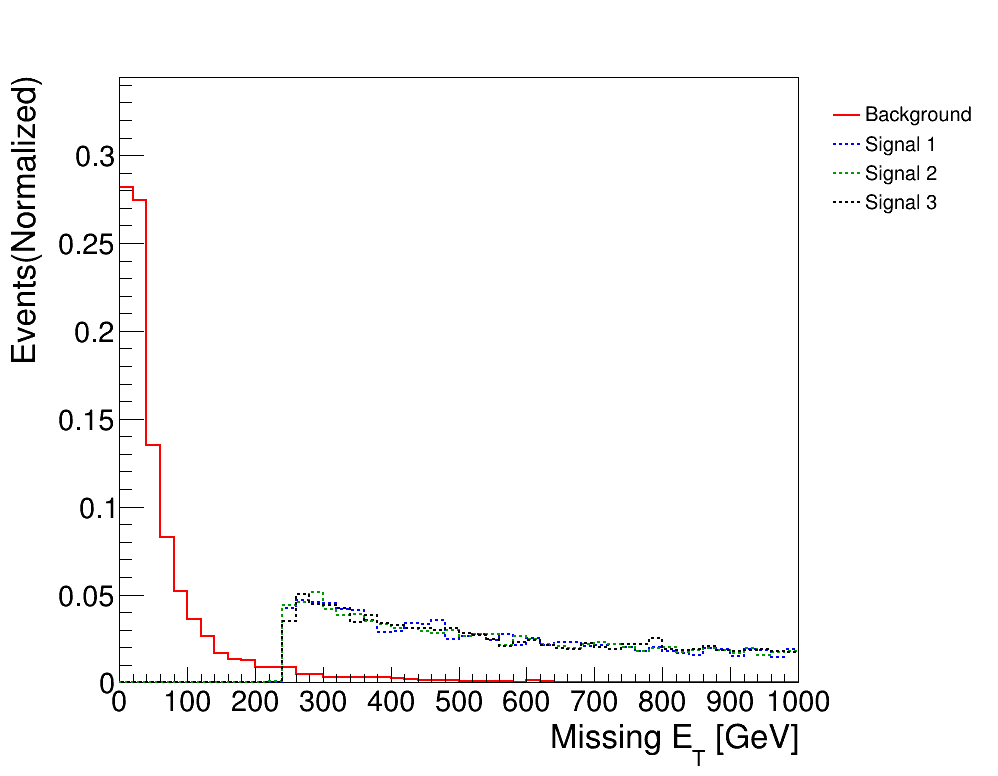}
\includegraphics[width=0.48\textwidth]{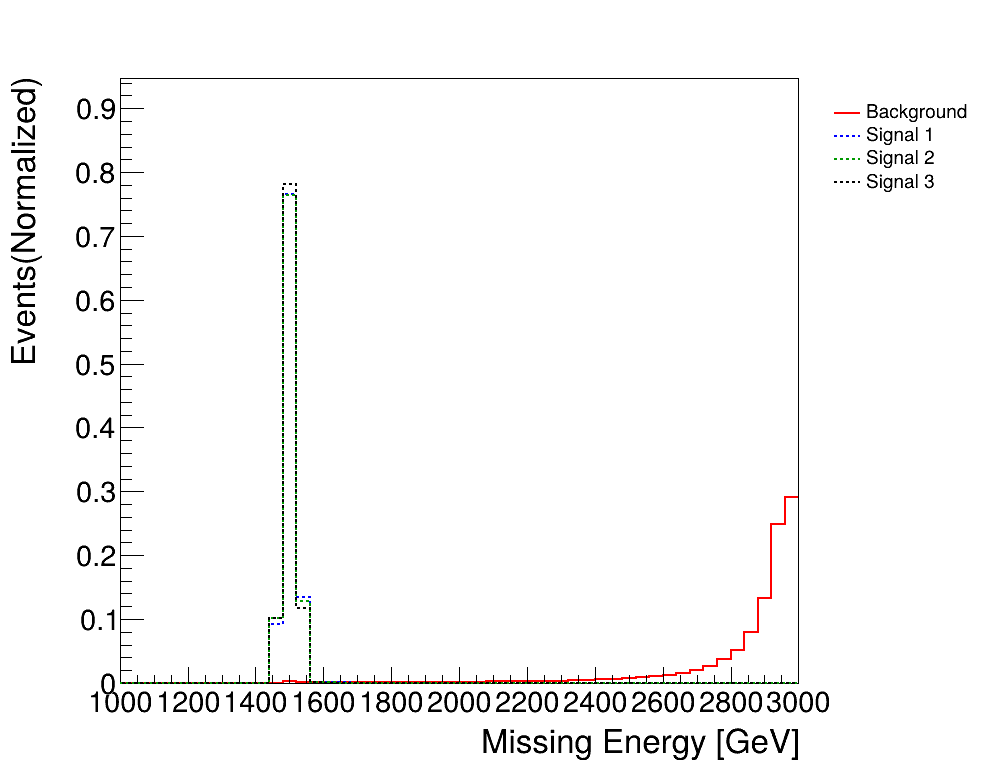}
\includegraphics[width=0.48\textwidth]{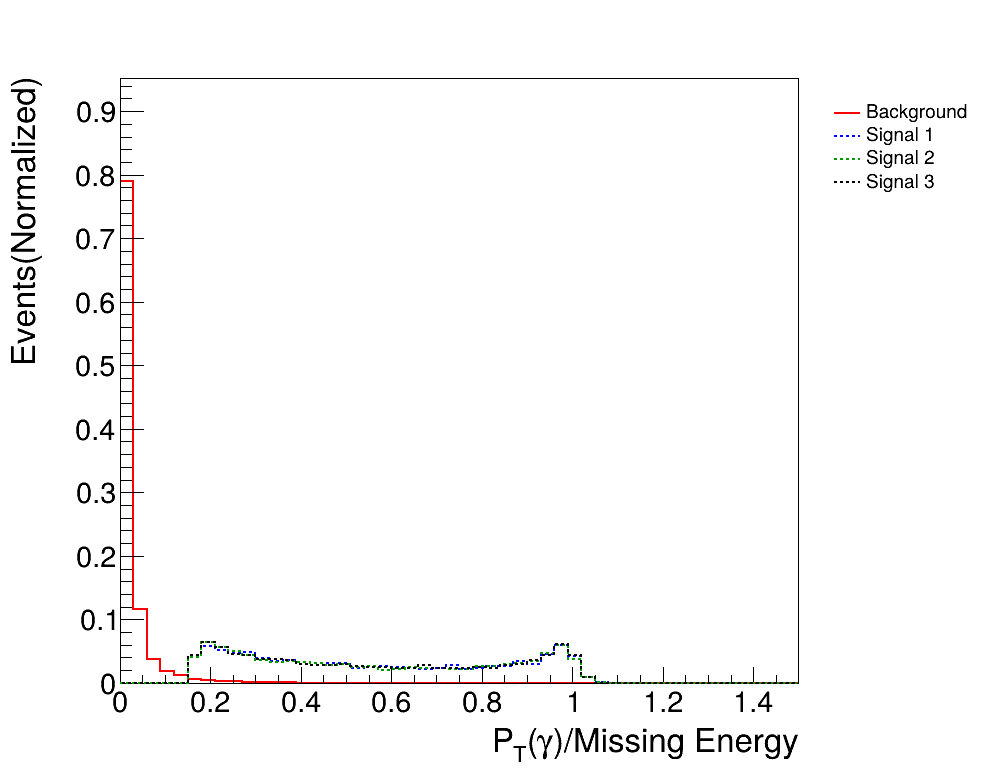}
\caption{Similar to Fig.~\ref{fig:L3_20_offshell}, but for the $\mathcal{L}_9$ model with $m_D=20$ GeV. } 
\label{fig:L9_20_offshell}
\end{figure}
 
\begin{figure}[h]
\centering 
\includegraphics[width=0.48\textwidth]{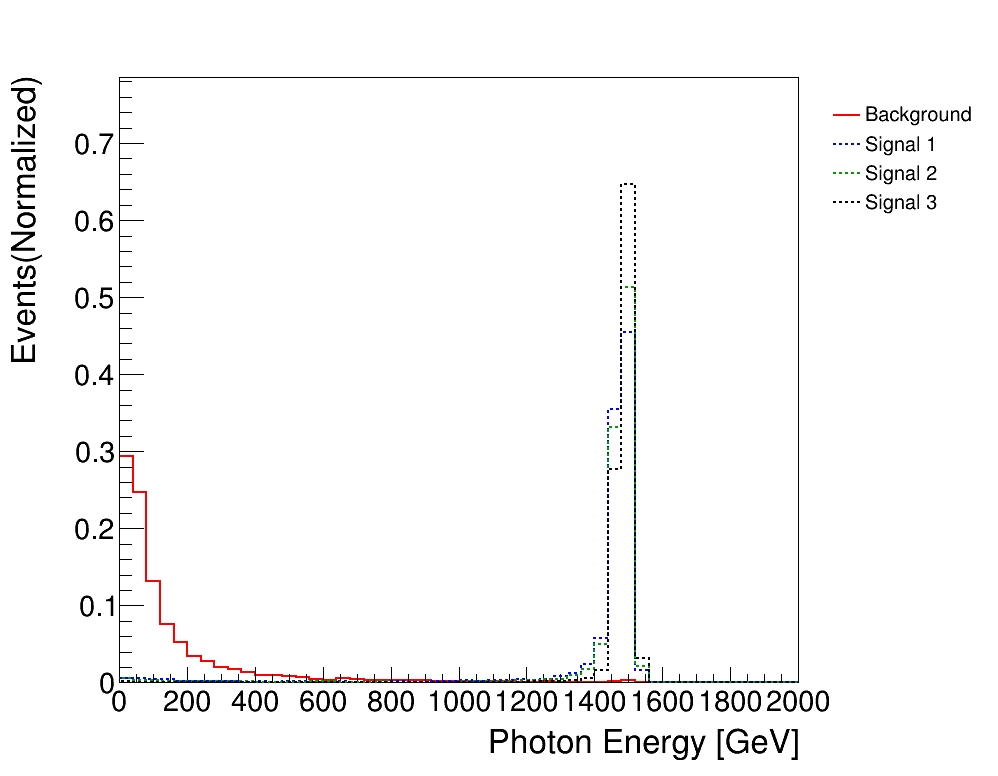} 
\includegraphics[width=0.48\textwidth]{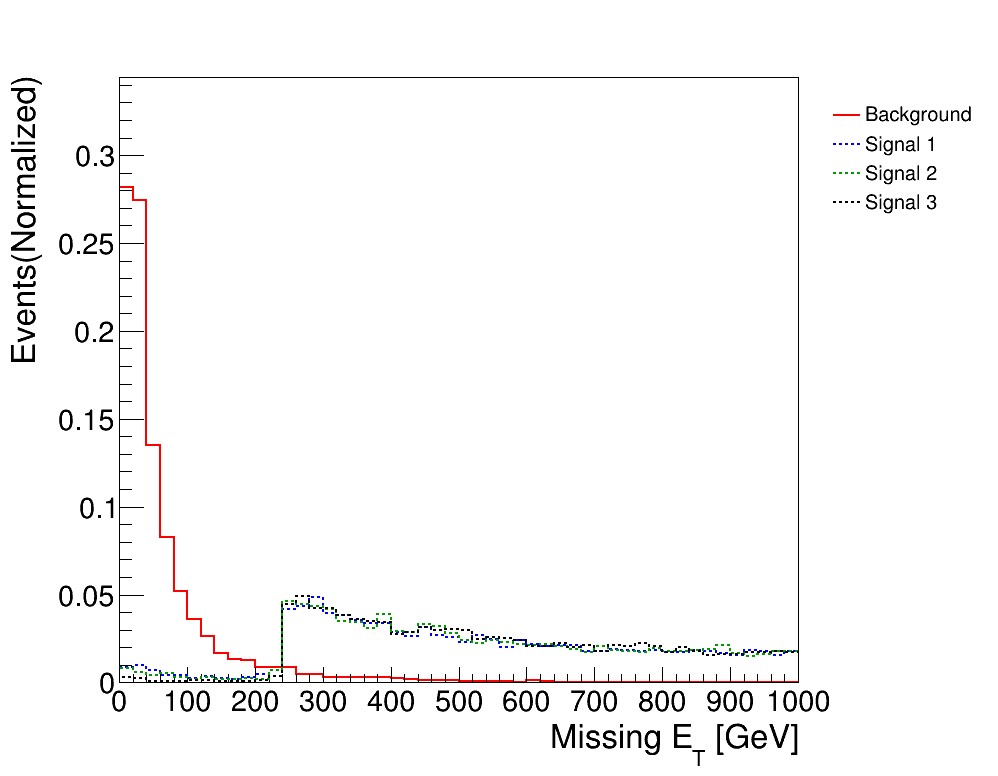}
\includegraphics[width=0.48\textwidth]{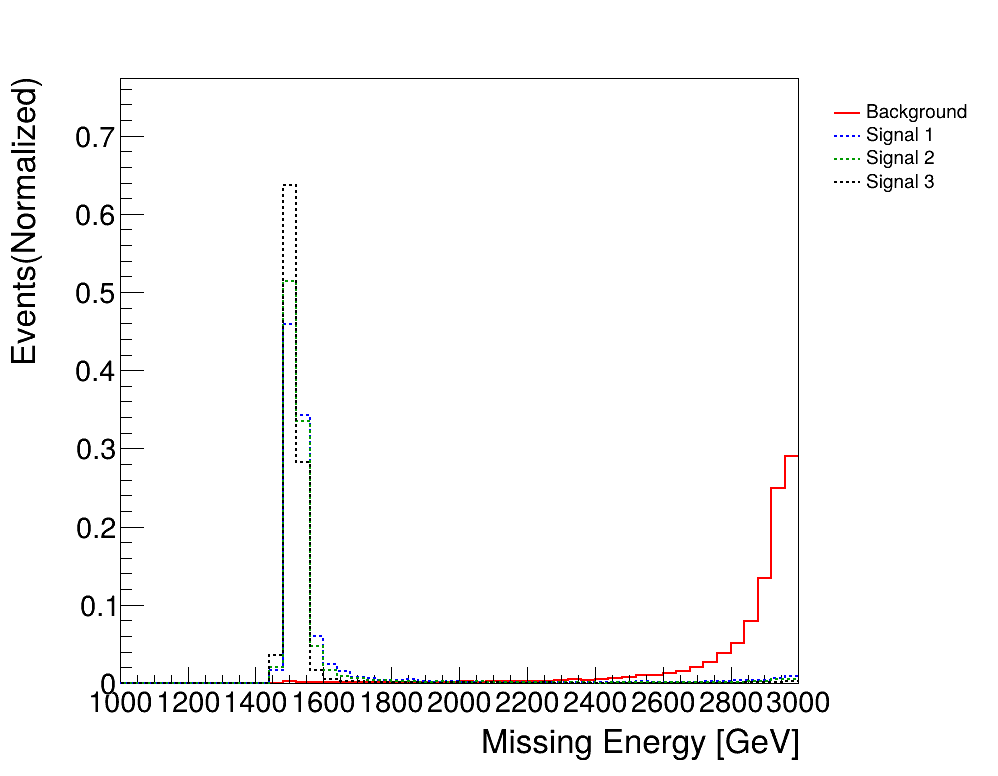}
\includegraphics[width=0.48\textwidth]{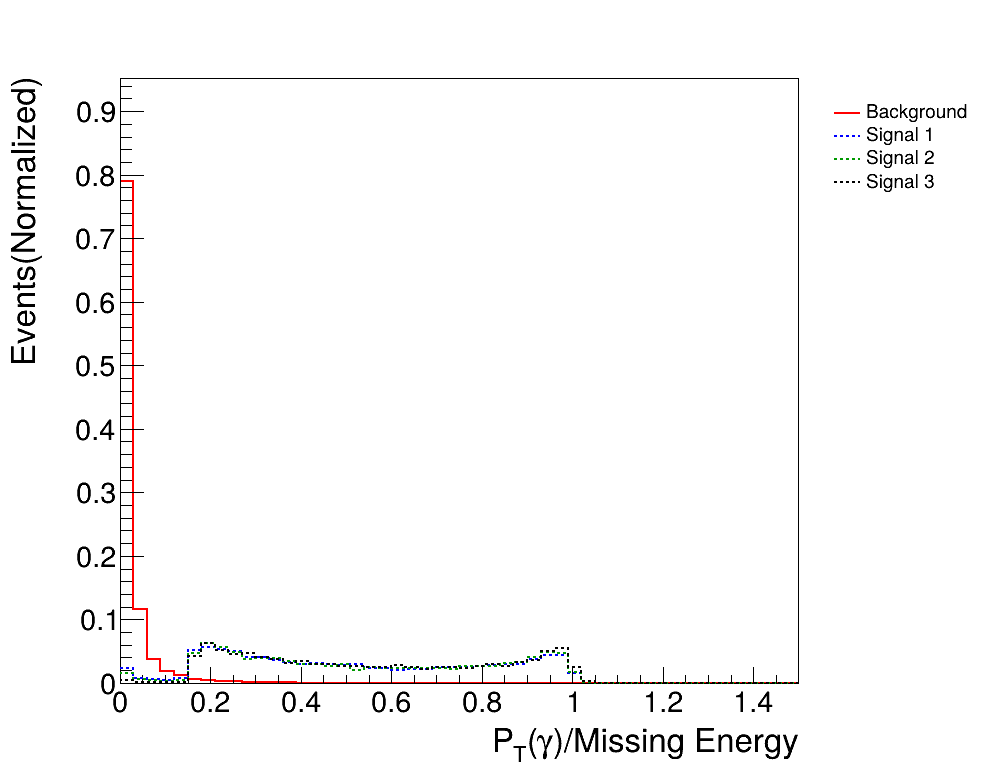}
\caption{Similar to Fig.~\ref{fig:L3_20_offshell}, but for the $\mathcal{L}_9$ model with $m_D = 100$ GeV.} 
\label{fig:L9_100_offshell}
\end{figure}

\begin{figure}[h]
\centering 
\includegraphics[width=0.48\textwidth]{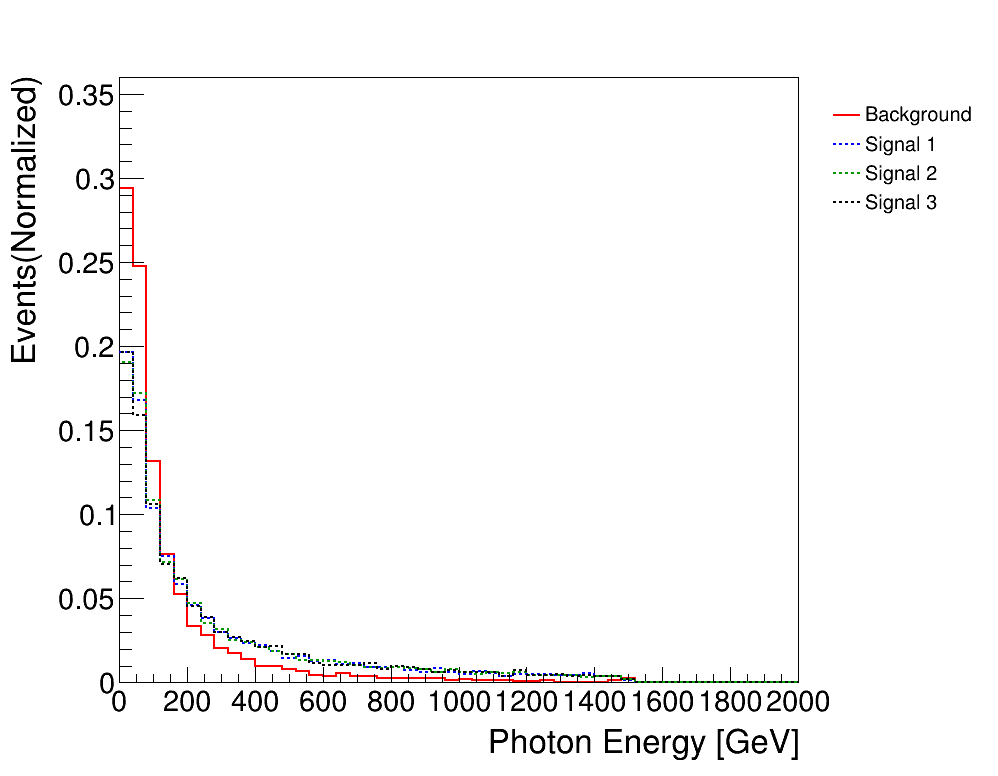} 
\includegraphics[width=0.48\textwidth]{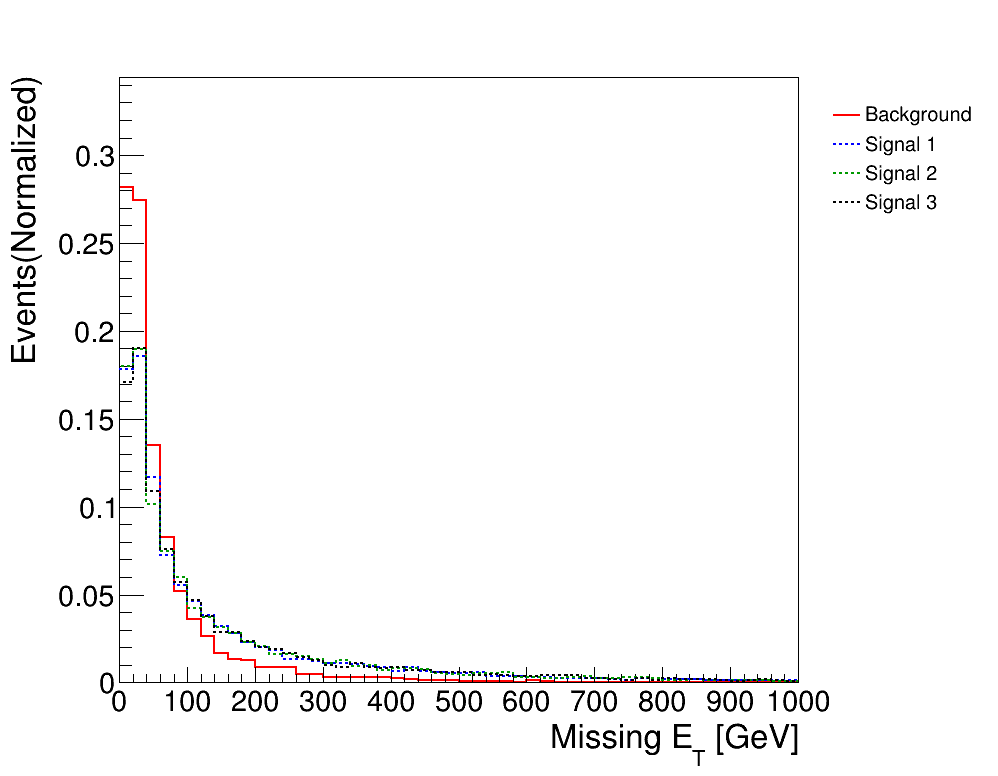}
\includegraphics[width=0.48\textwidth]{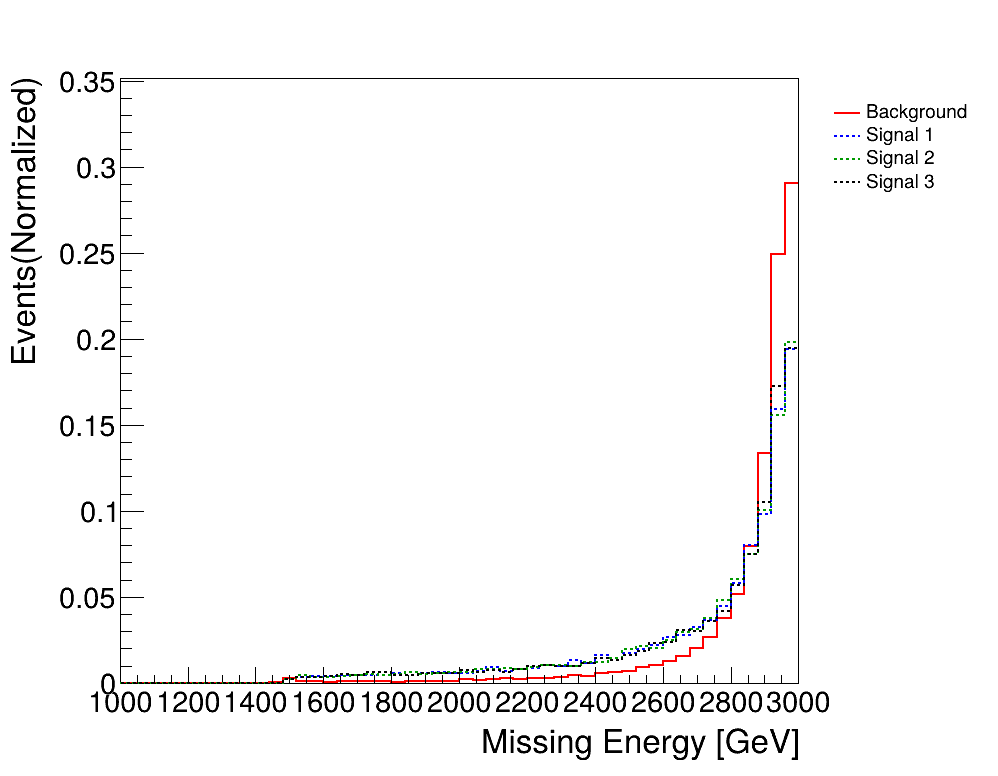}
\includegraphics[width=0.48\textwidth]{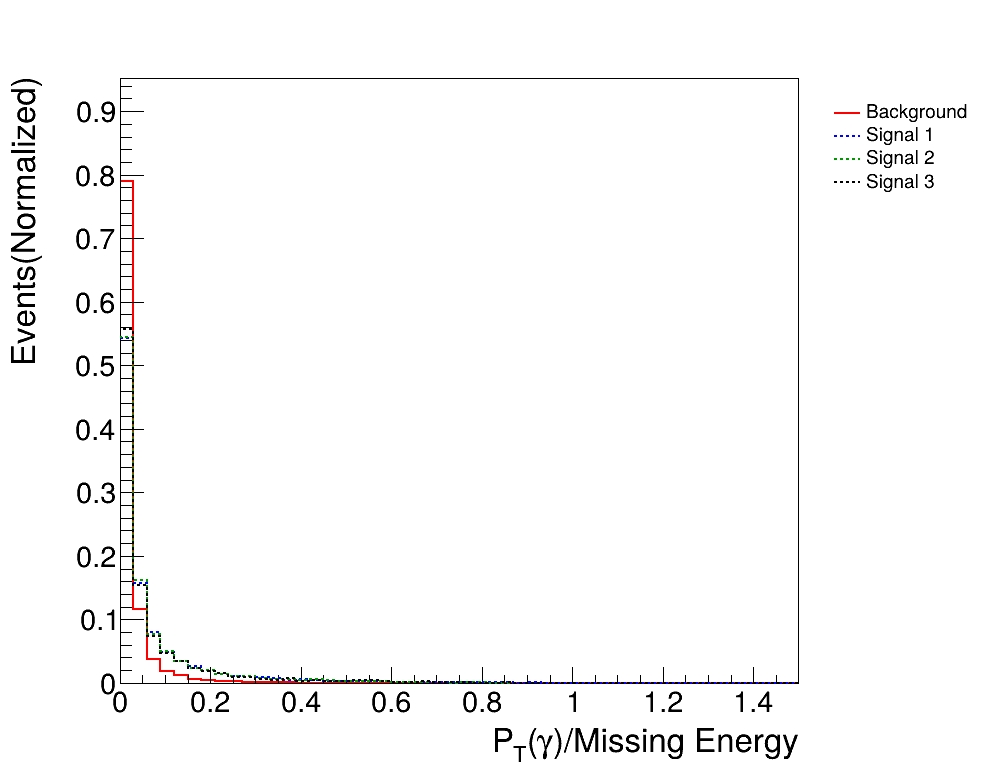}
\caption{Similar to Fig.~\ref{fig:L3_20_offshell}, but for the $\mathcal{L}_{13}$ model with $m_D=20$ GeV. } 
\label{fig:L13_20_offshell}
\end{figure}

\begin{figure}[h]
\centering 
\includegraphics[width=0.48\textwidth]{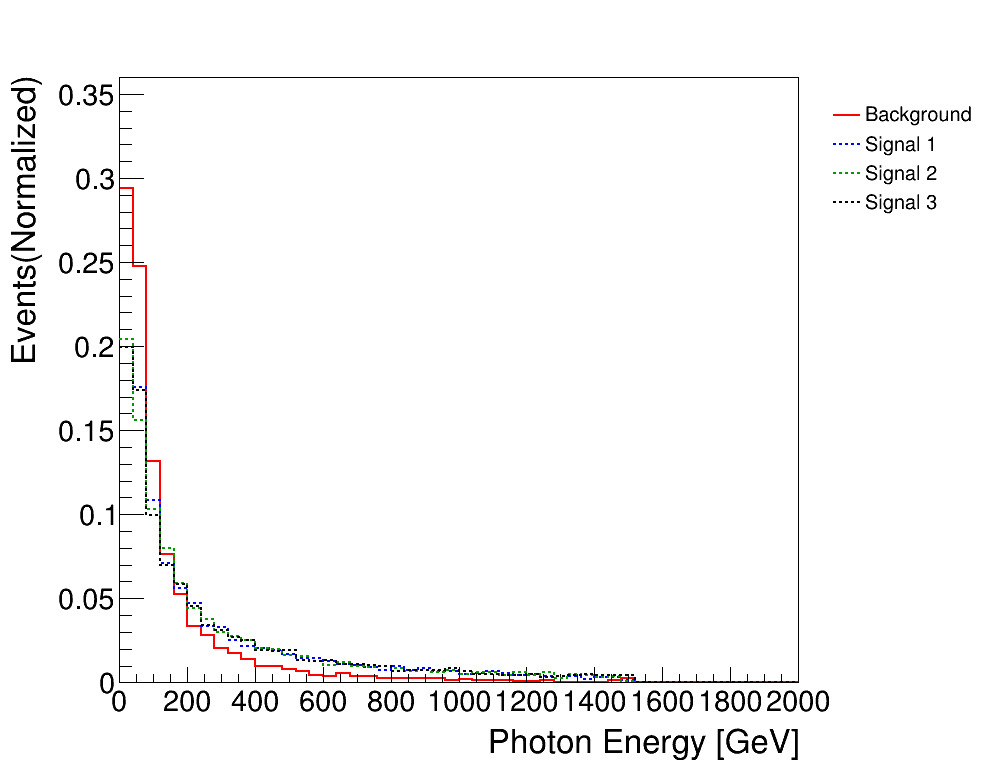} 
\includegraphics[width=0.48\textwidth]{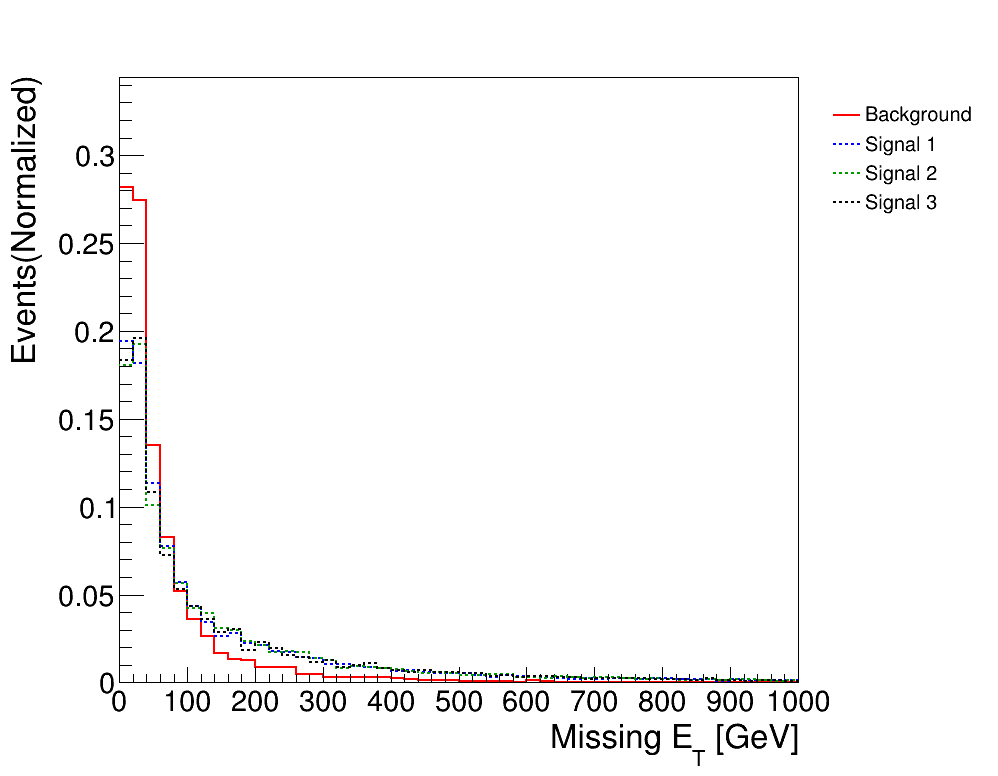}
\includegraphics[width=0.48\textwidth]{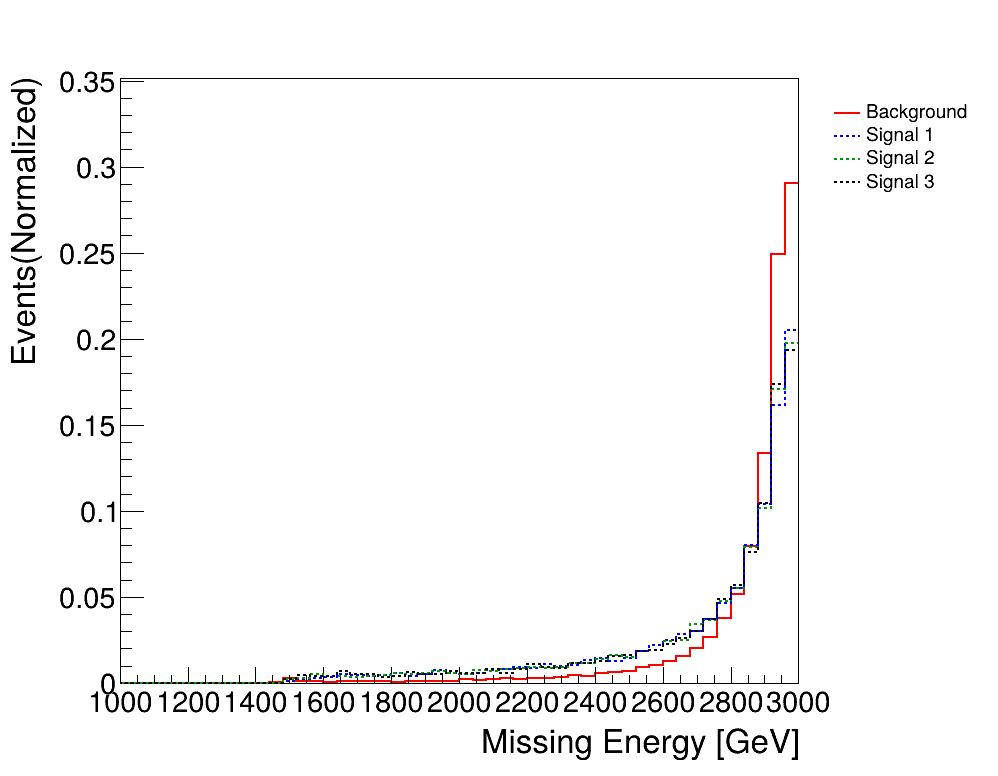}
\includegraphics[width=0.48\textwidth]{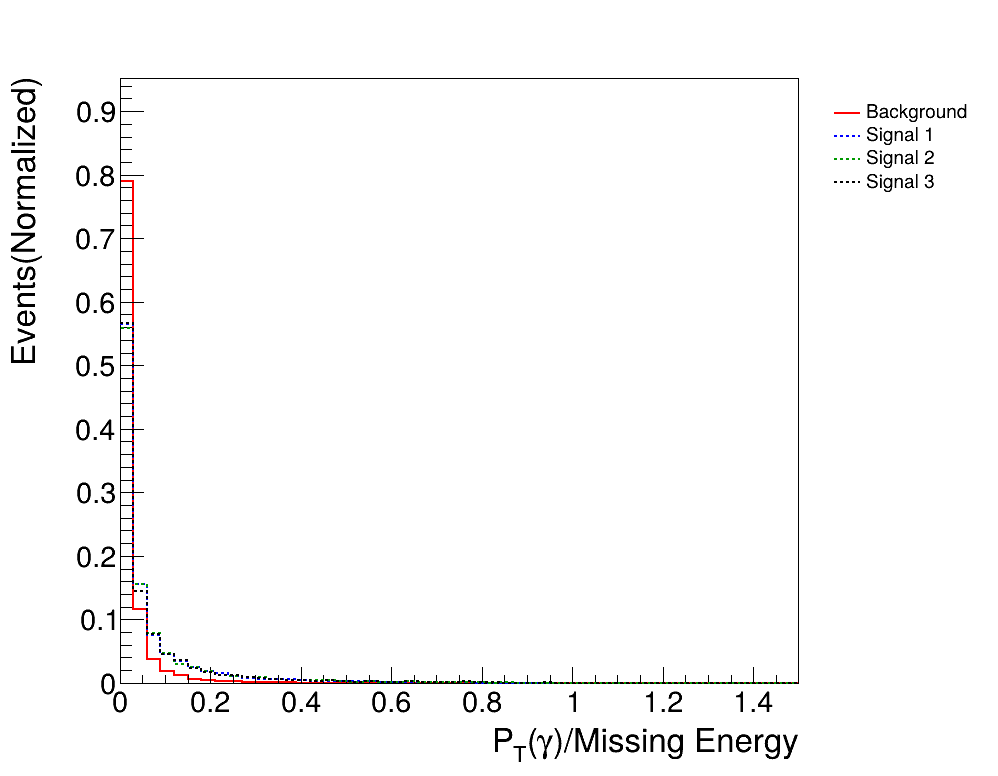}
\caption{Similar to Fig.~\ref{fig:L3_20_offshell}, but for the $\mathcal{L}_{13}$ model with $m_D = 100$ GeV.} 
\label{fig:L13_100_offshell}
\end{figure}

We select two representative DM masses, $m_D=20$ and $100$ GeV, with three benchmark mediator-to-DM mass ratio: (1) Signal-1: \( M/m_D = 1.1 \); (2) Signal-2: \( M/m_D = 1.5 \); (3) Signal-3: \( M/m_D = 1.9 \) for each scenario. The benchmark couplings are: $g_D = 0.5$, $g_f = 0.1$ for the $\mathcal{L}_3$ model; $M_{D\phi} = 100$ GeV, $g_f = 0.1$ for the $\mathcal{L}_9$ model; and $M_{D\phi} = 10$ GeV, $g_f = 0.1$ for the $\mathcal{L}_{13}$ model. Taking the $\mathcal{L}_3$ model as an example, the signal process is $\mu^+\mu^-\to\chi\bar{\chi}\gamma$ via off-shell mediator $\phi$, with irreducible SM background process $\mu^+\mu^-\to\nu\bar{\nu}\gamma$. The corresponding production cross-sections are listed in the second row of Tables~\ref{tab:L3_20_offshell}-\ref{tab:L13_100_offshell} for the respective models and masses, while kinematic distributions appear in Figs.~\ref{fig:L3_20_offshell}-\ref{fig:L13_100_offshell}.

Due to off-shell mediator production, significant overlaps may occur between signal and background distributions in $E(\gamma)$, ${\:/\!\!\!\! E}_T$, and ${\:/\!\!\!\! E}$, comparing with the on-shell case in Sec.~\ref{sec:invisible_onsell}. However, the $\mathcal{L}_9$ model shows obviously less overlap than the $\mathcal{L}_{13}$ model. Nevertheless, the ratio $P_T(\gamma)/{\:/\!\!\!\! E}$ and $M_T$ distributions (defined in Sec.~\ref{sec:invisible_onsell}) remain effective discriminators except for the $\mathcal{L}_{13}$ model. Applying the same event selections as Sec.~\ref{sec:invisible_onsell}, we summarize the selection efficiencies in Tables~\ref{tab:L3_20_offshell}-\ref{tab:L13_100_offshell} for the respective models and masses. The distinguishable kinematics for the $\mathcal{L}_9$ model yield higher signal efficiencies after all cuts, while the nearly identical kinematic distributions between signal and background events for the $\mathcal{L}_{13}$ model pose challenges, resulting in only factor-of-ten signal-background separation. Additionally, signal efficiency increases slightly with $M/m_D$ but remains largely independent of $m_D$ for its value below $100$ GeV. 

\subsection{Vector boson fusion production: Visible and invisible mediator decays} 
\label{sec:VBF}

Since the vector boson fusion (VBF) production at TeV muon
collider becomes more dominant, we further extend the search strategies to the VBF production at the 3 TeV muon collider, with the hard process defined as $\mu^+\mu^- \to \nu_\mu\bar{\nu}_\mu \mu^+\mu^- \mathrm{MED}$. The mediator, $\mathrm{MED}$, subsequently decays into either a muon pair ($\mathrm{MED} \to \mu^+\mu^-$, visible decay) or a DM pair ($\mathrm{MED} \to \mathrm{DM}+\mathrm{DM}$, invisible decay). This process is driven by the weak interaction of the initial-state muon pair, with the on-shell mediator emitted from one of the muon legs. Again, the fixed mass ratio $M/m_D=2.5$ is imposed to ensure the kinematic accessibility of both decay channels, and the coupling parameter ratios follow Table~\ref{tab:visibleL3_decay} to guarantee the branching ratio of either \(\mathcal{B}(\text{MED}\to \mu^+\mu^-) > 99\%\) or \(\mathcal{B}(\text{MED}\to\text{DM}+\text{DM}) > 99\%\).


For the visible decay channel: $\mathrm{MED} \to \mu^+\mu^-$, the full signal process is $\mu^+\mu^- \to \nu_\mu\bar{\nu}_\mu \mu^+\mu^- \mathrm{MED}$, where $\mathrm{MED} \to \mu^+\mu^-$. The major irreducible background is $\mu^+\mu^- \to \nu_l\bar{\nu}_l \mu^+\mu^-\mu^+\mu^-$. For the \(\mathcal{L}_3\) model, the benchmark coupling parameters are set to \(g_f = 0.01\) and \(g_D = 0.08\), corresponding to a coupling constant ratio of \(g_D/g_f = 1/8\); For the \(\mathcal{L}_8\) model, the benchmark coupling parameters are set to \(g_f = 0.01\) and \(g_D = 0.09\), with the corresponding coupling constant ratio being \(g_D/g_f = 1/9\). 

\begin{figure}[h]
\centering 
\includegraphics[width=0.48\textwidth]{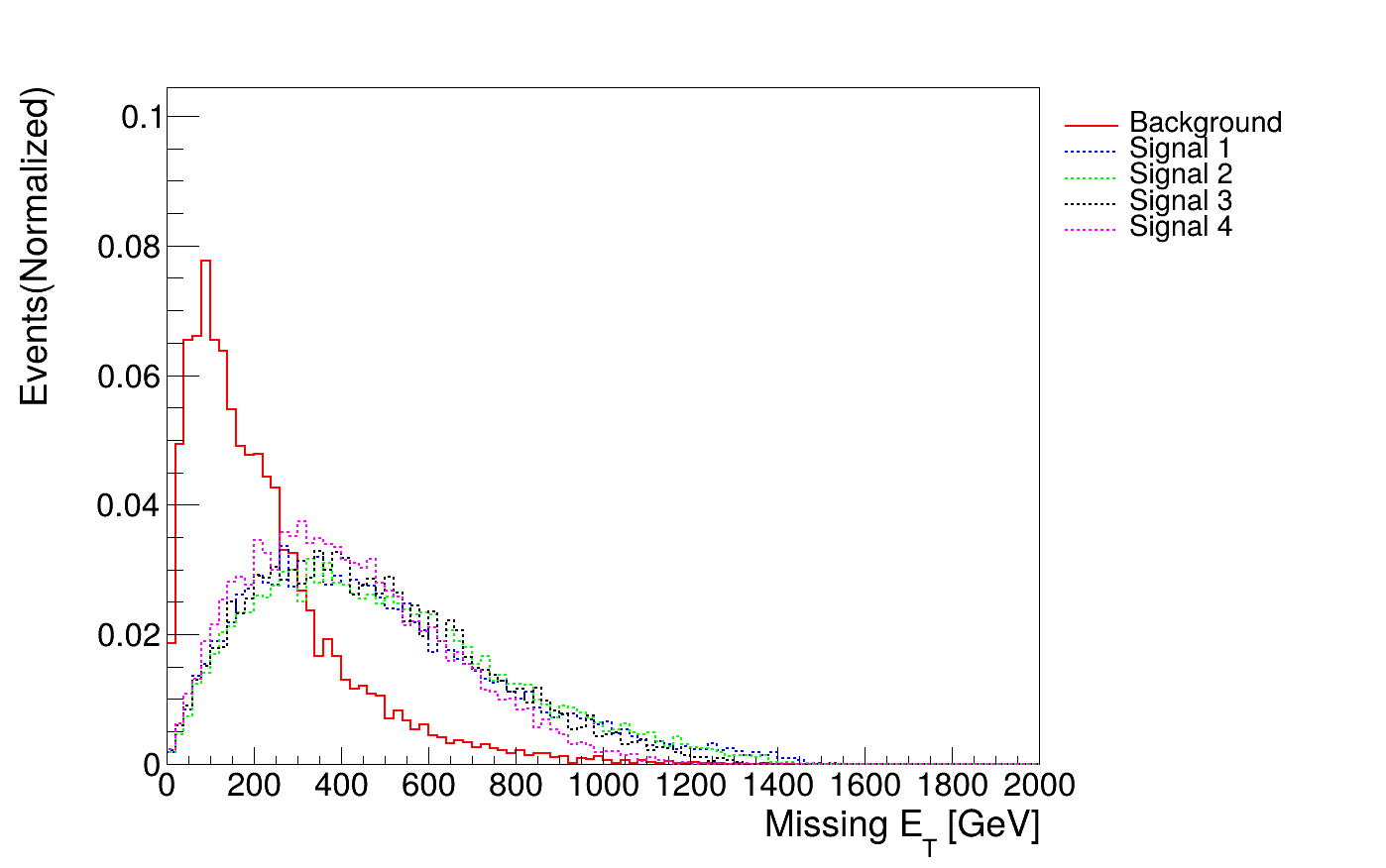} 
\includegraphics[width=0.48\textwidth]{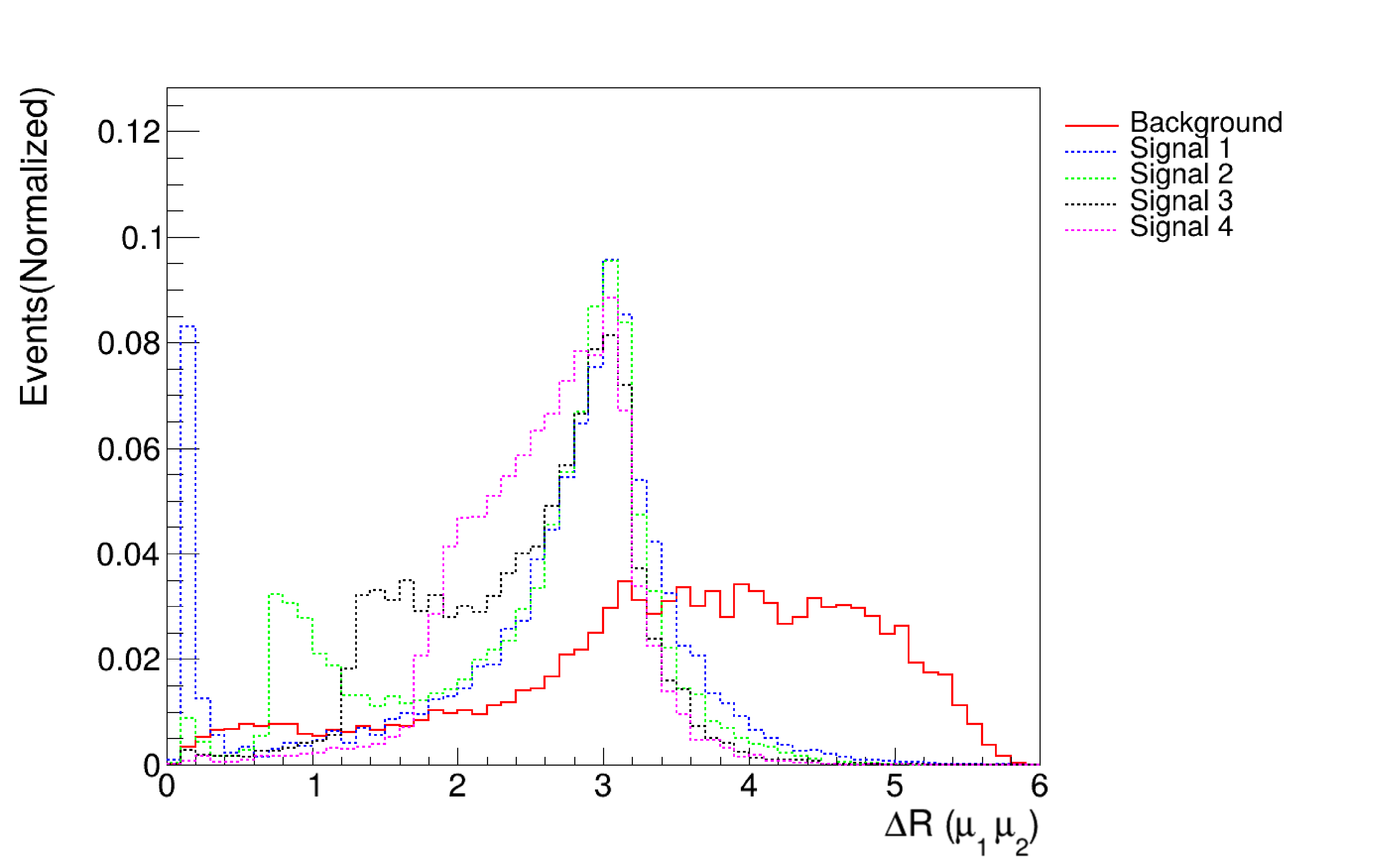}
\includegraphics[width=0.48\textwidth]{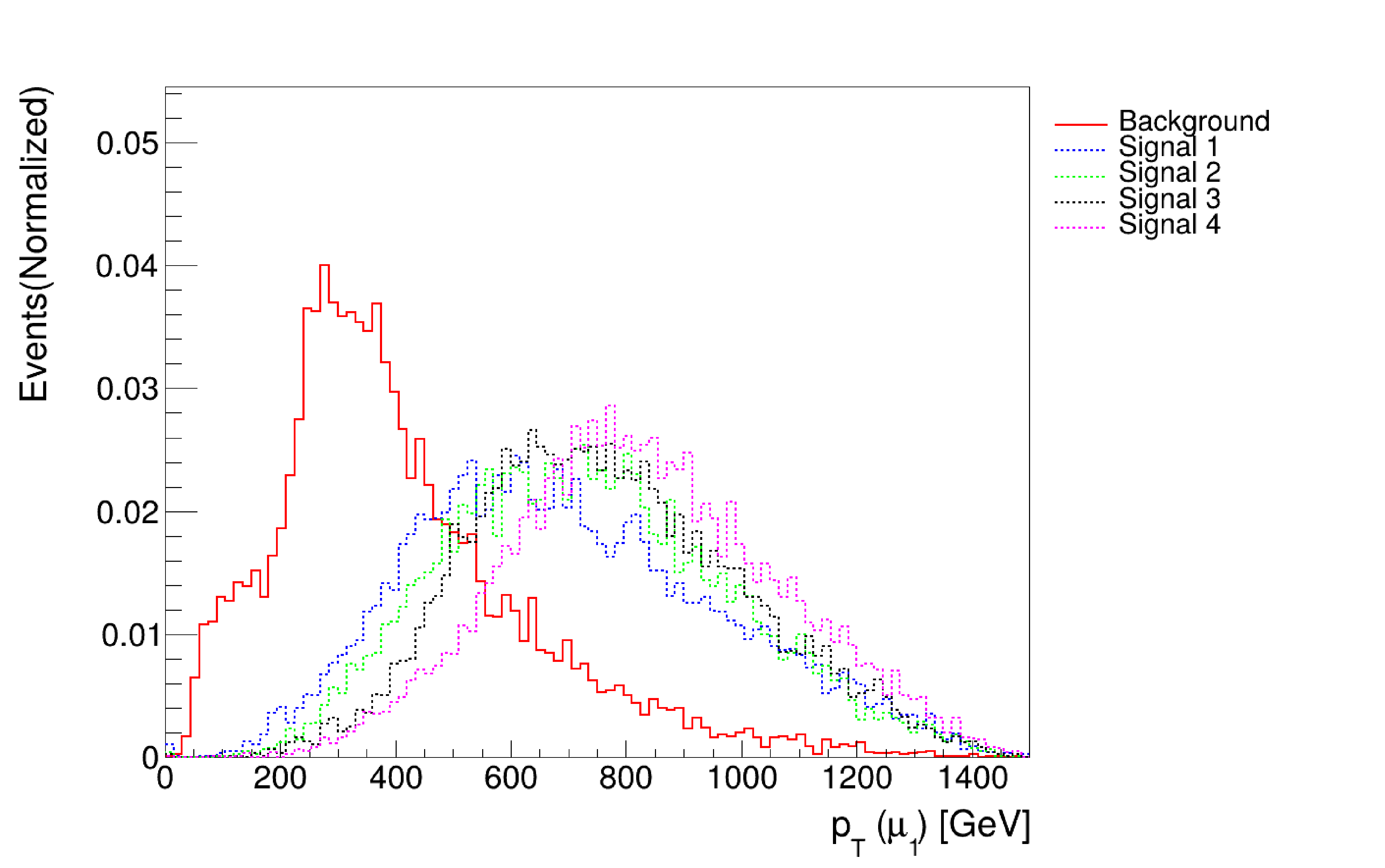}
\includegraphics[width=0.48\textwidth]{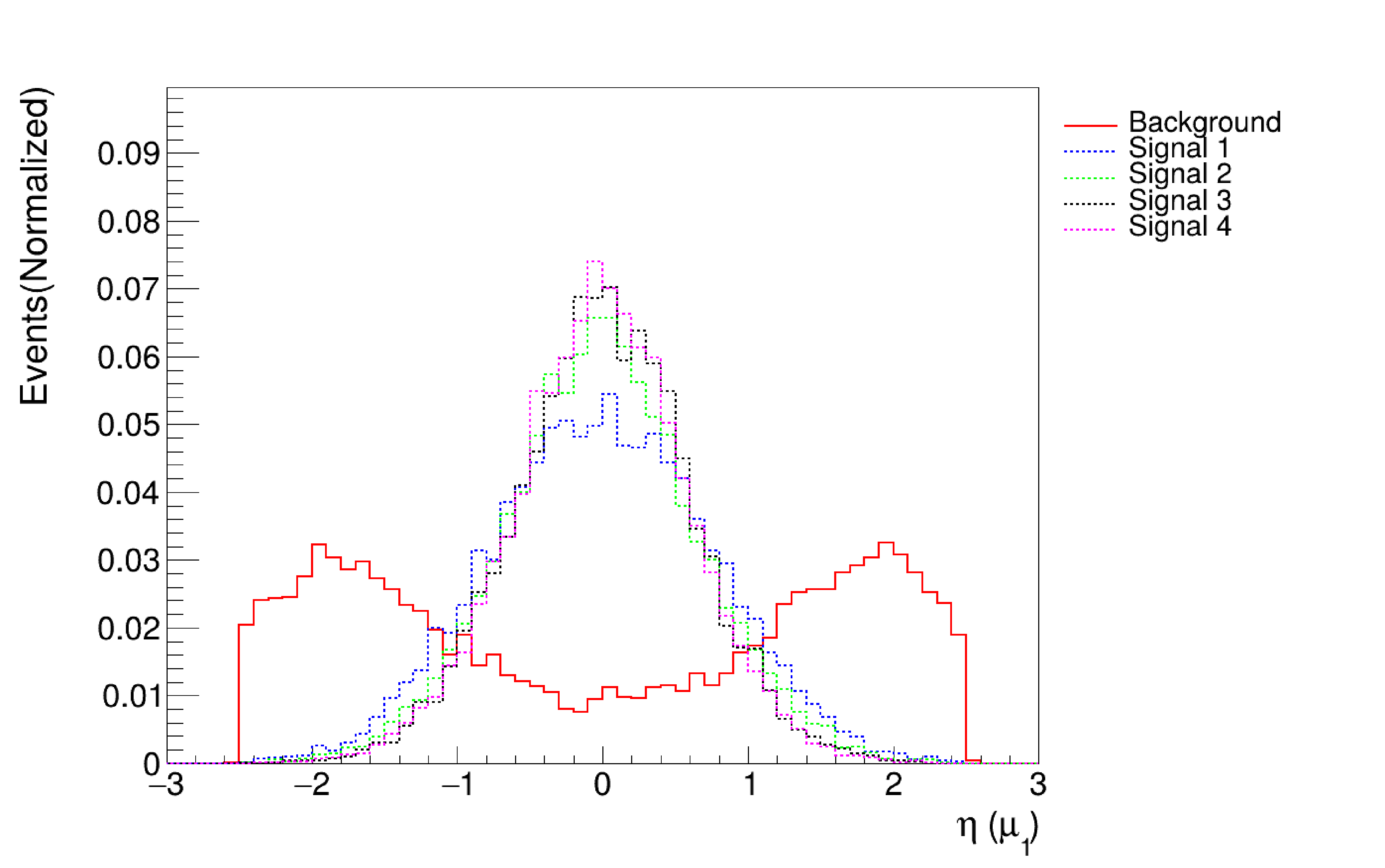}
\caption{Visible on-shell mediator decay scenario ($M/m_D=2.5$) of the $\mathcal{L}_3$ model in the VBF process, representative kinematic distributions include missing transverse energy $ {\:/\!\!\!\! E}_T$ (top-left), angular separation $\Delta R(\mu_1 \mu_2)$ (top-right), transverse momentum of leading muon $p_T(\mu_1)$ (bottom-left), and pseudorapidity of leading muon $|\eta(\mu_1)|$ (bottom-right) for signal-1 ($M = 50$ GeV, blue-dotted line), signal-2 ($M = 500$ GeV, green-dotted line), signal-3 ($M = 900$ GeV, black-dotted line), signal-4 ($M = 1300$ GeV, purple-dotted line) and background (red-solid line). } 
\label{fig:L3_visible}
\end{figure}

Some relevant kinematic distributions of signal and background processes are shown in Fig.~\ref{fig:L3_visible}. Based on the above kinematic distributions, we have optimized the event selection criteria for the multi-muon final state with missing energy, as follows:
\begin{itemize} 
    \item Cut-1 (basic cuts): $N_{\mu^+} \geq 2$, $N_{\mu^-} \geq 2$, ${\:/\!\!\!\! E}_T > 40\ \mathrm{GeV}$, $p_T(\mu) > 20\ \mathrm{GeV}$, and $|\eta(\mu)| < 2.5$;           
    \item Cut-2: $2.2 < \Delta R(\mu_1 \mu_2) < 3.3$;
    \item Cut-3: $p_T(\mu_1) \geq 130\ \mathrm{GeV}$, and $|\eta(\mu_1)| < 0.75$; 
    \item Cut-4: ${\:/\!\!\!\! E}_T > 350\ \mathrm{GeV}$; 
    \item Cut-5: $|M_{\mu\mu}^{\text{best}} - M| < 0.1 M$. 
\end{itemize}
where the ordering of the subscripts $\mu_1$ and $\mu_2$ is determined by their $p_T$ values, and $M_{\mu\mu}^{\text{best}}$ denotes the dimuon pair with invariant mass closest to the mediator mass, selected from all possible dimuon pairs formed by the four muons.

For the invisible decay channel: $\mathrm{MED} \to \text{DM}+\text{DM}$, the full signal process is $\mu^+\mu^- \to \nu_\mu\bar{\nu}_\mu \mu^+\mu^- \mathrm{MED}$, where $\mathrm{MED} \to \text{DM}+\text{DM}$, with missing energy from both neutrino and DM particles. The major background comes from $\mu^+\mu^- \to \nu_l\bar{\nu}_l \mu^+\mu^-$. For the \(\mathcal{L}_3\) model, the benchmark coupling parameters are set to \(g_f = 0.15\) and \(g_D = 0.01\), corresponding to a coupling constant ratio of \(g_D/g_f = 15\); For the \(\mathcal{L}_8\) model, the benchmark coupling parameters are set to \(g_f = 0.12\) and \(g_D = 0.01\), with the corresponding coupling constant ratio being \(g_D/g_f = 12\). 

\begin{figure}[h]
\centering 
\includegraphics[width=0.48\textwidth]{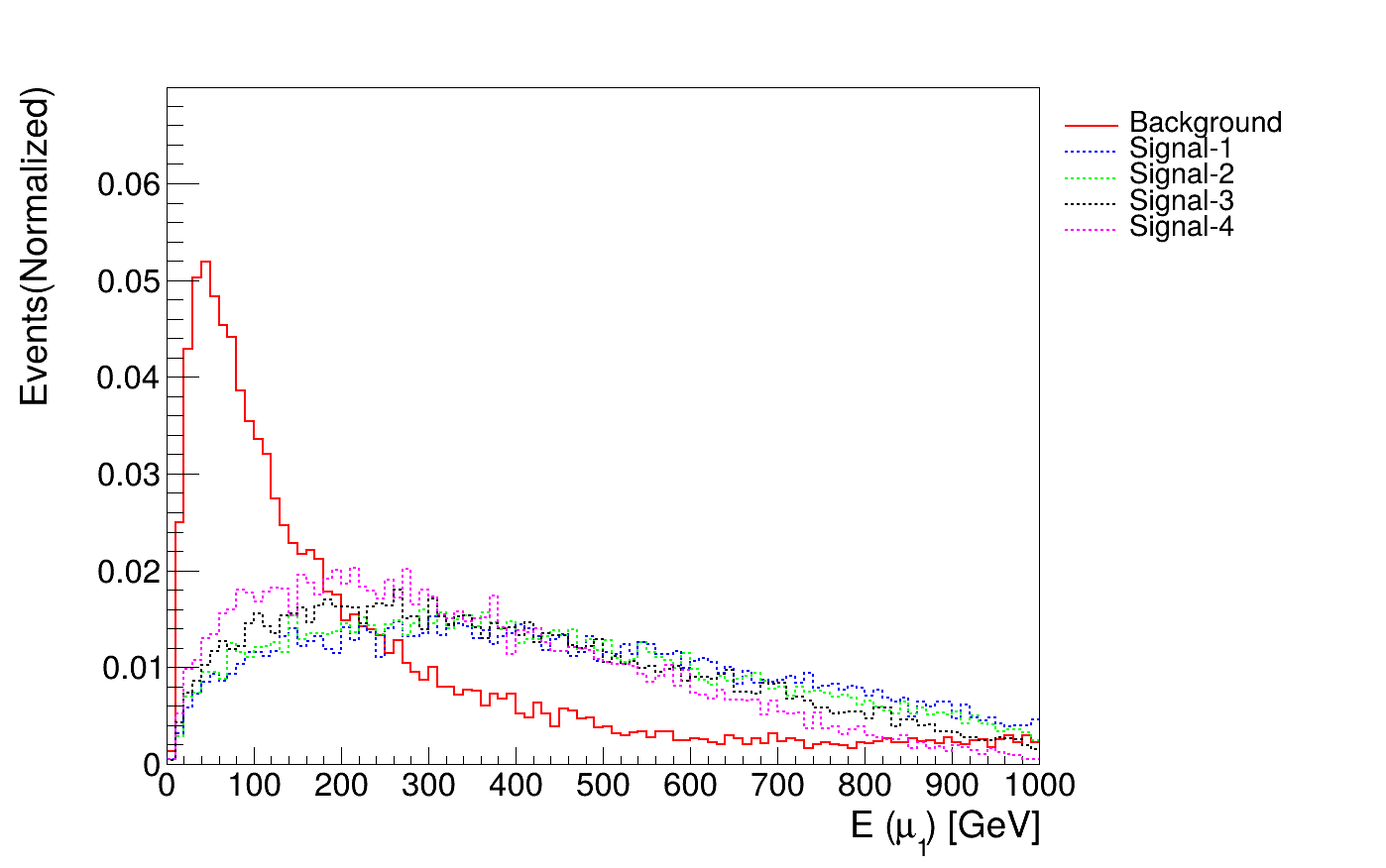} 
\includegraphics[width=0.48\textwidth]{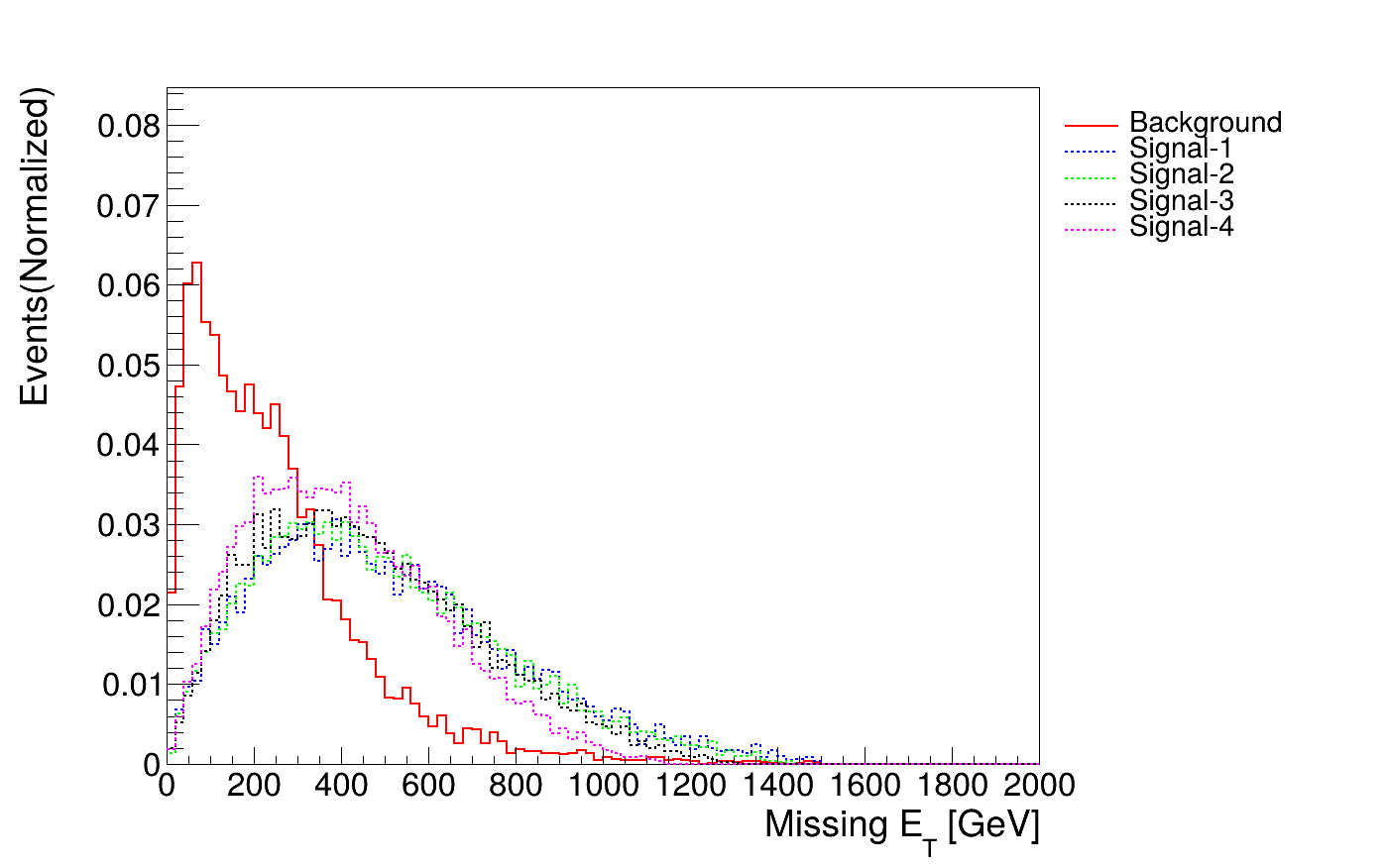}
\includegraphics[width=0.48\textwidth]{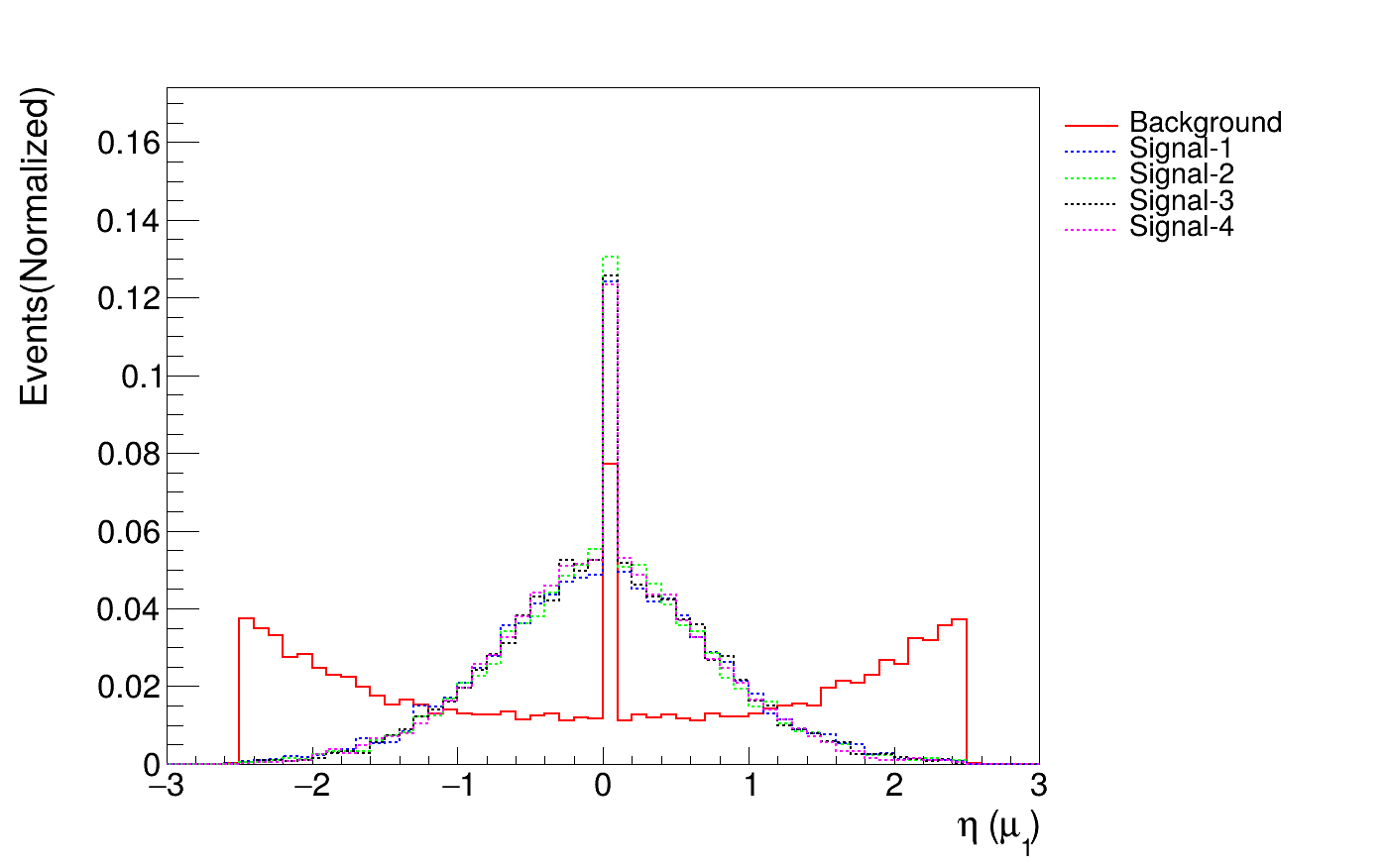}
\includegraphics[width=0.48\textwidth]{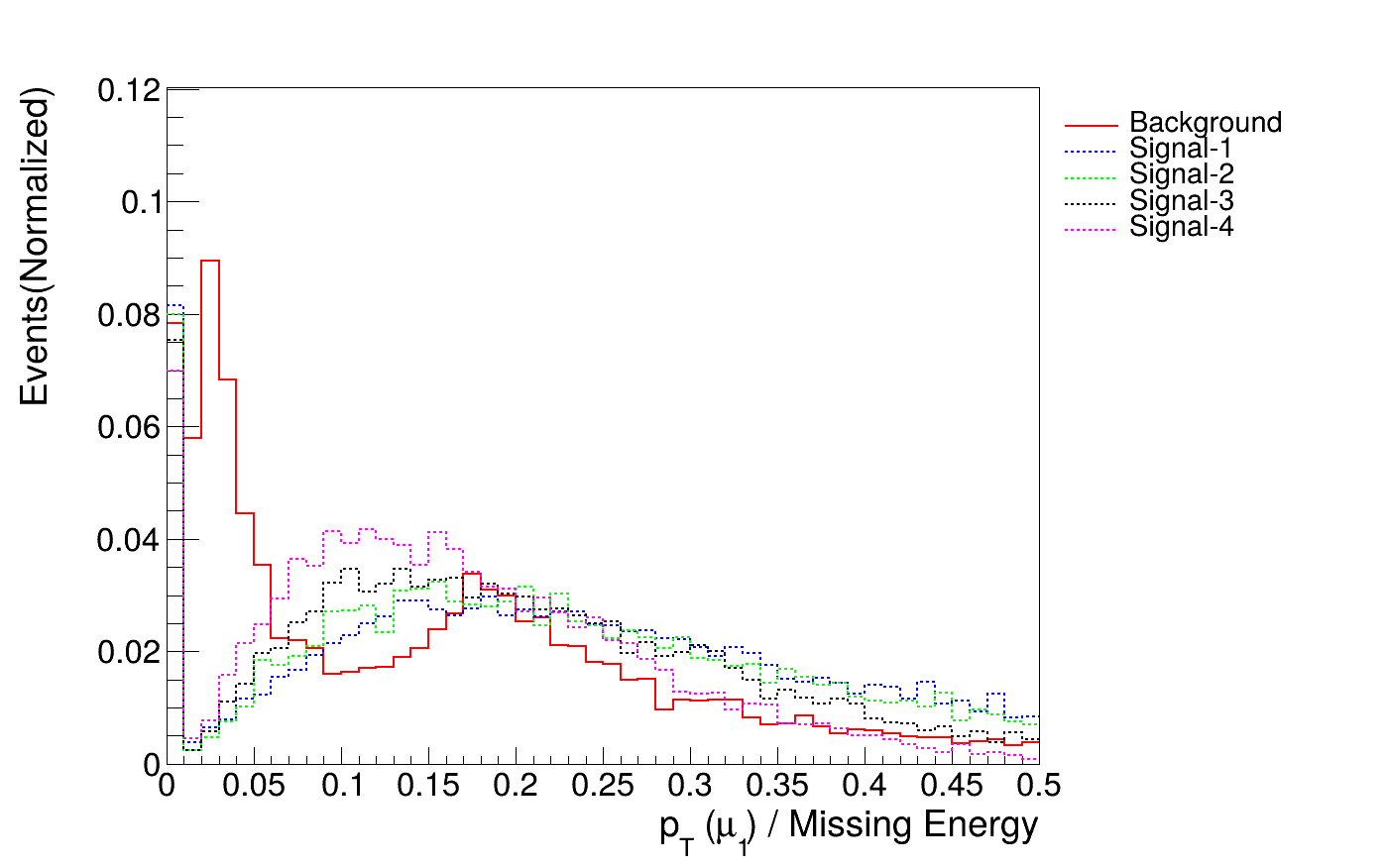}
\caption{Invisible on-shell mediator decay scenario ($M/m_D=2.5$) of the $\mathcal{L}_3$ model in the VBF process, representative kinematic distributions include leading muon energy $E(\mu_1)$ (top-left), missing transverse energy ${\:/\!\!\!\! E}_T$(top-right), pseudorapidity of leading muon $|\eta(\mu_1)|$ (bottom-left), and transverse momentum ratio of leading muon to missing energy (bottom-right) for signal-1 ($M = 50$ GeV, blue-dotted line), signal-2 ($M = 500$ GeV, green-dotted line), signal-3 ($M = 900$ GeV, black-dotted line), signal-4 ($M = 1300$ GeV, purple-dotted line) and background (red-solid line). } 
\label{fig:L3_invisible}
\end{figure}

Figure~\ref{fig:L3_invisible} shows the relevant kinematic distributions for signal and background processes. The event selection criteria are designed to enhance the discrimination between signal and background processes, as follows: 
\begin{itemize} 
    \item Cut-1 (basic cuts): $N_{\mu^+} \geq 1$, $N_{\mu^-} \geq 1$, $p_T(\mu) > 20\ \mathrm{GeV}$, $|\eta(\mu)| < 2.5$, and ${\:/\!\!\!\! E}_T > 40\ \mathrm{GeV}$; 
    \item Cut-2: $E(\mu_1) \geq 350\ \mathrm{GeV}$, and $|\eta(\mu_1)| < 0.8$; 
    \item Cut-3: ${\:/\!\!\!\! E}_T > 250\ \mathrm{GeV}$;  
    \item Cut-4: $p_T(\mu_1)/{\:/\!\!\!\! E}>0.12$. 
\end{itemize}

\begin{table}[htbp]
\centering
\footnotesize  
\begin{tabular}{|>{\centering\arraybackslash}p{1.2cm}|>{\centering\arraybackslash}p{4.8cm}|>{\centering\arraybackslash}p{2.1cm}|*{4}{>{\centering\arraybackslash}p{1.8cm}|}}
\hline
\multicolumn{2}{|c|}{\textbf{Cut description}} & \textbf{background} & \textbf{signal1} & \textbf{signal2} & \textbf{signal3} & \textbf{signal4}  \\
\hline
\multicolumn{2}{|c|}{\textbf{Cross-section [fb]}} & $0.428$ & $2.061$ & $2.078$ & $1.473$ & $0.899$  \\
\hline
\textbf{Cut-1} & $N_{\mu^+} \geq 2$ and $N_{\mu^-} \geq 2$, \newline $p_T(\mu) > 20\ \mathrm{GeV}$, 
$|\eta(\mu)| < 2.5$, \newline ${\:/\!\!\!\! E}_T > 40\ \mathrm{GeV}$ & {\footnotesize 0.65} & {\footnotesize 0.79} & {\footnotesize 0.84} & {\footnotesize 0.83} & {\footnotesize 0.82}  \\
\hline
\textbf{Cut-2} & $2.2 < \Delta R(\mu_1, \mu_2) < 3.3$& {\footnotesize 0.15} & {\footnotesize 0.48} & {\footnotesize 0.50} & {\footnotesize 0.49} & {\footnotesize 0.60}  \\
\hline
\textbf{Cut-3} & $p_T(\mu_1) \geq 130\ \mathrm{GeV}$, \newline 
$|\eta(\mu_1)| < 0.75$ & {\footnotesize 0.05} & {\footnotesize 0.38} & {\footnotesize 0.42} & {\footnotesize 0.41} & {\footnotesize 0.49}  \\
\hline
\textbf{Cut-4} & ${\:/\!\!\!\! E}_T > 350\ \mathrm{GeV}$ & {\footnotesize 0.01} & {\footnotesize 0.23} & {\footnotesize 0.26} & {\footnotesize 0.23} & {\footnotesize 0.25}  \\
\hline
\multirow{4}{*}{\textbf{Cut-5}} & \multirow{4}{*}{$|M_{\mu\mu}^{\text{best}} - M| < 0.1 M$} & {\footnotesize $6.0 \times 10^{-4}$} & {\footnotesize 0.22} & {\footnotesize /} & {\footnotesize /} & {\footnotesize /}  \\
\cline{3-7}
&& {\footnotesize $6.0 \times 10^{-3}$} & {\footnotesize /} & {\footnotesize 0.25} & {\footnotesize /} & {\footnotesize /}  \\
\cline{3-7}
&& {\footnotesize $5.4 \times 10^{-3}$} & {\footnotesize /} & {\footnotesize /} & {\footnotesize 0.22} & {\footnotesize /}  \\
\cline{3-7}
&& {\footnotesize $6.5 \times 10^{-3}$} & {\footnotesize /} & {\footnotesize /} & {\footnotesize /} & {\footnotesize 0.24}  \\
\hline
\end{tabular}
\caption{The cut-flow table for visible on-shell mediator decay scenario ($M/m_D=2.5$) of the $\mathcal{L}_3$ model in the VBF process and its relevant background with cumulative efficiencies from \textbf{Cut-1} to \textbf{Cut-5}. The four BPs are signal-1 ($M = 50$ GeV), signal-2 ($M = 500$ GeV), signal-3 ($M = 900$ GeV), and signal-4 ($M = 1300$ GeV). }
\label{tab:visibleL3_decay}
\end{table}

\begin{table}[htbp]
\centering
\footnotesize  
\begin{tabular}{|>{\centering\arraybackslash}p{1.2cm}|>{\centering\arraybackslash}p{4.8cm}|>{\centering\arraybackslash}p{2.1cm}|*{4}{>{\centering\arraybackslash}p{1.8cm}|}}
\hline
\multicolumn{2}{|c|}{\textbf{Cut description}} & \textbf{background} & \textbf{signal1} & \textbf{signal2} & \textbf{signal3} & \textbf{signal4}  \\
\hline
\multicolumn{2}{|c|}{\textbf{Cross-section [fb]}} & $0.428$ & $891.3$ & $20.3$ & $7.823$ & $3.748$  \\
\hline
\textbf{Cut-1} & $N_{\mu^+} \geq 2$ and $N_{\mu^-} \geq 2$, \newline $p_T(\mu) > 20\ \mathrm{GeV}$, 
$|\eta(\mu)| < 2.5$, \newline ${\:/\!\!\!\! E}_T > 40\ \mathrm{GeV}$ & {\footnotesize 0.65} & {\footnotesize 0.80} & {\footnotesize 0.83} & {\footnotesize 0.83} & {\footnotesize 0.83}  \\
\hline
\textbf{Cut-2} & $2.2 < \Delta R(\mu_1, \mu_2) < 3.3$& {\footnotesize 0.15} & {\footnotesize 0.53} & {\footnotesize 0.51} & {\footnotesize 0.49} & {\footnotesize 0.59}  \\
\hline
\textbf{Cut-3} & $p_T(\mu_1) \geq 130\ \mathrm{GeV}$, \newline
$|\eta(\mu_1)| < 0.75$ & {\footnotesize 0.05} & {\footnotesize 0.44} & {\footnotesize 0.45} & {\footnotesize 0.41} & {\footnotesize 0.46}  \\
\hline
\textbf{Cut-4} & ${\:/\!\!\!\! E}_T > 350\ \mathrm{GeV}$ & {\footnotesize 0.01} & {\footnotesize 0.28} & {\footnotesize 0.28} & {\footnotesize 0.23} & {\footnotesize 0.22}  \\
\hline
\multirow{4}{*}{\textbf{Cut-5}} & \multirow{4}{*}{$|M_{\mu\mu}^{\text{best}} - M| < 0.1 M$} & {\footnotesize $6.0 \times 10^{-4}$} & {\footnotesize 0.27} & {\footnotesize /} & {\footnotesize /} & {\footnotesize /}  \\
\cline{3-7}
&& {\footnotesize $6.0 \times 10^{-3}$} & {\footnotesize /} & {\footnotesize 0.27} & {\footnotesize /} & {\footnotesize /}  \\
\cline{3-7}
&& {\footnotesize $5.4 \times 10^{-3}$} & {\footnotesize /} & {\footnotesize /} & {\footnotesize 0.22} & {\footnotesize /}  \\
\cline{3-7}
&& {\footnotesize $6.5 \times 10^{-3}$} & {\footnotesize /} & {\footnotesize /} & {\footnotesize /} & {\footnotesize 0.20}  \\
\hline
\end{tabular}
\caption{Similar to Table~\ref{tab:visibleL3_decay}, but for the $\mathcal{L}_8$ model.}
\label{tab:visibleL8_decay}
\end{table}

\begin{table}[htbp]
\centering
\footnotesize  
\begin{tabular}{|>{\centering\arraybackslash}p{1.2cm}|>{\centering\arraybackslash}p{4.8cm}|>{\centering\arraybackslash}p{2.1cm}|*{4}{>{\centering\arraybackslash}p{1.8cm}|}}
\hline
\multicolumn{2}{|c|}{\textbf{Cut description}} & \textbf{background} & \textbf{signal1} & \textbf{signal2} & \textbf{signal3} & \textbf{signal4}  \\
\hline
\multicolumn{2}{|c|}{\textbf{Cross-section [fb]}} & $165$ & $4.20\times 10^{-2}$ & $3.36\times 10^{-2}$ & $2.36\times 10^{-2}$ & $1.46\times 10^{-2}$  \\
\hline
\textbf{Cut-1} & $N_{\mu^+} \geq 1$ and $N_{\mu^-} \geq 1$, \newline $p_T(\mu) > 20\ \mathrm{GeV}$, $|\eta(\mu)| < 2.5$ \newline ${\:/\!\!\!\! E}_T > 40\ \mathrm{GeV}$ & {\footnotesize 0.79} & {\footnotesize 0.90} & {\footnotesize 0.90} & {\footnotesize 0.90} & {\footnotesize 0.90}  \\
\hline
\textbf{Cut-2} & $E(\mu_1) \geq 350\ \mathrm{GeV}$, \newline
$|\eta(\mu_1)| < 0.8$ & {\footnotesize 0.05} & {\footnotesize 0.59} & {\footnotesize 0.59} & {\footnotesize 0.56} & {\footnotesize 0.52}  \\
\hline
\textbf{Cut-3} & ${\:/\!\!\!\! E}_T > 350\ \mathrm{GeV}$ & {\footnotesize 0.04} & {\footnotesize 0.46} & {\footnotesize 0.46} & {\footnotesize 0.44} & {\footnotesize 0.40}  \\
\hline
\textbf{Cut-4} & $P_T(\mu_1)/{\:/\!\!\!\! E}>0.12$ & {\footnotesize 0.03} & {\footnotesize 0.45} & {\footnotesize 0.45} & {\footnotesize 0.43} & {\footnotesize 0.39}  \\
\hline
\end{tabular}
\caption{The cut-flow table for invisible on-shell mediator decay scenario ($M/m_D=2.5$) of the $\mathcal{L}_3$ model in the VBF process and its relevant background with cumulative efficiencies from \textbf{Cut-1} to \textbf{Cut-4}. The four BPs are signal-1 ($M = 50$ GeV), signal-2 ($M = 500$ GeV), signal-3 ($M = 900$ GeV), and signal-4 ($M = 1300$ GeV). }
\label{tab:invisibleL3_decay}
\end{table}

\begin{table}[htbp]
\centering
\footnotesize  
\begin{tabular}{|>{\centering\arraybackslash}p{1.2cm}|>{\centering\arraybackslash}p{4.8cm}|>{\centering\arraybackslash}p{2.1cm}|*{4}{>{\centering\arraybackslash}p{1.8cm}|}}
\hline
\multicolumn{2}{|c|}{\textbf{Cut description}} & \textbf{background} & \textbf{signal1} & \textbf{signal2} & \textbf{signal3} & \textbf{signal4}  \\
\hline
\multicolumn{2}{|c|}{\textbf{Cross-section [fb]}} & $165$ & $19.28$ & $0.26$ & $9.99\times 10^{-2}$ & $4.83\times 10^{-2}$  \\
\hline
\textbf{Cut-1} & $N_{\mu^+} \geq 1$ and $N_{\mu^-} \geq 1$, \newline $p_T(\mu) > 20\ \mathrm{GeV}$, $|\eta(\mu)| < 2.5$ \newline ${\:/\!\!\!\! E}_T > 40\ \mathrm{GeV}$ & {\footnotesize 0.79} & {\footnotesize 0.91} & {\footnotesize 0.91} & {\footnotesize 0.90} & {\footnotesize 0.91}  \\
\hline
\textbf{Cut-2} & $E(\mu_1) \geq 350\ \mathrm{GeV}$, \newline
$|\eta(\mu_1)| < 0.8$ & {\footnotesize 0.05} & {\footnotesize 0.66} & {\footnotesize 0.64} & {\footnotesize 0.59} & {\footnotesize 0.55}  \\
\hline
\textbf{Cut-3} & ${\:/\!\!\!\! E}_T > 350\ \mathrm{GeV}$ & {\footnotesize 0.04} & {\footnotesize 0.58} & {\footnotesize 0.53} & {\footnotesize 0.46} & {\footnotesize 0.39}  \\
\hline
\textbf{Cut-4} & $P_T(\mu_1)/{\:/\!\!\!\! E}>0.12$ & {\footnotesize 0.03} & {\footnotesize 0.57} & {\footnotesize 0.52} & {\footnotesize 0.45} & {\footnotesize 0.38}  \\
\hline
\end{tabular}
\caption{Similar to Table~\ref{tab:invisibleL3_decay}, but for the $\mathcal{L}_8$ model.}
\label{tab:invisibleL8_decay}
\end{table}

Finally, we show the cut-flow in Tables~\ref{tab:visibleL3_decay},~\ref{tab:visibleL8_decay} and Tables~\ref{tab:invisibleL3_decay},~\ref{tab:invisibleL8_decay} for various signal BPs of $\mathcal{L}_3$ and $\mathcal{L}_8$ models as well as the background in the VBF process under the visible and invisible on-shell mediator decay scenarios ($M/m_D=2.5$), respectively. 

\section{Results and discussions}
\label{sec:exclusion}

After establishing the four search strategies in Sec.~\ref{sec:search}, we determine the efficiencies ($\epsilon_{s,b}$) for both signal and background events. Two specific scenarios required adjustments to the selection criteria. First, for mediator masses near $1500$ GeV, the initial requirement of $E(\gamma) > 1200$ GeV in Cut-2 for the search of invisible on-shell mediator decays proved overly restrictive; we therefore relaxed it to $E(\gamma) > 1000$ GeV. Similarly, for the mediator mass close to $20$ GeV in the analysis of visible on-shell mediator decays, the pseudorapidity requirements in Cut-2, Cut-3, and the original $p_T(\mu^\pm) > 150$ GeV in Cut-3 were found to be overly stringent. Therefore, we relaxed them to $p_T(\mu^\pm) > 50$ GeV, $|\eta(\gamma)| < 2.2$, and $|\eta(\mu^\pm)| < 2.2$, respectively. Moreover, signal and background production cross-sections ($\sigma_{s,b}$) are computed using \texttt{MadGraph5\_aMC@NLO}. Additionally, we adopt an optimistic integrated luminosity of \(L = 4400\ \text{fb}^{-1}\)~\cite{Delahaye:2019omf,Long:2020wfp}. The expected numbers of signal and background events are then given by $N_{s,b} = \sigma_{s,b}\times\epsilon_{s,b}\times L$. 

After obtaining the signal and background event counts, the signal significance is calculated as~\cite{Read:2002hq,Arhrib:2019ywg}
\begin{equation}
Z = \sqrt{2 \cdot [ (N_s + N_b) \cdot \ln(1 + \frac{N_s}{N_b}) - N_s ]}.  
\end{equation}
To account for systematic uncertainties in the background, this formula is modified to~\cite{Arhrib:2019ywg,Cheung:2023nzg}
\begin{equation}
Z = \sqrt{2 \cdot [ (N_s + N_b) \cdot \ln[ \frac{(N_s + N_b)(N_b + \sigma_b^2)}{N_b^2 + (N_s + N_b)\sigma_b^2} ] - \frac{N_s^2}{\sigma_b^2} \cdot \ln(1 + \frac{N_b\sigma_b^2}{N_s^2 + N_b\sigma_b^2}) ]}. 
\end{equation}
In this work, we adopt a conservative estimate for the systematic uncertainty, setting \(\sigma_b = 0.05 N_b\), which corresponds to flat $5\%$ of the total background event. This flat $5\%$ systematic uncertainty is determined by referring to the typical estimation range of systematic uncertainties in muon collider studies.
Here we use a significance threshold of \(Z = 1.96\), corresponding to a $95\%$ confidence level, to project the future exclusion limits for $\mathcal{L}_3$, $\mathcal{L}_4$, $\mathcal{L}_8$, $\mathcal{L}_9$, $\mathcal{L}_{10}$, $\mathcal{L}_{13}$, and $\mathcal{L}_{14}$ at a 3 TeV muon collider.

\begin{table}[htbp]
\centering
\footnotesize  
\begin{tabular}{|>{\centering\arraybackslash}p{1.5cm}|>{\centering\arraybackslash}p{3cm}|>{\centering\arraybackslash}p{4.5cm}|>{\centering\arraybackslash}p{4.5cm}|}
\hline
\textbf{Models} & \textbf{Coupling parameter} & \textbf{Visible on-shell mediator decays} & \textbf{Invisible on-shell mediator decays} \\
\hline
$\mathcal{L}_3$ & $g_D/g_f$ & $1/8$ & $15$ \\
\hline
$\mathcal{L}_4$ &  $g_D/g_f$ & $1/8$ & $15$ \\
\hline
$\mathcal{L}_8$ &  $g_D/g_f$ & $1/9$ & $12$ \\
\hline
$\mathcal{L}_9$ & $M_{D\phi}\ (\text{TeV})/g_f$ & $3.66\times 10^{-3}$ -- $2.75\times 10^{-1}$ & $0.91$-- $27.3$\\
\hline
$\mathcal{L}_{10}$ & $M_{D\phi}\ (\text{TeV})/g_f$ & $3.66\times 10^{-3}$ -- $2.75\times 10^{-1}$ & $0.91$-- $27.3$ \\
\hline
$\mathcal{L}_{13}$ & $M_{D\phi}\ (\text{TeV})/g_f$ & $1.43\times 10^{-3}$ -- $1.07\times 10^{-1}$ & $0.356$-- $10.68$ \\
\hline
$\mathcal{L}_{14}$ & $M_{D\phi}\ (\text{TeV})/g_f$ & $1.43\times 10^{-3}$ -- $1.07\times 10^{-1}$ & $0.356$-- $10.68$ \\
\hline
\end{tabular}
\caption{The ratio of coupling parameters \(g_D/g_f\) (for $\mathcal{L}_3$, $\mathcal{L}_4$, $\mathcal{L}_8$) and \(M_{D\phi}\ (\text{TeV})/g_f\) (for $\mathcal{L}_9$, $\mathcal{L}_{10}$, $\mathcal{L}_{13}$, $\mathcal{L}_{14}$) used in this analysis for visible on-shell mediator decays with the condition \(\mathcal{B}(\text{MED}\to \mu^+\mu^-) > 99\%\) and invisible on-shell mediator decays with the condition \(\mathcal{B}(\text{MED}\to\text{DM}+\text{DM}) > 99\%\), respectively. }
\label{tab:coupling_parameters}
\end{table}


\begin{figure}[h]
\centering 
\includegraphics[width=0.48\textwidth]{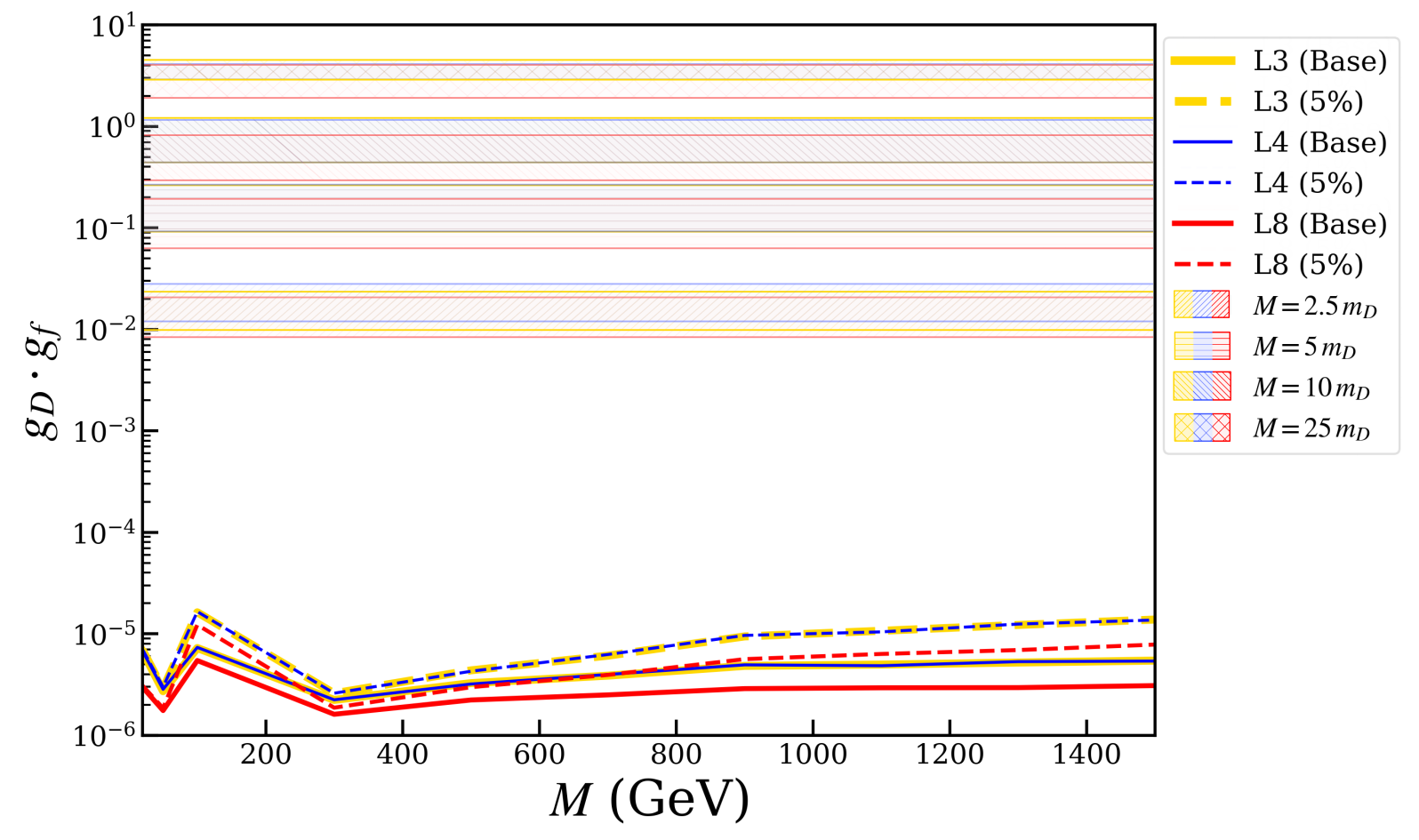} 
\includegraphics[width=0.48\textwidth]{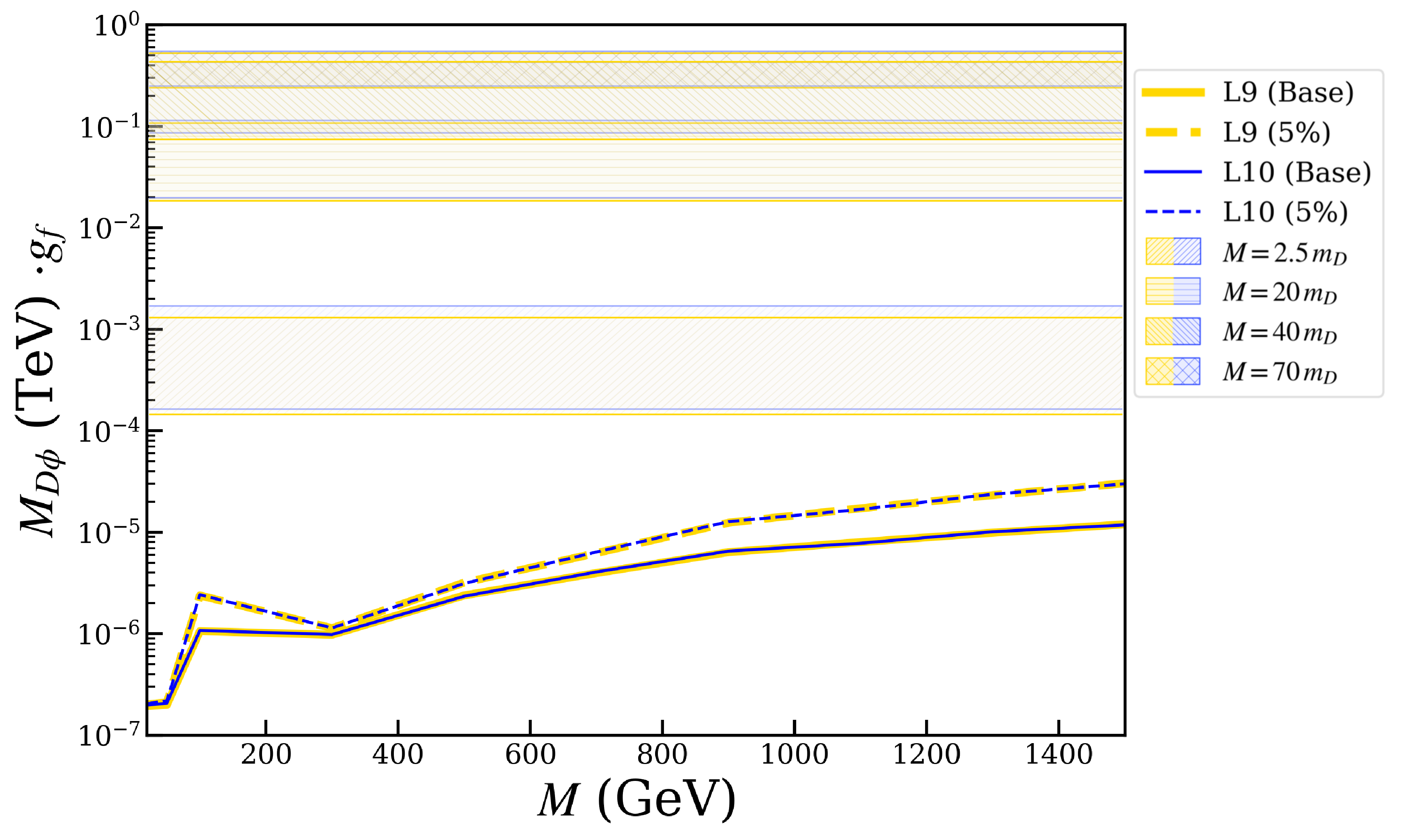}
\includegraphics[width=0.48\textwidth]{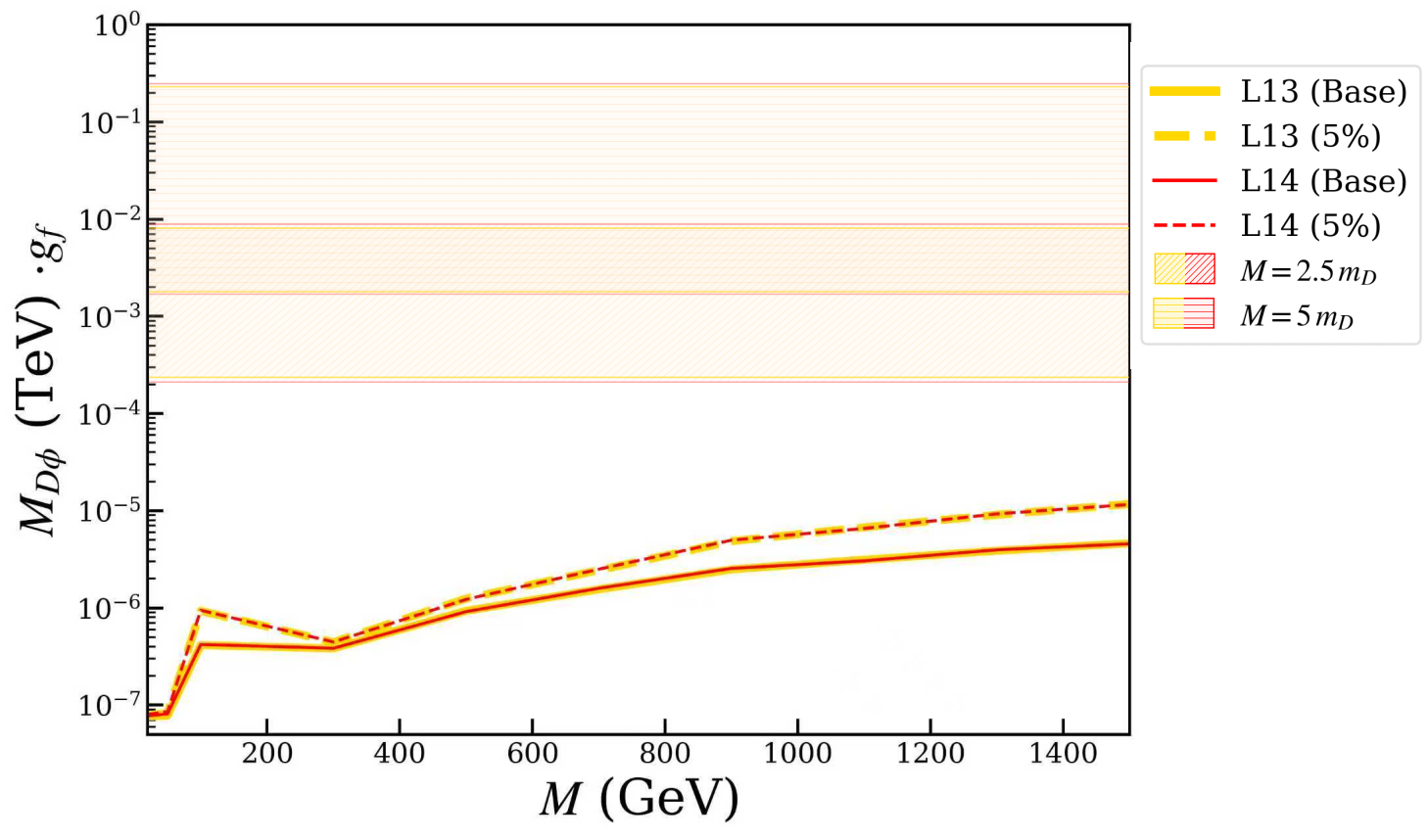}
\caption{Projected exclusion limits at $95\%$ confidence level on the parameter space for visible on-shell mediator decays in seven models. The solid and dashed lines correspond to the cases without and with a flat $5\%$ background systematic uncertainty, respectively. The condition \(\mathcal{B}(\text{MED}\to \mu^+\mu^-) > 99\%\) is imposed to determine the ratio of coupling parameters \(g_D/g_f\) (for $\mathcal{L}_3$, $\mathcal{L}_4$, $\mathcal{L}_8$) and \(M_{D\phi}\ (\text{TeV})/g_f\) (for $\mathcal{L}_9$, $\mathcal{L}_{10}$, $\mathcal{L}_{13}$, $\mathcal{L}_{14}$) as shown in Table~\ref{tab:coupling_parameters}. Colored bands denote the GCE-favored parameter regions consistent with Ref.~\cite{Abdughani:2021oit}, where distinct colors correspond to individual models and various patterns within the bands distinguish the mediator-to-DM mass ratios $M/m_D$ (see the main text for details).}
\label{fig:exclusion_A}
\end{figure}

For visible on-shell mediator decays, we impose the condition \(\mathcal{B}(\text{MED}\to \mu^+\mu^-) > 99\%\) to determine the coupling parameter ratios, which in turn fixes \(g_D/g_f\) for the $\mathcal{L}_3$, $\mathcal{L}_4$, and $\mathcal{L}_8$ models and \(M_{D\phi}/g_f\) for the $\mathcal{L}_9$, $\mathcal{L}_{10}$, $\mathcal{L}_{13}$, and $\mathcal{L}_{14}$ models. These ratios are listed in the third column of Table~\ref{tab:coupling_parameters}. We can find that the coupling parameter ratios for four models $\mathcal{L}_9$, $\mathcal{L}_{10}$, $\mathcal{L}_{13}$, and $\mathcal{L}_{14}$ span a range of values rather than taking a single fixed one. This behavior occurs because the dimension-1 coupling \(M_{D\phi}\), which is closely related to the mediator mass for the mediator branching ratio calculations. Specifically, the ratio $M_{D\phi}/g_f$ required to satisfy \(\mathcal{B}(\text{MED}\to \mu^+\mu^-) > 99\%\) increases monotonically with the mediator mass. Furthermore, relaxing the branching ratio condition to values below $99\%$ would lead to a decrease in both the $g_D/g_f$ and $M_{D\phi}/g_f$ ratios.

Using a mass ratio of \(M/m_D = 2.5\) (ensuring \(M > 2m_D\)) and benchmark mediator masses $M$ from $20$ GeV to $1500$ GeV, we derive the projected exclusion limits at significance \(Z = 1.96\). Our analysis shows that once the condition \(M > 2m_D\) is met, the results depend only moderately on the specific value of the $M/m_D$ ratio. The solid and dashed lines in Fig.~\ref{fig:exclusion_A} show these projected exclusion limits without and with flat $5\%$ systematic uncertainty, respectively.

The exclusion lines of all seven models exhibit an overall increasing trend with the increase of the mediator mass $M$, yet a distinct peak emerges at $M = 100\ \text{GeV}$ for all models. Specifically, the exclusion lines corresponding to the $\mathcal{L}_3$, $\mathcal{L}_4$, and $\mathcal{L}_8$ models show a monotonically decreasing behavior within the mass interval of 20–50 GeV, while those associated with the $\mathcal{L}_9$, $\mathcal{L}_{10}$, $\mathcal{L}_{13}$, and $\mathcal{L}_{14}$ models increase monotonically as $M$ grows.

The origin of this characteristic behavior is as follows: for the mediator with $M = 20\ \text{GeV}$, relaxed kinematic selection criteria are adopted in this work, which results in the background detection efficiency being enhanced to twice its nominal value. To satisfy the statistical significance threshold of $Z = 1.96$, the required production cross-section for $M = 20\ \text{GeV}$ is correspondingly increased. In addition, under the standard selection criteria, the background detection efficiency at $M = 100\ \text{GeV}$ is significantly higher than that at adjacent mass points. This phenomenon is mainly attributed to the fact that the rest mass of the $Z$ boson is approximately $91\ \text{GeV}$, a large number of $Z$ bosons are resonantly produced in this energy regime, and the energy spectrum broadening effect in experiments extends this resonance peak to the vicinity of $100\ \text{GeV}$. The $Z$ bosons are abundantly produced and decay near this mass interval, generating a considerable number of detectable final-state particles, which in turn leads to a significant enhancement of the background detection efficiency within the 100 GeV mass window. This is the core reason why a distinct peak appears at $M = 100\ \text{GeV}$ for all models. 

\begin{figure}[h]
\centering 
\includegraphics[width=0.48\textwidth]{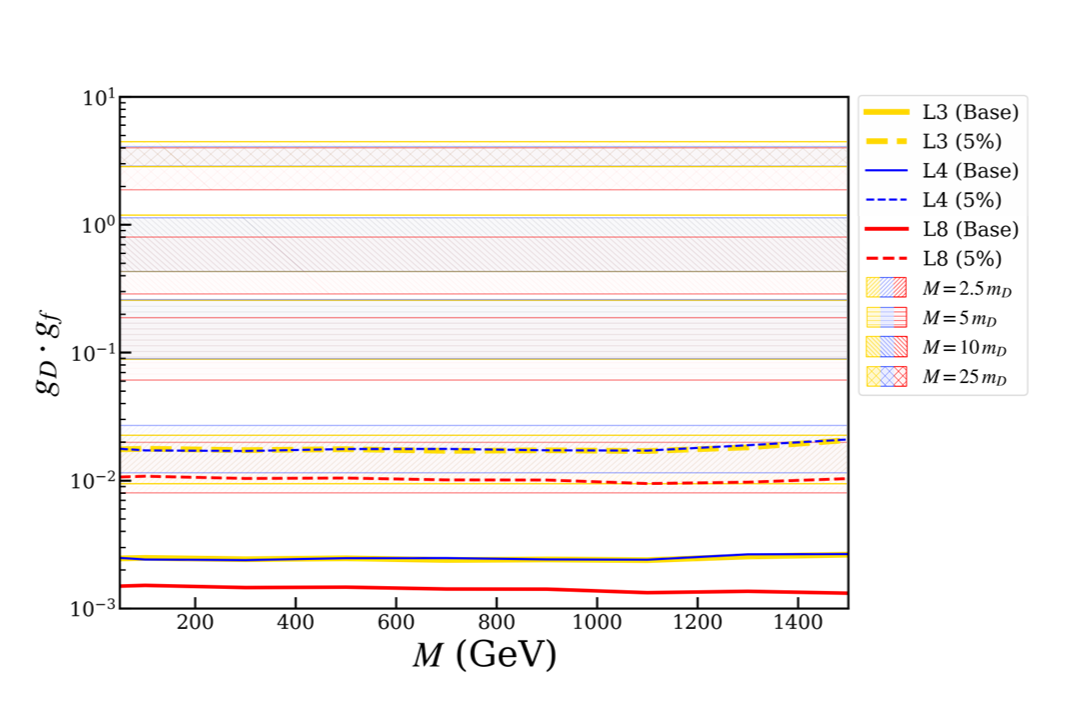} 
\includegraphics[width=0.48\textwidth]{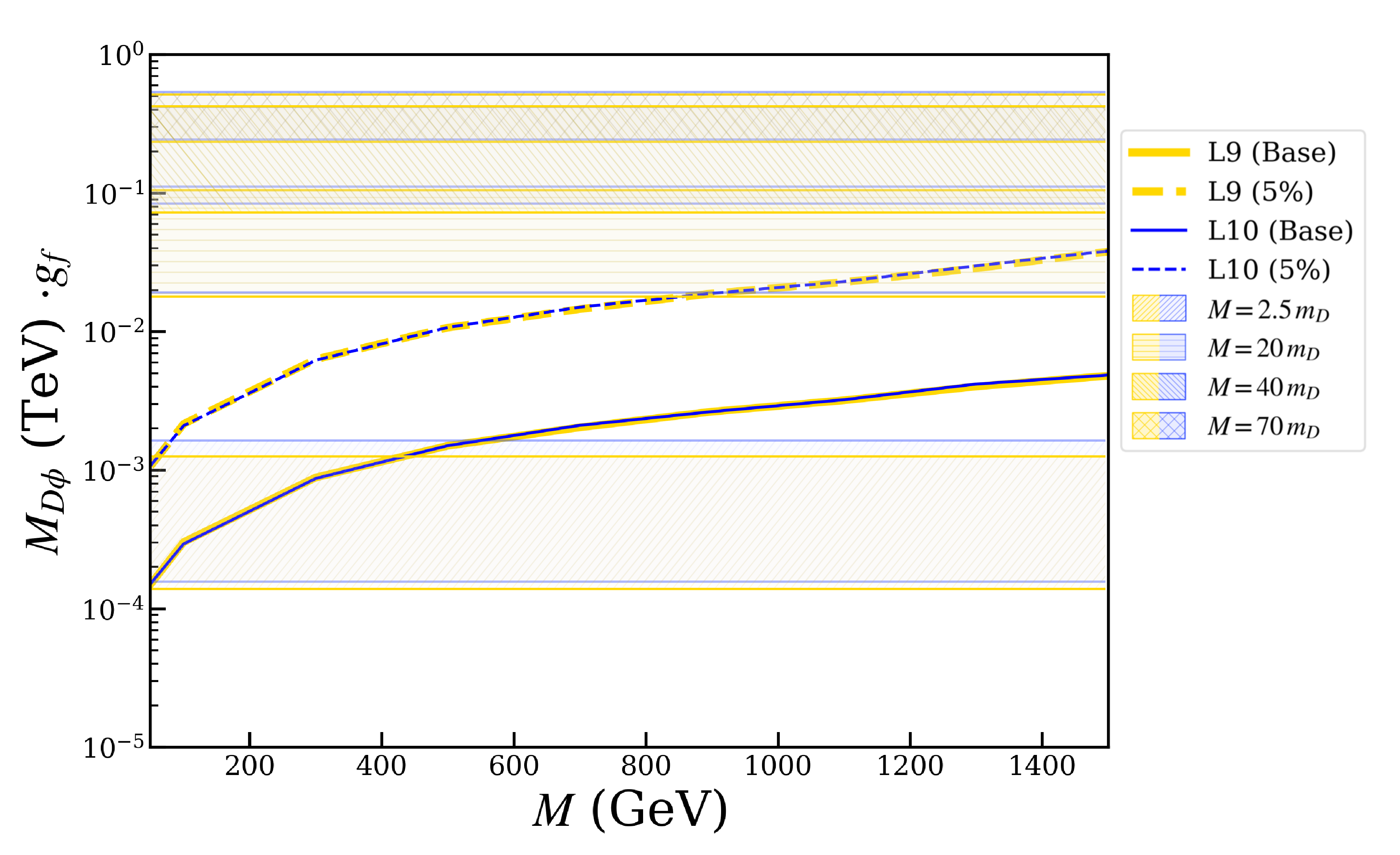}
\includegraphics[width=0.48\textwidth]{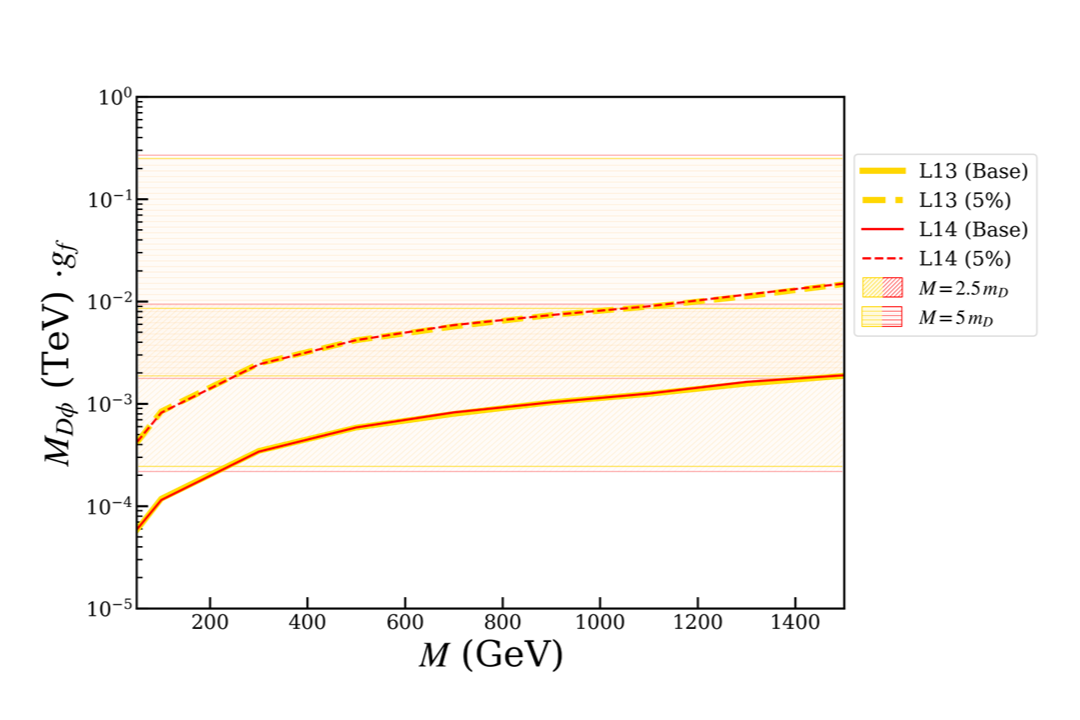}
\caption{Similar to Fig.~\ref{fig:exclusion_A}, but for invisible on-shell mediator decays in seven models and applying the condition \(\mathcal{B}(\text{MED}\to\text{DM}+\text{DM}) > 99\%\) to determine the ratio of coupling parameters. } 
\label{fig:exclusion_B}
\end{figure}

To assess the consistency of our results with a GCE interpretation, we overlay the GCE-favored parameter regions, subject to the relevant constraints identified in Ref.~\cite{Abdughani:2021oit}, onto Figs.~\ref{fig:exclusion_A} and~\ref{fig:exclusion_B}. These regions are shown as colored bands corresponding to different mediator-to-DM mass ratios \(M/m_D\). For the \(\mathcal{L}_3\), \(\mathcal{L}_4\), and \(\mathcal{L}_8\) models we use \(M/m_D = \{2.5, 5, 10, 25\}\); for the \(\mathcal{L}_9\) and \(\mathcal{L}_{10}\) models, \(M/m_D = \{2.5, 20, 40, 70\}\). Although the \(\mathcal{L}_{13}\) and \(\mathcal{L}_{14}\) models are studied with \(M/m_D = \{2.5, 5, 7.5, 10\}\), only the bands for \(\{2.5, 5\}\) are displayed, as they fully encompass the parameter space covered by the higher ratios, making the latter redundant for visualization. Remarkably, in the visible decay case, all projected exclusion lines lie consistently below the GCE-favored bands across the investigated mass range, indicating a robust probe of the GCE explanation.

To quantify the dependence of our final exclusion limits on the ratio of $M/m_D$, we have performed a supplementary comparative analysis with different $M/m_D$ values. In the baseline analysis of this work, we fix $M/m_D=2.5$, and we extend the study to two alternative values $M/m_D=5$ and $M/m_D=10$. 
For the visible on-shell mediator decay channel, in the $\mathcal{L}_3$ model with the fixed baseline coupling ratio $g_D/g_f = 1/8$ of this work, the mediator decay branching ratio $\mathcal{B}({\rm MED} \to \mu^+ \mu^-)$ is lower than the 99\% threshold set in the previous assumption, and slightly decreases with the increase of $M/m_D$: it drops to 98.6\% when $M/m_D = 5$, and further decreases to 98.5\% when $M/m_D = 10$. In the $\mathcal{L}_8$ model with the fixed baseline coupling ratio $g_D/g_f = 1/9$, the branching ratio is also lower than the 99\% threshold, showing a slight fluctuation: it is 98.79\% when $M/m_D = 5$, and slightly drops to 98.78\% when $M/m_D = 10$. On the other hand, the variation of $M/m_D$ has negligible impact on the signal detection efficiency. Taking the representative $\mathcal{L}_3$ and $\mathcal{L}_8$ models as examples: for the $\mathcal{L}_3$ model, the relative deviations of the exclusion limits for $M/m_D=5$ and $M/m_D=10$ with respect to the baseline $M/m_D=2.5$ are about 11\% and 19\%, respectively; for the $\mathcal{L}_8$ model, the above relative deviations are about 9\% and 25\%, respectively. 

In contrast, for the invisible on-shell mediator decay channel, in the $\mathcal{L}_3$ model with the fixed baseline coupling ratio $g_D/g_f = 15$, the mediator decay branching ratio $\mathcal{B}({\rm MED} \to {\rm DM}+{\rm DM})$ is higher than the 99\% threshold set in the previous assumption, and slightly increases with the increase of $M/m_D$: it is 99.359\% when $M/m_D = 5$, and rises to 99.399\% when $M/m_D = 10$; in the $\mathcal{L}_8$ model with the fixed baseline coupling ratio $g_D/g_f = 12$, the branching ratio is also higher than the 99\% threshold, showing a slight increase with the larger $M/m_D$: it is 99.201\% when $M/m_D = 5$, and rises to 99.209\% when $M/m_D = 10$. Again, the detection efficiencies for different $M/m_D$ values show almost no difference for the same model and mediator mass. As a result, for both the $\mathcal{L}_3$ and $\mathcal{L}_8$ models, the relative deviations of the exclusion limits for $M/m_D=5$ and $M/m_D=10$ with respect to the baseline $M/m_D=2.5$ are all less than 3\%.

The analysis for invisible on-shell mediator decays follows the same workflow but uses the condition \(\mathcal{B}(\text{MED}\to\text{DM}+\text{DM}) > 99\%\) to prioritize DM pair production. The corresponding coupling parameter ratios \(g_D/g_f\) and \(M_{D\phi}/g_f\) are listed in the fourth column of Table~\ref{tab:coupling_parameters}. Again, the ratio \(M_{D\phi}/g_f\) increases monotonically with the mediator mass for the fixed condition \(\mathcal{B}(\text{MED}\to\text{DM}+\text{DM}) > 99\%\). Since the results depend only mildly on the $M/m_D$ ratio once $M > 2m_D$ is satisfied, we fix \(M/m_D = 2.5\) and use the same range of benchmark mediator masses. The resulting projected exclusion limits, with and without systematic uncertainty, are shown by the solid and dashed lines in Fig.~\ref{fig:exclusion_B}, respectively. In contrast to the visible decay process, the exclusion lines of the $\mathcal{L}_3$, $\mathcal{L}_4$, and $\mathcal{L}_8$ models in the invisible decay process exhibit a relatively flat trend with the increase of the mediator mass $M$,
when systematic uncertainties are neglected (solid lines), the projected limits remain entirely below the GCE-preferred bands. For the $\mathcal{L}_9, \mathcal{L}_{10}, \mathcal{L}_{13}$, and $\mathcal{L}_{14}$ models, the exclusion lines exhibit a monotonic increase with the mediator mass $M$. In the lower mass regime (almost $M < 200$~GeV), the projected limits consistently lie below the GCE-preferred bands, indicating a robust exclusion of the parameter space favored by the GCE explanation. However, as $M$ increases beyond 200~GeV, the rising exclusion lines eventually surpass the preferred bands. Consequently, in this higher mass region, the GCE-favored parameter space can only be effectively probed if systematic uncertainties are neglected (as represented by the solid lines). This behavior suggests that for these specific models, a stronger exclusion of the GCE-favored regions is primarily achieved in the $M < 200$~GeV regime when systematic errors are not taken into account.

The exclusion limits derived from visible on-shell mediator decays are consistently stronger than those from invisible on-shell mediator decays across all models. This behavior can be understood for two main reasons. First, the relevant SM background for the visible decay signal process is more than an order of magnitude smaller than for the invisible decay signal process. Second, although the production cross-sections for both signal processes are proportional to $g^2_f$, the requirement of a larger than $99\%$ branching ratio to either $\mu^+\mu^-$ or a pair of DM particles results in distinctly different values for the coupling parameter ratios $g_D/g_f$ and $M_{D\phi}/g_f$, as shown in Table~\ref{tab:coupling_parameters}. 

\begin{figure}[h]
\centering 
\includegraphics[width=0.48\textwidth]{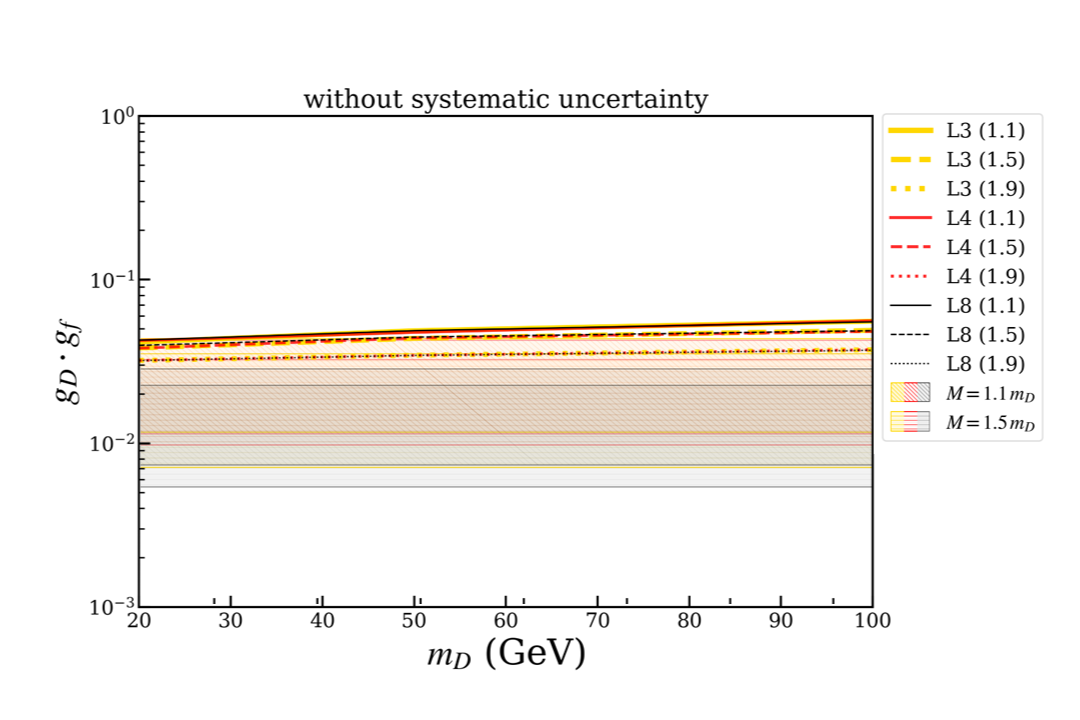}
\includegraphics[width=0.48\textwidth]{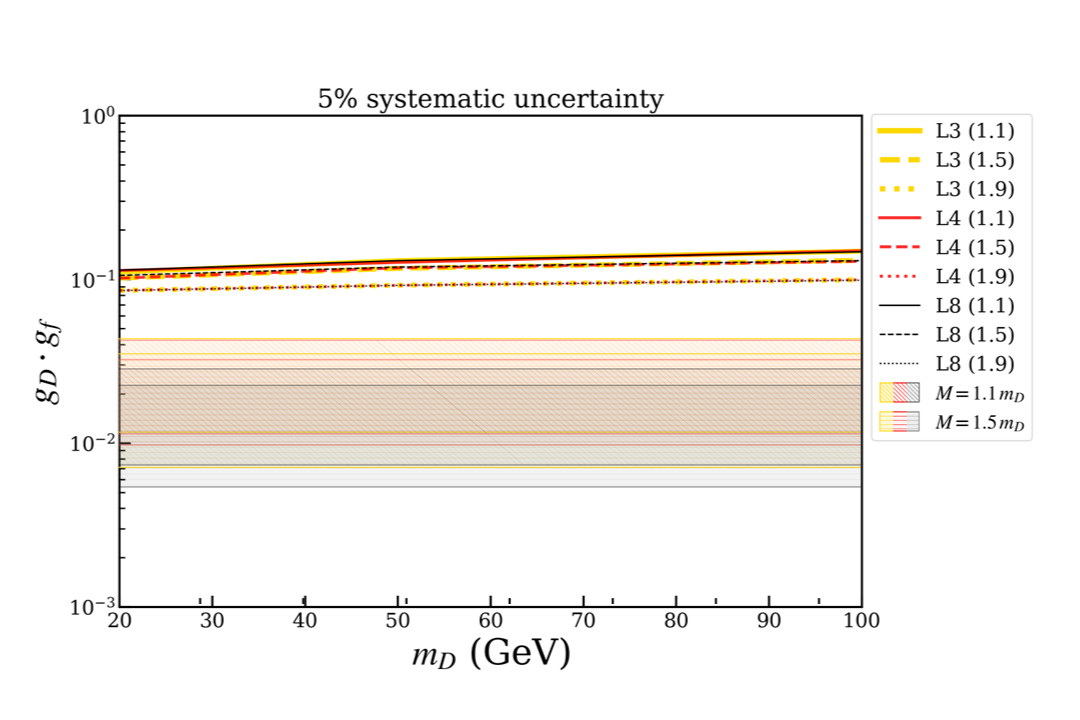}
\includegraphics[width=0.48\textwidth]{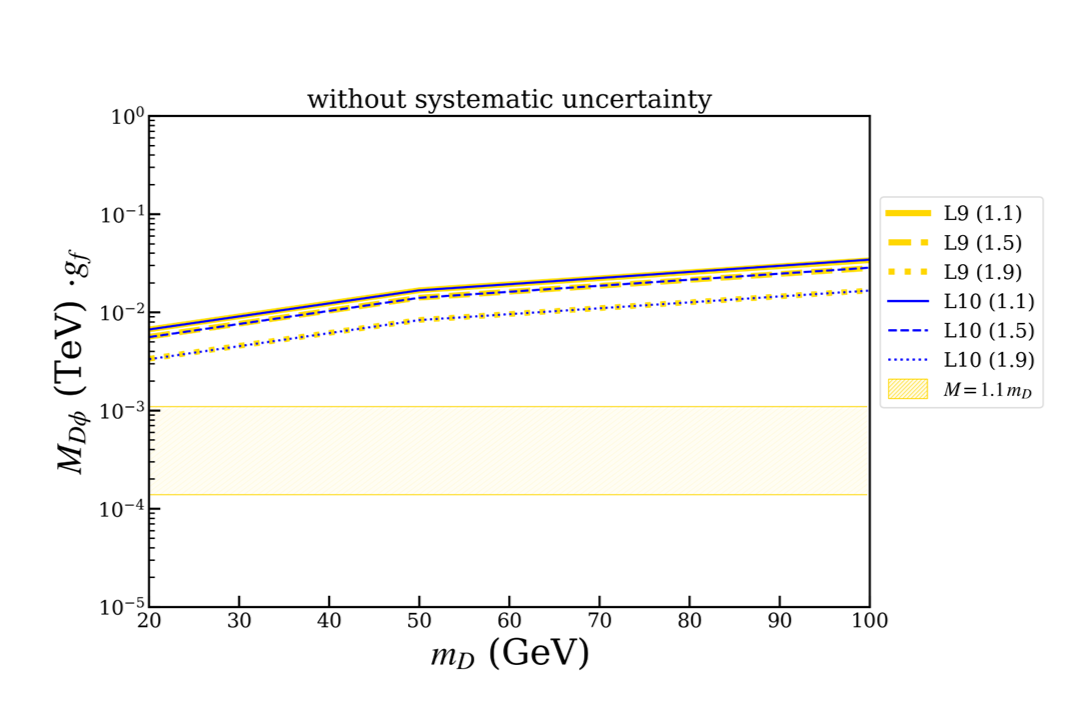}
\includegraphics[width=0.48\textwidth]{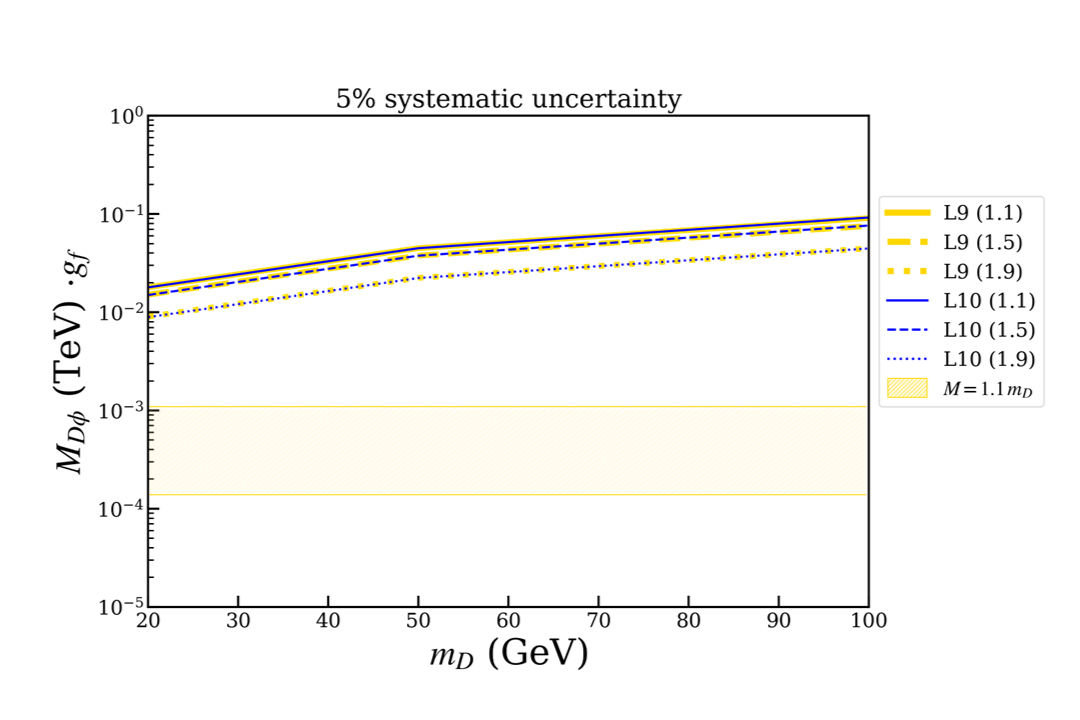}
\includegraphics[width=0.48\textwidth]{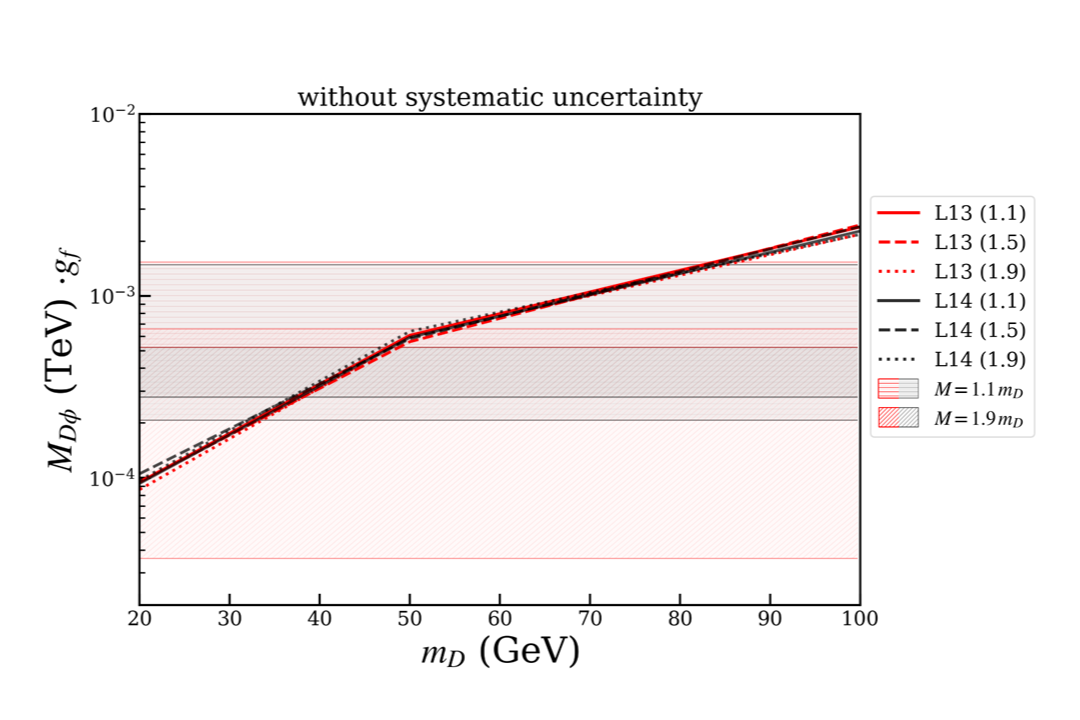}
\includegraphics[width=0.48\textwidth]{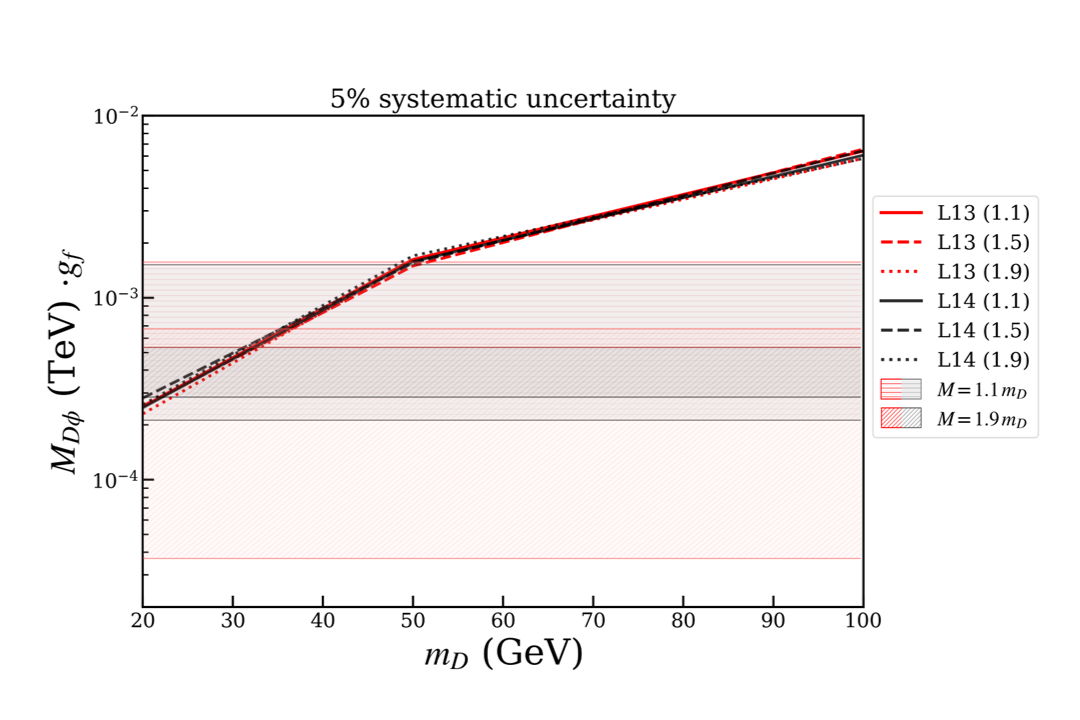}
\caption{Projected exclusion limits at $95\%$ confidence level on the parameter space for mono-photon channel with an off-shell mediator in seven models without (top panels) and with a flat $5\%$ background systematic uncertainty (bottom panels). Colored bands denote the GCE-favored parameter regions consistent with Ref.~\cite{Abdughani:2021oit}, where distinct colors correspond to individual models and various patterns within the bands distinguish the mediator-to-DM mass ratios $M/m_D$ (see the main text for details).} 
\label{fig:exclusion_C}
\end{figure}

For the mono-photon channel with an off-shell mediator, the production cross-sections are directly proportional to \(g_D \cdot g_f\) for $\mathcal{L}_3$, $\mathcal{L}_4$, $\mathcal{L}_8$ models and to \(M_{D\phi} \cdot g_f\) for $\mathcal{L}_9$, $\mathcal{L}_{10}$, $\mathcal{L}_{13}$, $\mathcal{L}_{14}$ models. We consider mass ratios \(M/m_D = 1.1, 1.5, 1.9\) (satisfying \(M < 2m_D\)) and benchmark DM masses from $20$ GeV to $100$ GeV. 
To further evaluate the sensitivity to the GCE explanation in this context, we similarly overlay the GCE-preferred regions from Ref.~\cite{Abdughani:2021oit} onto Fig.~\ref{fig:exclusion_C}. The colored bands denote the favored regions for the respective models across mass ratios $M/m_D = \{1.1, 1.5, 1.9\}$. For visual clarity, we only display the $M/m_D = 1.1$ bands for the $\mathcal{L}_9$ and $\mathcal{L}_{10}$ models, as they fully encompass the parameter space of the higher ratios. Following a similar reason, we present only the $M/m_D = \{1.1, 1.5\}$ bands for the $\mathcal{L}_3, \mathcal{L}_4, \mathcal{L}_8$ models, and the $M/m_D = \{1.1, 1.9\}$ bands for the $\mathcal{L}_{13}, \mathcal{L}_{14}$ models.
The projected exclusion limits for this scenario, without and with systematic uncertainty, are presented in Fig.~\ref{fig:exclusion_C}. In contrast to the on-shell decays, the exclusion lines of all models show a monotonic upward trend with the increase of mediator mass $M$. Moreover, the larger the mass ratio \(M/m_D\) is, the lower the exclusion lines lie, indicating higher detection sensitivity. 
Notably, all exclusion lines lie either above or within the GCE-preferred bands, indicating that current sensitivity remains insufficient to fully probe or constrain the parameter space favored by the GCE explanation in the off-shell regime.

\begin{figure}[h]
\centering 
\includegraphics[width=0.48\textwidth]{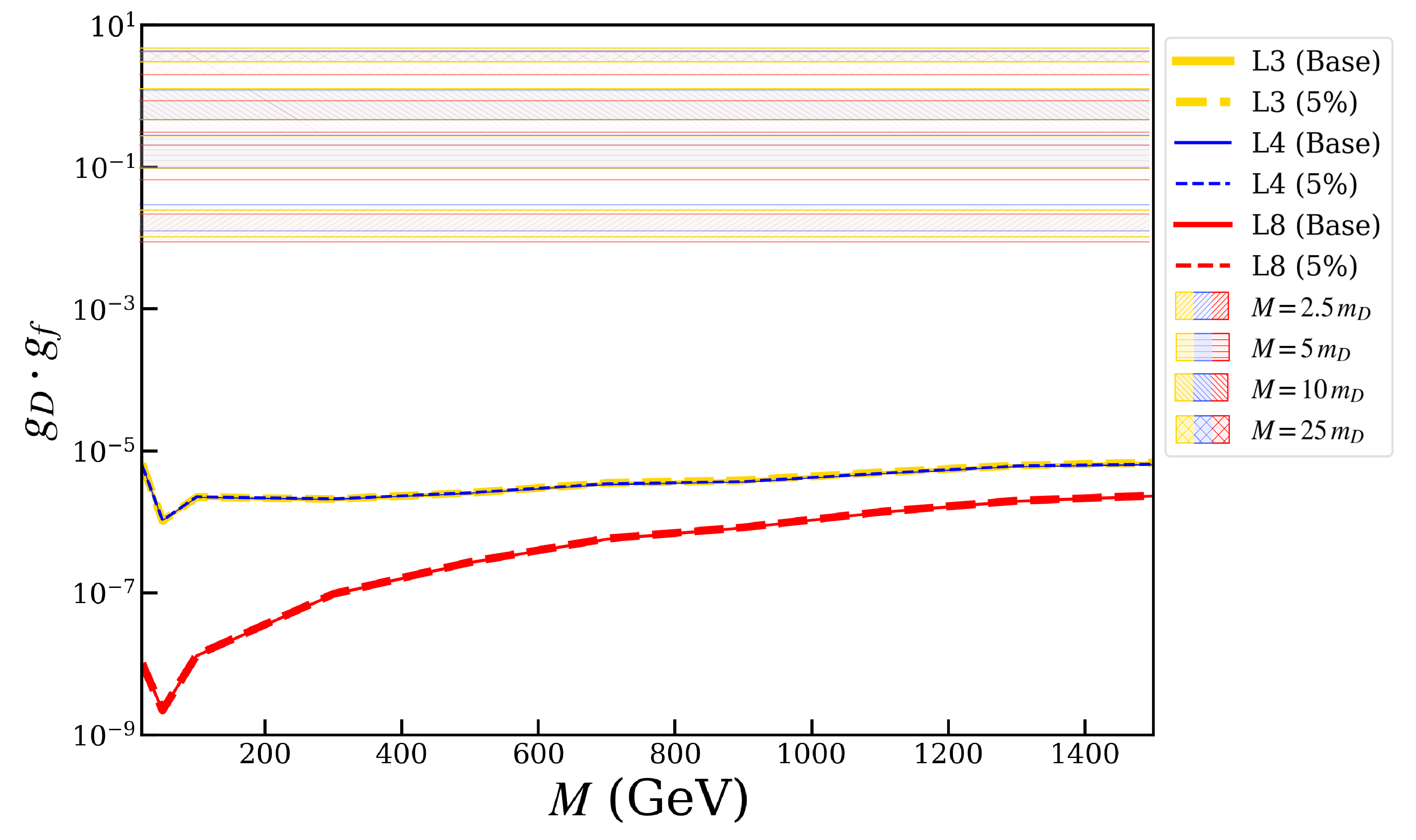}
\includegraphics[width=0.48\textwidth]{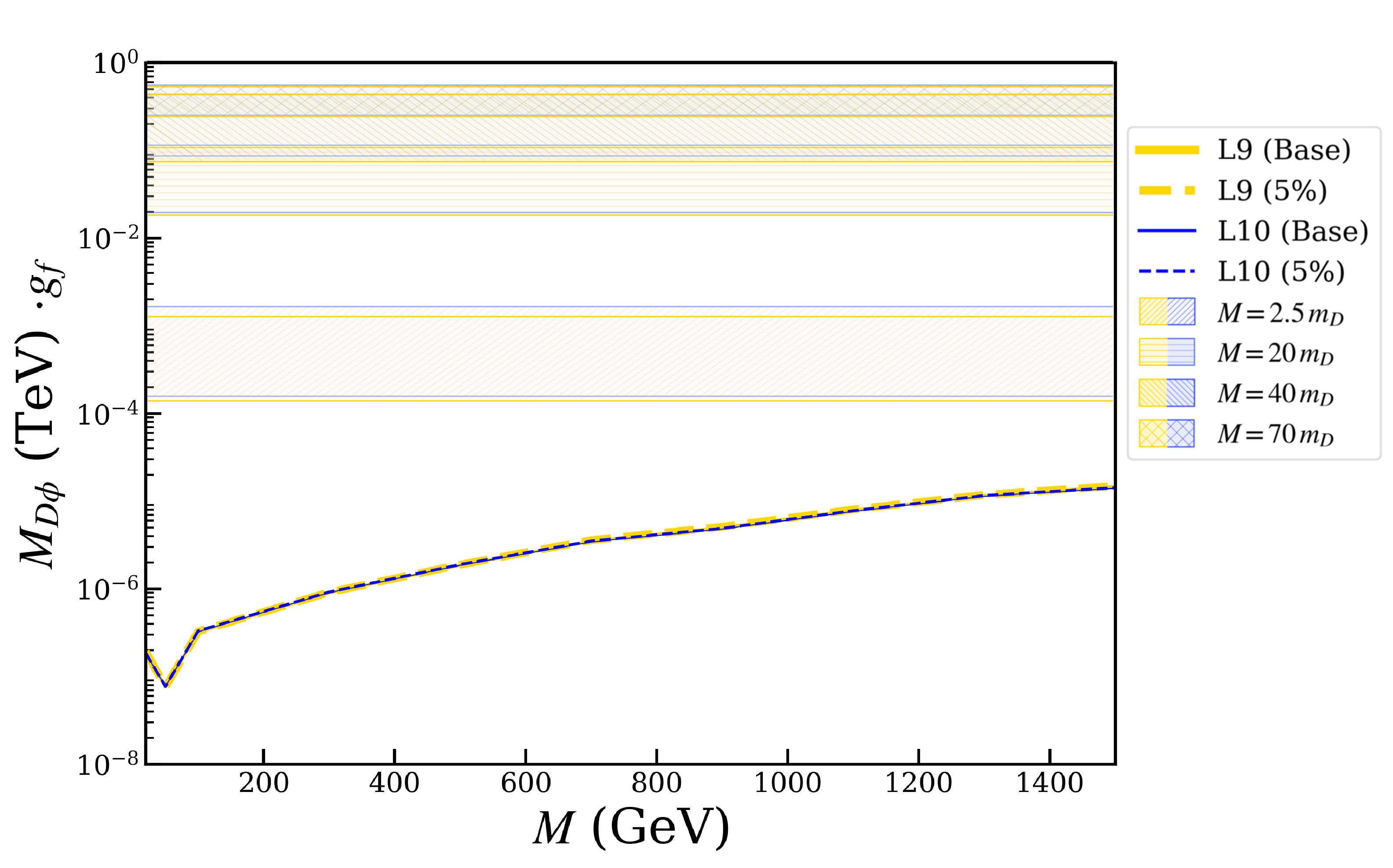}
\includegraphics[width=0.48\textwidth]{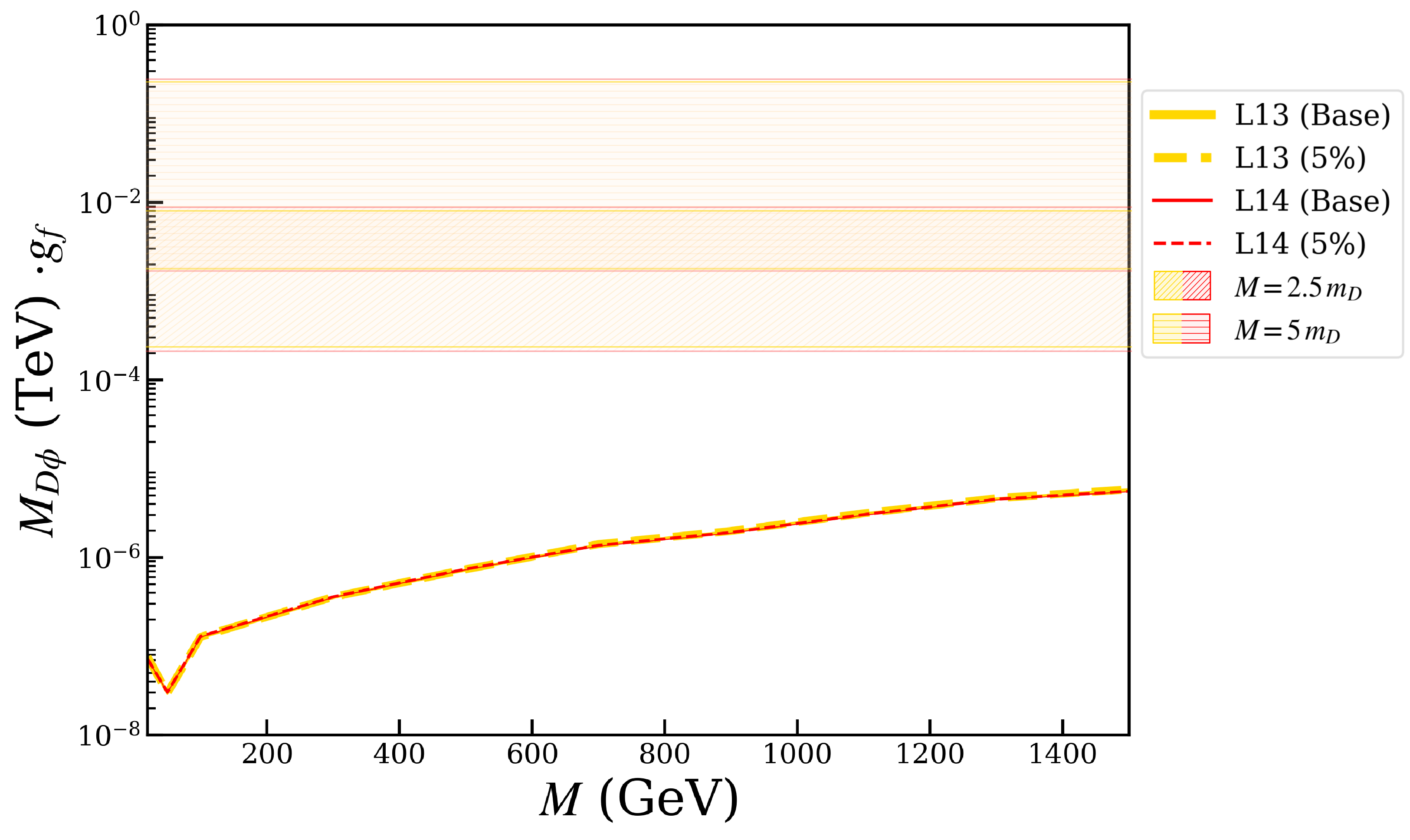}
\caption{Similar to Fig.~\ref{fig:exclusion_A}, but for visible on-shell mediator decays via VBF process in seven models.} 
\label{fig:exclusion_visible}
\end{figure}

\begin{figure}[h]
\centering 
\includegraphics[width=0.48\textwidth]{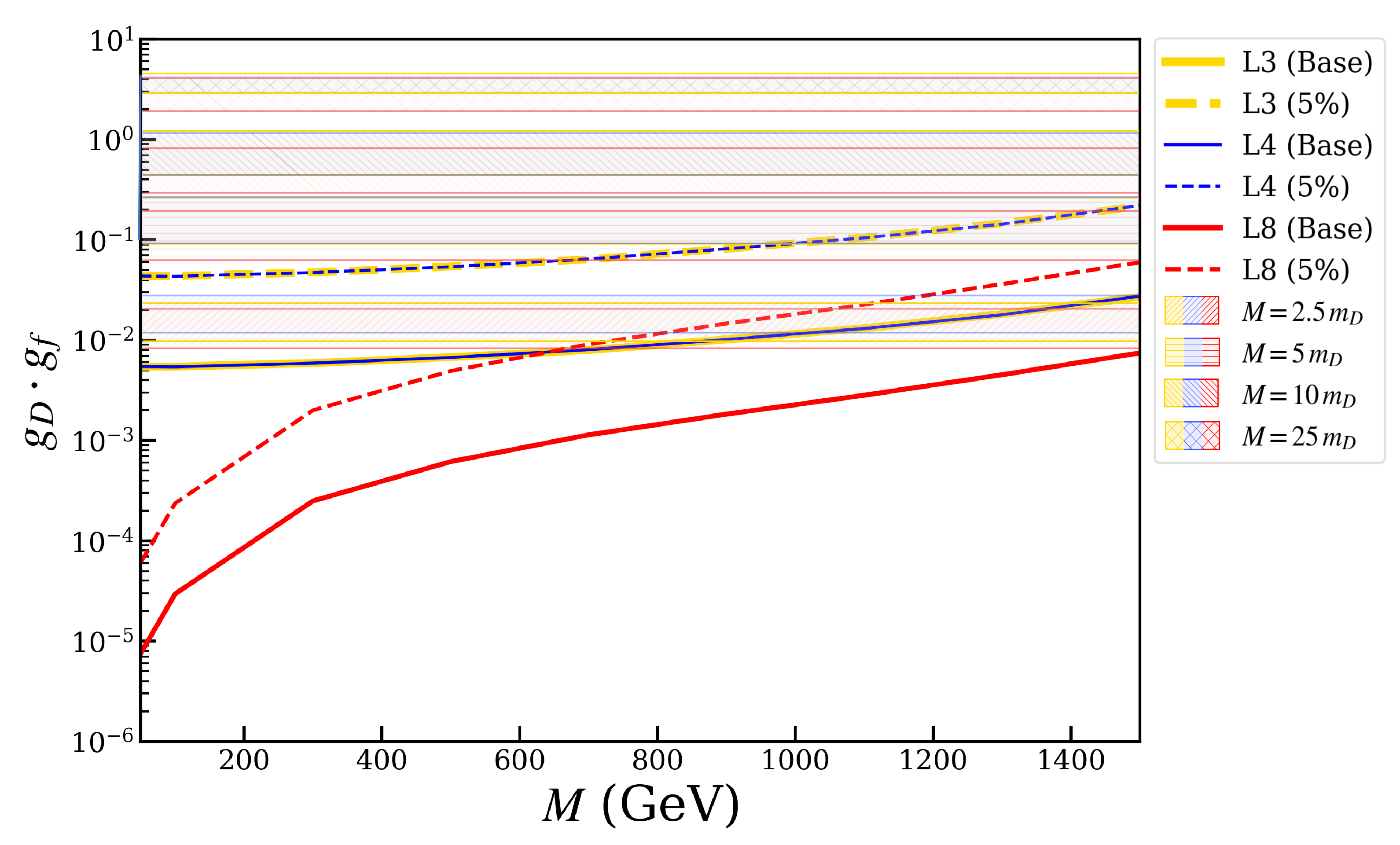}
\includegraphics[width=0.48\textwidth]{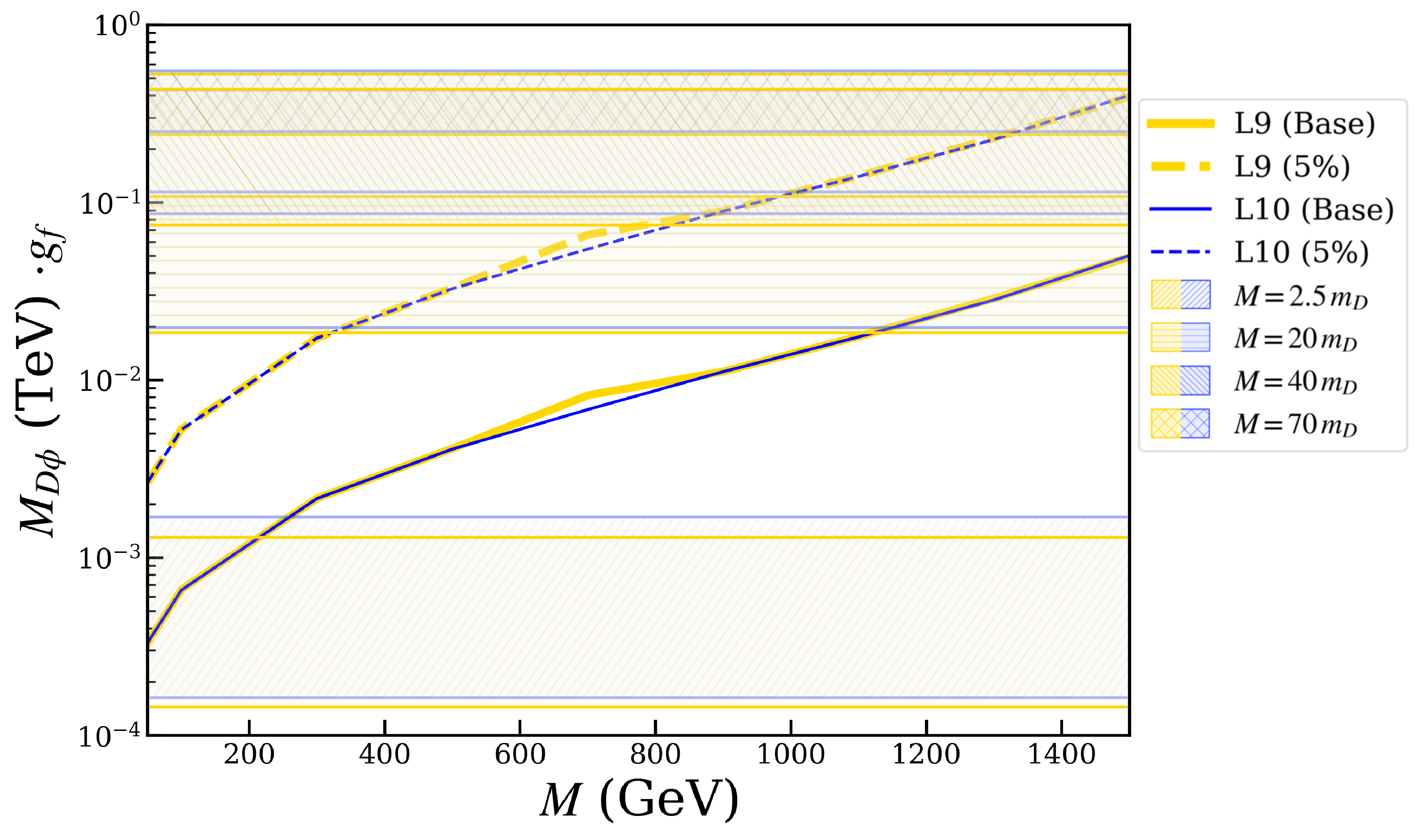}
\includegraphics[width=0.48\textwidth]{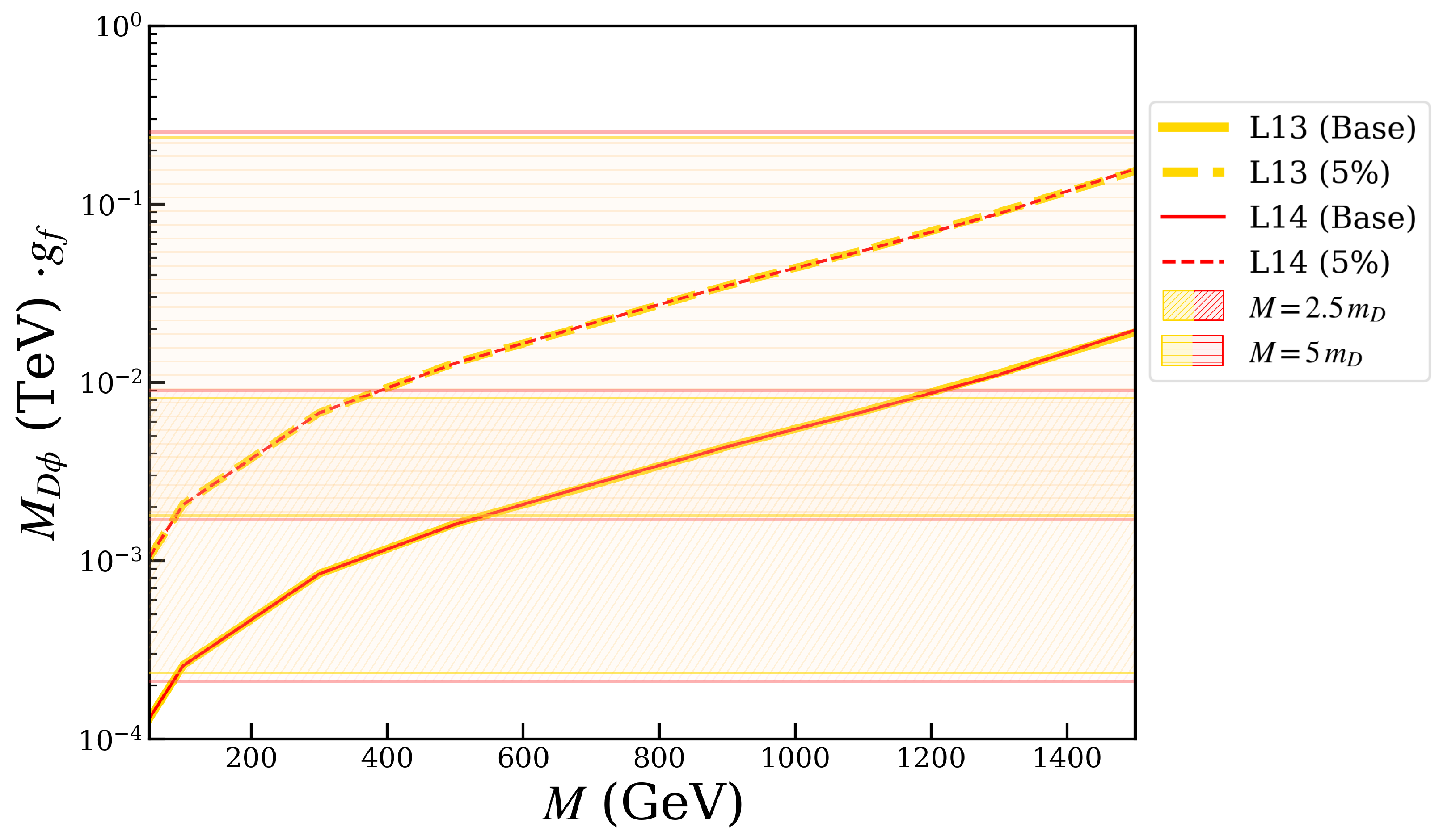}
\caption{Similar to Fig.~\ref{fig:exclusion_visible}, but for invisible on-shell mediator decays via VBF process in seven models.}
\label{fig:exclusion_invisible}
\end{figure}

Among them, the visible process exhibits a monotonically decreasing trend in the mass range of 20 $\mathrm{GeV}$ to 50 $\mathrm{GeV}$, while showing a monotonically increasing trend in all other mass ranges. This is mainly due to the excessively small background cross section and detection efficiency. Additionally, under the event selection of $20~\mathrm{GeV}$ (imposed by the kinematic cut on the mediator mass window), the number of background events is less than $1$, so the process at $20~\mathrm{GeV}$ is naively analyzed as a background-free process. For the invisible process, its results show a monotonically increasing trend in the mass range of $50~\mathrm{GeV}$ to $1500~\mathrm{GeV}$. Both processes are characterized by better exclusion limits in the low-mass range than in the high-mass range.

On the other hand, for the exclusion limits of the VBF process, it can be observed that for the invisible on-shell mediator decay channel, the exclusion capability of the VBF process is significantly weaker than that of the muon pair annihilation process. In contrast, for the visible on-shell mediator decay channel, the exclusion capability of the VBF process is comparable to that of the muon pair annihilation process. 
The core reason for this phenomenon is as follows: For the VBF process, the detection efficiencies of the signal and background are significantly different from those of the aforementioned muon pair annihilation process. In the visible on-shell mediator decay channel, the background detection efficiency of the VBF process is more stable across different mediator mass windows, especially in the high-mass regime, where the background detection efficiency is only $1/2$ to $1/3$ of that of the muon pair annihilation process. For the signal detection efficiency, the two processes are basically similar when the mediator mass is below $500\ \mathrm{GeV}$. However, above $500\ \mathrm{GeV}$, the signal detection efficiency of the VBF process becomes lower than that of the muon pair annihilation process, with the gap widening as the mediator mass increases. In the invisible on-shell mediator decay channel, the background detection efficiency of the VBF process is approximately twice that of the muon pair annihilation process, while there is no significant difference in the signal detection efficiency between the two processes. Although the signal cross-section of the VBF visible decay channel is lower than that of the muon pair annihilation process, the background cross-section is greatly suppressed thanks to the unique four muons signature in its final state. The gain from this background suppression far outweighs the combined loss from the reduced signal cross-section and decreased signal efficiency, ultimately making its exclusion capability comparable to that of the muon pair annihilation process.

To clarify the influence of mediator decay branching ratios on detection performance, we further compare the exclusion performance characteristics at branching ratios of 90\% and 50\% with the case of 99\% as the reference benchmark. For the visible mediator decay processes, the detection sensitivity at $\mathcal{B}(\text{MED}\to\mu^+\mu^-) > 50\%$ is 20 times lower than that at $\mathcal{B}(\text{MED}\to\mu^+\mu^-) > 99\%$, while the detection sensitivity at $\mathcal{B}(\text{MED}\to\mu^+\mu^-) > 90\%$ is 3.6 times lower than that at $\mathcal{B}(\text{MED}\to\mu^+\mu^-) > 99\%$. This study explicitly states that the exclusion limits drawn for specific mediator decay channels are all based on the premise of $\mathcal{B}(\text{MED}\to\mu^+\mu^-) > 99\%$.

In contrast, the exclusion limits of the invisible mediator decay processes exhibit significant regularity with the variation of branching ratios: all models show the weakest exclusion performance at $\mathcal{B}(\text{MED}\to\text{DM}+\text{DM}) > 99\%$ and the strongest exclusion performance at $\mathcal{B}(\text{MED}\to\text{DM}+\text{DM}) > 50\%$. Quantitative analysis results indicate that the exclusion limit at $\mathcal{B}(\text{MED}\to\text{DM}+\text{DM}) > 90\%$ is approximately 3 times that at $\mathcal{B}(\text{MED}\to\text{DM}+\text{DM}) > 99\%$; the exclusion limit at $\mathcal{B}(\text{MED}\to\text{DM}+\text{DM}) > 50\%$ is approximately 5 times that at $\mathcal{B}(\text{MED}\to\text{DM}+\text{DM}) > 99\%$. It is worth noting that for the same process and the same model, the variation trends of exclusion lines remain consistent across different mediator decay branching ratios.

In summary, the exclusion limits for the fermionic DM models with scalar or vector mediators ($\mathcal{L}_3$, $\mathcal{L}_4$, $\mathcal{L}_8$) reach $\mathcal{O}(10^{-5})$ for $g_D \cdot g_f$ in visible on-shell mediator decays, $\mathcal{O}(10^{-3}-10^{-2})$ for invisible on-shell mediator decays, and $\mathcal{O}(10^{-2}-0.1)$ for the mono-photon channel with an off-shell mediator. The projected exclusion contours cover most of the allowed non-resonance regions with $g_D \cdot g_f \gtrsim 10^{-2}$ and mediator masses $M$ between $20$ GeV and $1500$ GeV for these muonphilic DM models, as shown in Ref.~\cite{Abdughani:2021oit}. Similarly, for scalar DM models with a scalar mediator ($\mathcal{L}_9$, $\mathcal{L}_{10}$) and vector DM models with a scalar mediator ($\mathcal{L}_{13}$, $\mathcal{L}_{14}$), the exclusion limits on $M_{D\phi}~(\text{TeV}) \cdot g_f$ reach $\mathcal{O}(10^{-6}-10^{-5})$ for visible on-shell mediator decays, $\mathcal{O}(10^{-4}-10^{-2})$ for invisible on-shell mediator decays, and $\mathcal{O}(10^{-4}-10^{-2})$ for the mono-photon channel with an off-shell mediator. Again, the projected limits cover most non-resonance regions with $M_{D\phi}~(\text{TeV}) \cdot g_f \gtrsim 10^{-4}$ across the same mediator mass range. Therefore, a future $3$ TeV muon collider would be critical for testing the muonphilic DM explanation of the GCE puzzle. 

\section{Conclusion}
\label{sec:conclusion} 

In this work, we have performed a detailed and systematic study of the prospects for discovering muonphilic dark matter (DM) at a future 3 TeV muon collider, focusing on simplified models with a \(Z_2\)-even mediator. Motivated by the model's ability to explain the Galactic Center Excess (GCE) and observed relic density while evading multi-messenger, direct detection and collider constraints, we have explored its viable parameter space through four distinct search channels: visible on-shell mediator decays, invisible on-shell mediator decays, mono-photon signatures with off-shell mediators, and vector boson fusion production.

Our analysis demonstrates that a muon collider possesses exceptional sensitivity to probe the non-resonant regions of these models. For visible on-shell decays, the projected exclusion limits on the fermionic DM model couplings \(g_D \cdot g_f\) can reach \( \mathcal{O}(10^{-5}) \), while for scalar and vector DM models, the limits on \(M_{D\phi}~(\text{TeV}) \cdot g_f\) can be as strong as \( \mathcal{O}(10^{-6} - 10^{-5}) \). The invisible on-shell decay channel, while slightly less sensitive due to a larger irreducible background and coupling parameter ratios shown in Table~\ref{tab:coupling_parameters}, still provides powerful constraints of \( \mathcal{O}(10^{-3} - 10^{-2}) \) and \( \mathcal{O}(10^{-4} - 10^{-2}) \) for the respective model classes. Furthermore, the mono-photon channel with off-shell mediators offers a complementary probe, especially in parameter regions where \(M < 2m_D\), with exclusion limits reaching \( \mathcal{O}(10^{-2} - 0.1) \) and \( \mathcal{O}(10^{-4} - 10^{-2}) \). We have also quantified the impact of a conservative flat $5\%$ systematic uncertainty, demonstrating that while it shifts the exclusion limits upward, the projected sensitivity remains substantial. 
This confirms the robustness of our results. The resulting exclusion contours cover most of the allowed parameter space, specifically for \(g_D \cdot g_f \gtrsim 10^{-2}\) and \(M_{D\phi} \cdot g_f \gtrsim 10^{-4}~\text{TeV}\), across the mediator mass ranges of $20$ GeV to $1.5$ TeV in the non-resonant regions.  

In conclusion, a high-energy muon collider serves as a decisive machine to directly test the muonphilic DM hypothesis as an explanation for the GCE. Its clean collision environment and high luminosity provide a unique opportunity to perform precision measurements of the DM-muon coupling and mediator mass, directly complementing and exceeding the reach of indirect and direct detection experiments as well as conventional collider searches. Our work firmly establishes that such a facility is critical for advancing our understanding of the DM particle nature.

\appendix
\section{Collider Constraints from LEP and LHC on Muonphilic DM Models}
\label{app:collider_constraints}

\subsection{LEP Constraints on Visible and Invisible Mediator Decays} 

At LEP, which operated at $\sqrt{s}=209\ \mathrm{GeV}$, we focus on $e^+e^-$ collision-induced production of the mediator $\text{MED}$ (denoted in the main text) for muonphilic DM models. Below we present the detailed analysis of LEP constraints for both visible and invisible mediator decay processes.

For the visible mediator decay scenario ($\text{MED} \to \mu^+\mu^-$), the signal process is defined as:
\begin{equation}
e^+e^- \to \mu^+\mu^- \text{MED},\quad \text{MED} \to \mu^+\mu^-,
\end{equation}
and the dominant SM background process is:
\begin{equation}
e^+e^- \to \mu^+\mu^- \mu^+\mu^-.
\end{equation}
Similarly, for the invisible mediator decay scenario ($\text{MED} \to \text{DM}+\text{DM}$), the signal process is:
\begin{equation}
e^+e^- \to \mu^+\mu^- \text{MED},\quad \text{MED} \to \text{DM}+\text{DM},
\end{equation}
and the dominant SM background process is:
\begin{equation}
e^+e^- \to \mu^+\mu^- \nu_l\bar{\nu}_l. 
\end{equation} 

\begin{table}[htbp]
\centering
\footnotesize  
\begin{tabular}{|>{\centering\arraybackslash}p{1.2cm}|>{\centering\arraybackslash}p{4.8cm}|>{\centering\arraybackslash}p{2.1cm}|*{3}{>{\centering\arraybackslash}p{1.8cm}|}}
\hline
\multicolumn{2}{|c|}{\textbf{Cut description}} & \textbf{background} & \textbf{signal1} & \textbf{signal2} & \textbf{signal3}  \\
\hline
\multicolumn{2}{|c|}{\textbf{Cross-section [fb]}} & $238.10$ & $1.20$ & $1.64$ & $1.41$  \\
\hline
\textbf{Cut-1} & $N_{\mu^+} \geq 1$, $N_{\mu^-} \geq 1$, \newline 
$p_T(\mu) > 20\ \mathrm{GeV}$, $|\eta(\mu)| < 2.5$, \newline
${\:/\!\!\!\! E}_T > 20\ \mathrm{GeV}$ & {\footnotesize 0.60} & {\footnotesize 0.63} & {\footnotesize 0.66} & {\footnotesize 0.66}  \\
\hline
\textbf{Cut-2} & $p_T(\mu_1) \geq 80\ \mathrm{GeV}$, 
$|\eta(\mu_1)| < 0.8$ & {\footnotesize 0.02} & {\footnotesize 0.27} & {\footnotesize 0.20} & {\footnotesize 0.09}  \\
\hline
\textbf{Cut-3} & $p_T(\mu_1)/{\:/\!\!\!\! E}_T > 1.1$ & {\footnotesize 0.01} & {\footnotesize 0.23} & {\footnotesize 0.15} & {\footnotesize 0.04}  \\
\hline
\end{tabular}
\caption{The cut-flow table for invisible on-shell mediator decay of the $\mathcal{L}_3$ model and its relevant background at the LEP. The three BPs are signal-1 ($M = 10$ GeV), signal-2 ($M = 25$ GeV), and signal-3 ($M = 50$ GeV). }
\label{tab:lepinvisibleL3_decay}
\end{table}

\begin{table}[htbp]
\centering
\footnotesize  
\begin{tabular}{|>{\centering\arraybackslash}p{1.2cm}|>{\centering\arraybackslash}p{4.8cm}|>{\centering\arraybackslash}p{2.1cm}|*{3}{>{\centering\arraybackslash}p{1.8cm}|}}
\hline
\multicolumn{2}{|c|}{\textbf{Cut description}} & \textbf{background} & \textbf{signal1} & \textbf{signal2} & \textbf{signal3}  \\
\hline
\multicolumn{2}{|c|}{\textbf{Cross-section [fb]}} & $238.10$ & $59.39$ & $14.15$ & $1.18$  \\
\hline
\textbf{Cut-1} & $N_{\mu^+} \geq 1$, $N_{\mu^-} \geq 1$, \newline 
$p_T(\mu) > 20\ \mathrm{GeV}$, $|\eta(\mu)| < 2.5$, \newline
${\:/\!\!\!\! E}_T > 20\ \mathrm{GeV}$ & {\footnotesize 0.60} & {\footnotesize 0.65} & {\footnotesize 0.66} & {\footnotesize 0.63}  \\
\hline
\textbf{Cut-2} & $p_T(\mu_1) \geq 80\ \mathrm{GeV}$, 
$|\eta(\mu_1)| < 0.8$ & {\footnotesize 0.02} & {\footnotesize 0.17} & {\footnotesize 0.09} & {\footnotesize 0.04}  \\
\hline
\textbf{Cut-3} & $p_T(\mu_1)/{\:/\!\!\!\! E}_T > 1.1$ & {\footnotesize 0.01} & {\footnotesize 0.13} & {\footnotesize 0.05} & {\footnotesize 0.02}  \\
\hline
\end{tabular}
\caption{Similar to Table~\ref{tab:lepinvisibleL3_decay}, but for the $\mathcal{L}_{9}$ model.}
\label{tab:lepinvisibleL9_decay}
\end{table}

\begin{table}[htbp]
\centering
\footnotesize  
\begin{tabular}{|>{\centering\arraybackslash}p{1.2cm}|>{\centering\arraybackslash}p{4.8cm}|>{\centering\arraybackslash}p{2.1cm}|*{3}{>{\centering\arraybackslash}p{1.8cm}|}}
\hline
\multicolumn{2}{|c|}{\textbf{Cut description}} & \textbf{background} & \textbf{signal1} & \textbf{signal2} & \textbf{signal3}  \\
\hline
\multicolumn{2}{|c|}{\textbf{Cross-section [fb]}} & $238.10$ & $28420$ & $1028$ & $23.54$  \\
\hline
\textbf{Cut-1} & $N_{\mu^+} \geq 1$, $N_{\mu^-} \geq 1$, \newline 
$p_T(\mu) > 20\ \mathrm{GeV}$, $|\eta(\mu)| < 2.5$, \newline
${\:/\!\!\!\! E}_T > 20\ \mathrm{GeV}$ & {\footnotesize 0.60} & {\footnotesize 0.63} & {\footnotesize 0.57} & {\footnotesize 0.51}  \\
\hline
\textbf{Cut-2} & $p_T(\mu_1) \geq 80\ \mathrm{GeV}$, 
$|\eta(\mu_1)| < 0.8$ & {\footnotesize 0.02} & {\footnotesize 0.06} & {\footnotesize 0.02} & {\footnotesize 0.005}  \\
\hline
\textbf{Cut-3} & $p_T(\mu_1)/{\:/\!\!\!\! E}_T > 1.1$ & {\footnotesize 0.01} & {\footnotesize 0.03} & {\footnotesize $4.1\times 10^{-3}$} & {\footnotesize $9.0\times 10^{-4}$}  \\
\hline
\end{tabular}
\caption{Similar to Table~\ref{tab:lepinvisibleL3_decay}, but for the $\mathcal{L}_{13}$ model.}
\label{tab:lepinvisibleL13_decay}
\end{table}

For the invisible on-shell mediator decay process, we consider the mediator mass $M = \{10, 25, 50\}\ \mathrm{GeV}$; for the visible process, the mediator masses are set to $M = \{5, 10, 25, 50\}\ \mathrm{GeV}$. With the integrated luminosity fixed at $\mathcal{L}=233.4\ \mathrm{pb}^{-1}$, the event selection criteria for both processes are simply optimized as detailed below:
\paragraph{Invisible Mediator Decay ($\text{MED} \to \text{DM}+\text{DM}$)}
\begin{itemize} 
    \item Cut-1 (basic cuts): $N_{\mu^+} \geq 1$, $N_{\mu^-} \geq 1$,  $p_T(\mu) > 20\ \mathrm{GeV}$, $|\eta(\mu)| < 2.5$ and ${\:/\!\!\!\! E}_T > 20\ \mathrm{GeV}$; 
    \item Cut-2: $p_T(\mu_1) \geq 80\ \mathrm{GeV}$;  
    \item Cut-3: $p_T(\mu_1)/{\:/\!\!\!\! E}_T > 1.1$;  
\end{itemize}
Based on these event selections, the corresponding cut-flow tables for various signal BPs in the $\mathcal{L}_9$ and $\mathcal{L}_{13}$ models as well as background are obtained in Tables~\ref{tab:lepinvisibleL9_decay} and~\ref{tab:lepinvisibleL13_decay}. It can be seen that for the invisible mediator decay process, the product of the background cross section, integrated luminosity and detection efficiency is much less than $1$. Therefore, we naively treat this analysis as a background-free scenario in the subsequent discussion.

\begin{table}[htbp]
\centering
\footnotesize  
\begin{tabular}{|>{\centering\arraybackslash}p{1.2cm}|>{\centering\arraybackslash}p{4.8cm}|*{4}{>{\centering\arraybackslash}p{1.8cm}|}}
\hline
\multicolumn{2}{|c|}{\textbf{Cut description}} & \textbf{signal1} & \textbf{signal2} & \textbf{signal3} & \textbf{signal4}  \\
\hline
\multicolumn{2}{|c|}{\textbf{Cross-section [fb]}} & $21.56$ & $212.10$ & $113.40$ & $46.29$  \\
\hline
\textbf{Cut-1} & $N_{\mu^+} \geq 2$ and $N_{\mu^-} \geq 2$, \newline
$p_T(\mu) > 20\ \mathrm{GeV}$, $|\eta(\mu)| < 2.5$ & {\footnotesize 0.19} & {\footnotesize 0.14} & {\footnotesize 0.18} & {\footnotesize 0.29}  \\
\hline
\textbf{Cut-2} & $|M_{\mu\mu}^{\text{best}} - M| < 0.15 M$ & {\footnotesize 0.002} & {\footnotesize 0.08} & {\footnotesize 0.11} & {\footnotesize 0.24}  \\
\hline
\end{tabular}
\caption{The cut-flow table for visible on-shell mediator decay of the $\mathcal{L}_3$ model. The four BPs are signal-1 ($M = 5$ GeV), signal-2 ($M = 10$ GeV), signal-3 ($M = 25$ GeV), and signal-4 ($M = 50$ GeV). }
\label{tab:lepvisibleL3_decay}
\end{table}

\begin{table}[htbp]
\centering
\footnotesize  
\begin{tabular}{|>{\centering\arraybackslash}p{1.2cm}|>{\centering\arraybackslash}p{3.2cm}|*{6}{>{\centering\arraybackslash}p{1.8cm}|}}
\hline
\multicolumn{2}{|c|}{\textbf{Cut description}} & \textbf{signal1} & \textbf{signal2} & \textbf{signal3} & \textbf{signal4} & \textbf{signal5} & \textbf{signal6}  \\
\hline
\multicolumn{2}{|c|}{\textbf{Cross-section [fb]}} & $171.40$ & $2538$ & $899.10$ & $252.40$ & $78.24$ & $23.32$  \\
\hline
\textbf{Cut-1} & $N_{\mu^+} \geq 2$, $N_{\mu^-} \geq 2$, \newline
$p_T(\mu) > 20\ \mathrm{GeV}$, \newline
$|\eta(\mu)| < 2.5$ & {\footnotesize 0.08} & {\footnotesize 0.05} & {\footnotesize 0.09} & {\footnotesize 0.25} & {\footnotesize 0.44} & {\footnotesize 0.50}  \\
\hline
\textbf{Cut-2} & $|M_{\mu\mu}^{\text{best}} - M| < 0.15 M$ & {\footnotesize 0.003} & {\footnotesize 0.04} & {\footnotesize 0.07} & {\footnotesize 0.22} & {\footnotesize 0.43} & {\footnotesize 0.48}  \\
\hline
\end{tabular}
\caption{The cut-flow table for visible on-shell mediator decay of the $\mathcal{L}_8$ model. The six BPs are signal-1 ($M = 5$ GeV), signal-2 ($M = 10$ GeV), signal-3 ($M = 25$ GeV), signal-4 ($M = 50$ GeV), signal-5 ($M = 75$ GeV), and signal-6 ($M = 100$ GeV). }
\label{tab:lepvisibleL8_decay}
\end{table}

\begin{table}[htbp]
\centering
\footnotesize  
\begin{tabular}{|>{\centering\arraybackslash}p{1.2cm}|>{\centering\arraybackslash}p{4.8cm}|*{4}{>{\centering\arraybackslash}p{1.8cm}|}}
\hline
\multicolumn{2}{|c|}{\textbf{Cut description}} & \textbf{signal1} & \textbf{signal2} & \textbf{signal3} & \textbf{signal4}  \\
\hline
\multicolumn{2}{|c|}{\textbf{Cross-section [fb]}} & $286.60$ & $212.00$ & $113.80$ & $46.01$  \\
\hline
\textbf{Cut-1} & $N_{\mu^+} \geq 2$ and $N_{\mu^-} \geq 2$, \newline
$p_T(\mu) > 20\ \mathrm{GeV}$, $|\eta(\mu)| < 2.5$ & {\footnotesize 0.13} & {\footnotesize 0.14} & {\footnotesize 0.18} & {\footnotesize 0.28}  \\
\hline
\textbf{Cut-2} & $|M_{\mu\mu}^{\text{best}} - M| < 0.15 M$ & {\footnotesize 0.09} & {\footnotesize 0.10} & {\footnotesize 0.13} & {\footnotesize 0.24}  \\
\hline
\end{tabular}
\caption{Similar to Table~\ref{tab:lepvisibleL3_decay}, but for the $\mathcal{L}_9$ model.}
\label{tab:lepvisibleL9_decay}
\end{table}

\paragraph{Visible Mediator Decay ($\text{MED} \to \mu^+\mu^-$)}
\begin{itemize} 
    \item Cut-1 (basic cuts): $N_{\mu^+} \geq 2$ and $N_{\mu^-} \geq 2$,   $p_T(\mu) > 20\ \mathrm{GeV}$ and $|\eta(\mu)| < 2.5$;   
    \item Cut-2: $|M_{\mu\mu}^{\text{best}} - M| < 0.15 M$
\end{itemize}
where $M_{\mu\mu}^{\text{best}}$ is defined as in the main text. For the visible mediator decay process, its background cross section is only $2.11\ \mathrm{fb}$, so the product of the background cross section and integrated luminosity alone is already less than $1$. Based on this, we also treat this analysis at the LEP as a background-free scenario in the subsequent discussion. Meanwhile, in the analysis of this process, we no longer apply kinematic cuts to the background nor perform relevant background analysis. Since the number of signal events passing the basic kinematic cuts is already at a low level, we only select two kinematic cuts to filter the signal events. The cut-flow tables for the $\mathcal{L}_3$, $\mathcal{L}_8$ and $\mathcal{L}_9$ models are shown in Tables~\ref{tab:lepvisibleL3_decay},~\ref{tab:lepvisibleL8_decay}, and~\ref{tab:lepvisibleL9_decay}, respectively.

For the $\mathcal{L}_8$ model, the number of events remains at a high level at the mediator mass of 50 $\mathrm{GeV}$. Thus, we additionally analyzed two larger mediator mass BPs until the number of signal events at the corresponding mass points is no more than 3 (no exclusion capability at 95\% confidence level of background-free assumption). Additionally, it can be seen that in the low-mass regime, the detection efficiency after the mediator mass window cut is significantly reduced. This is because the window range corresponding to low-mass mediators is extremely small, and the number of effective events that can pass this kinematic cut is correspondingly reduced, which leads to the decrease of the detection efficiency.

\begin{figure}[h]
\centering 
\includegraphics[width=0.48\textwidth]{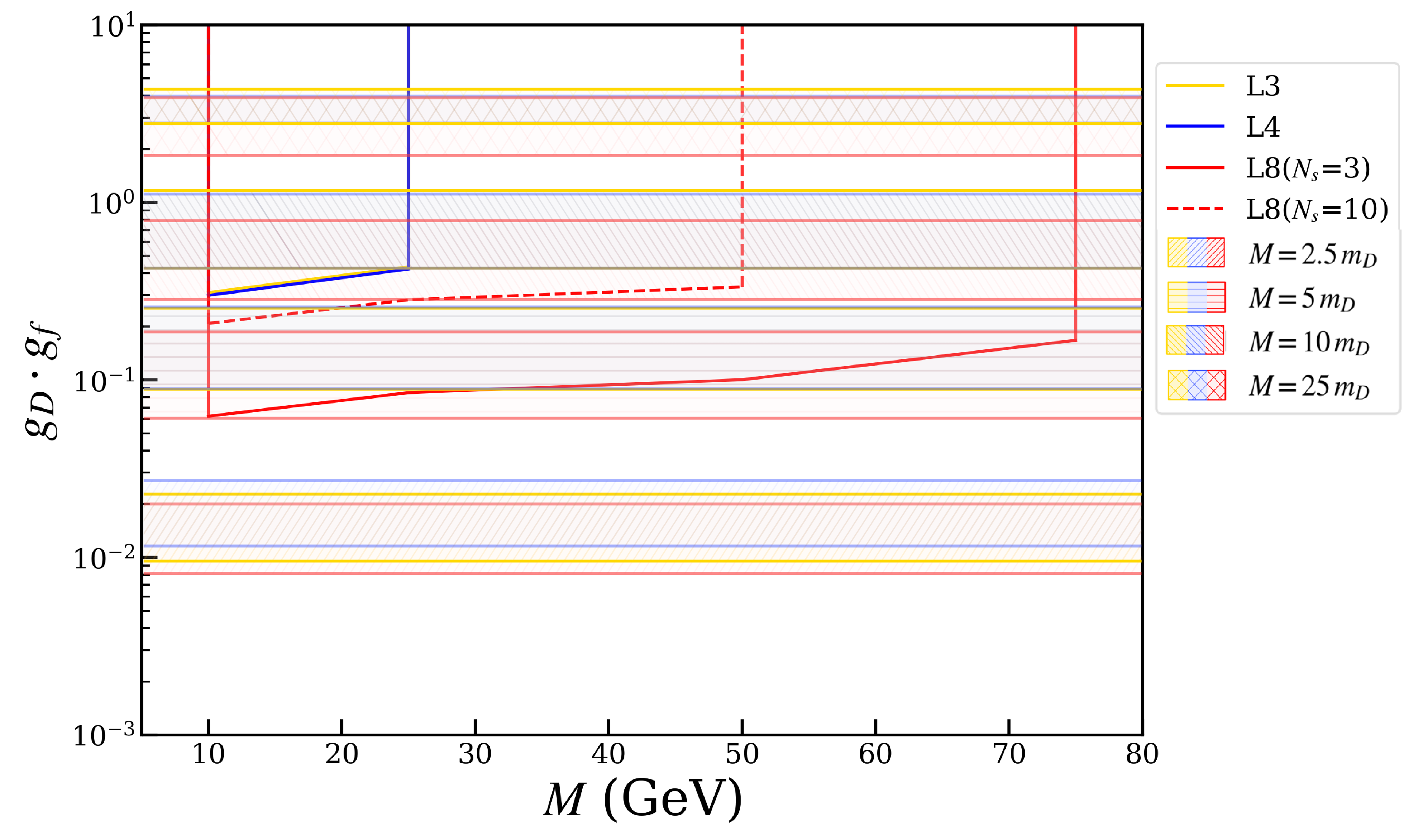} 
\includegraphics[width=0.48\textwidth]{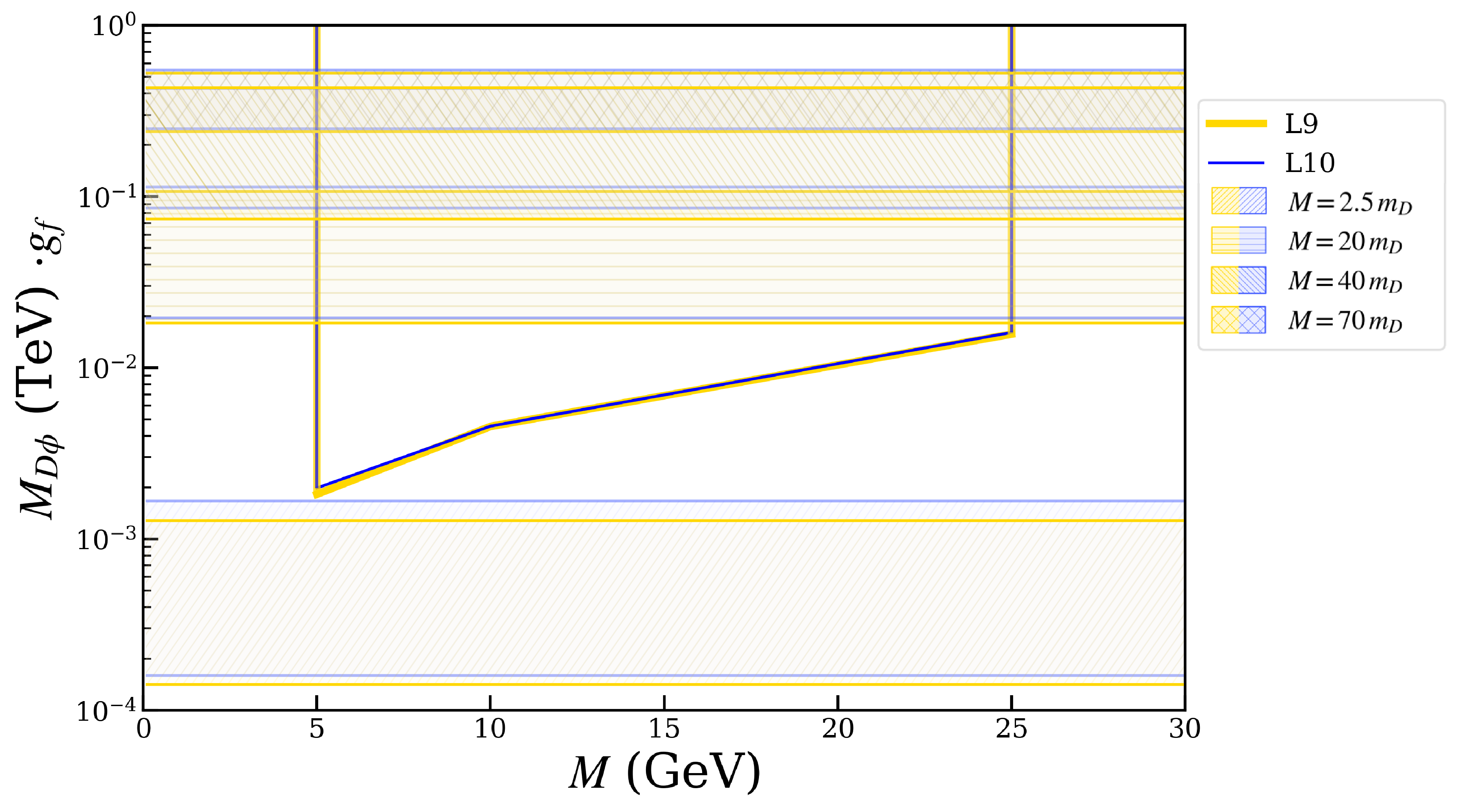}
\includegraphics[width=0.48\textwidth]{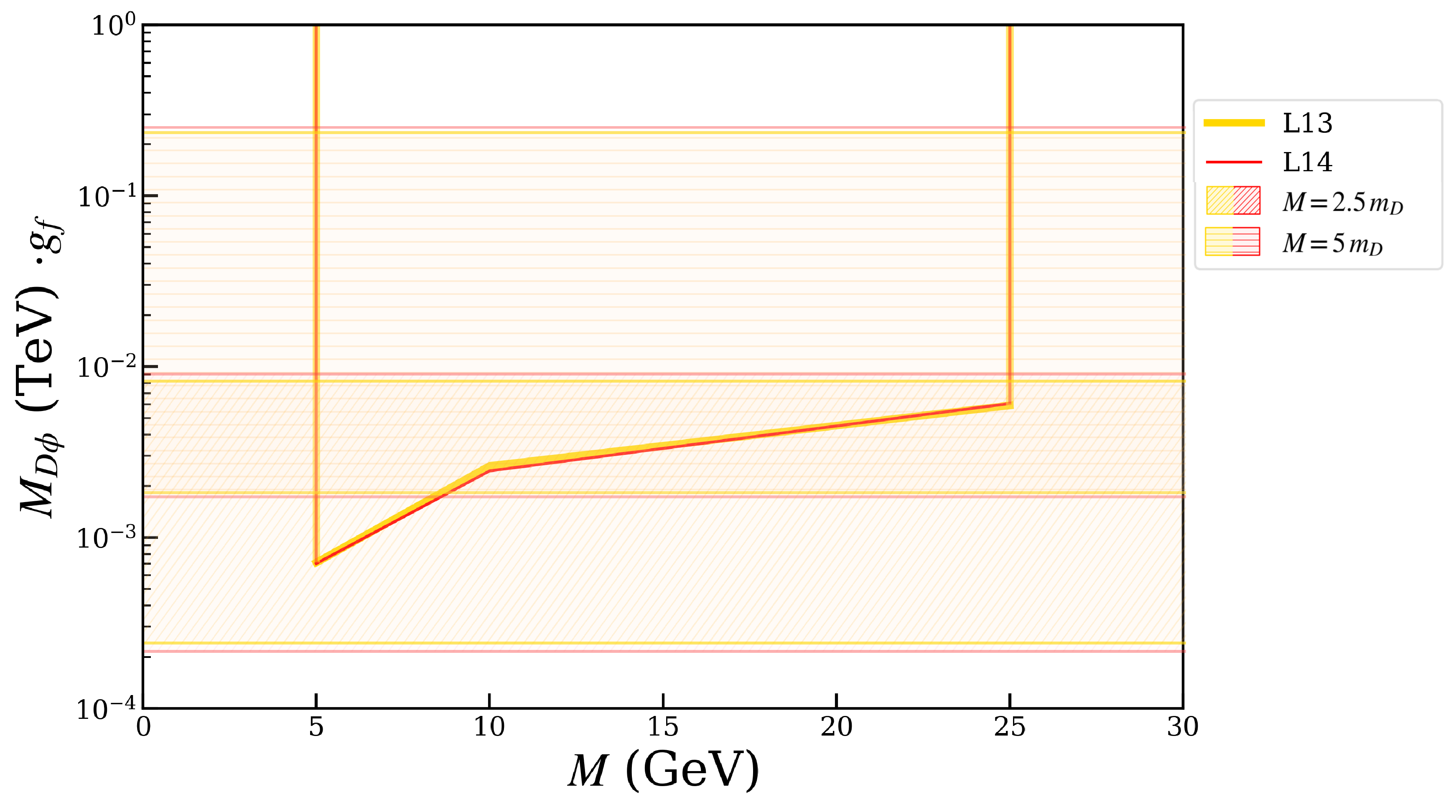}
\caption{Projected exclusion limits at 95\% confidence level for seven models in the visible mediator decay channel at LEP, derived from a background-free assumption with the signal event count criterion $N_s = 3$, with the corresponding exclusion line for the $\mathcal{L}_8$ model under $N_s = 10$ additionally labeled.}
\label{fig:lepvisible_exclusion}
\end{figure}

Figure~\ref{fig:lepvisible_exclusion} shows the projected exclusion limits from the LEP for seven $Z_2$-even muonphilic DM models in the visible on-shell mediator decay channel, with the mediator mass interval covered by $M = 5-75\ \mathrm{GeV}$. For the invisible on-shell mediator decay channel, the DM mass is required to be larger than $20\ \mathrm{GeV}$ in our previous analysis~\cite{Abdughani:2021oit}. However, our calculation shows that the parameter space with exclusion power for this process is restricted to the mediator mass interval $M \lesssim 10\ \mathrm{GeV}$. 
Consequently, this channel cannot impose valid exclusion constraints on the interested model parameter space. 
In addition, to maintain consistency with the analysis assumptions adopted in the main text (coupling constants within a reasonable range and the mediator decay branching ratio exceeding 99\%), some mass intervals are excluded from the figure as they cannot yield physically meaningful exclusion limits.

We also calculated the corresponding exclusion limits using the signal event count criterion $N_s = 10$. However, under the aforementioned analysis assumptions, except for the visible mediator decay process of the $\mathcal{L}_8$ model, other models have no valid parameter points satisfying this criterion; so the results under this criterion only show the case of $N_s = 10$ for the $\mathcal{L}_8$ model in Fig.~\ref{fig:lepvisible_exclusion}. Hence, the projected exclusion limits presented in this work may be optimistic compared to those from a full experimental data analysis, especially in the presence of reducible backgrounds or unaccounted detector effects.

The results show that for the visible mediator decay process, except for the $\mathcal{L}_8$ model, the LEP can only impose weak constraints on the low-mass mediator interval with $M \leq 25\ \mathrm{GeV}$. For the $\mathcal{L}_8$ model, it can only provide relatively weak constraints throughout the entire covered mass interval of $10-75\ \mathrm{GeV}$. For the invisible mediator decay process, as mentioned above, the parameter space with exclusion power is available only for $M \lesssim 10\ \mathrm{GeV}$, while the DM mass is required to be larger than $20\ \mathrm{GeV}$ in this analysis. Therefore, none of the models have valid exclusion limits in this channel.

\subsection{LHC Constraints}

For the LHC with $\sqrt{s}=13\ \mathrm{TeV}$, a simple recasting demonstrates that it cannot impose effective constraints on the studied $Z_2$-even muonphilic DM models, regardless of whether the on-shell mediator undergoes visible or invisible decay processes.

To quantitatively verify the above conclusion, we first adopt the signal production cross-sections corresponding to the exclusion limits of the 2$\ell$ (dimuon final state) and 4$\ell$ (tetramuon final state) channels for the integrated luminosity $\mathcal{L} = 35.9\ \mathrm{fb}^{-1}$ from Ref.~\cite{Drees:2018hhs}. For the 2$\ell$ channel, the mediator mass ranges from 10 to 250 GeV, with the excluded cross-section decreasing monotonically with increasing mass, spanning approximately from 110 pb to 0.06 pb; for the 4$\ell$ channel, the mediator mass ranges from 1 to 350 GeV, and the exclued cross-section also decreases with increasing mass, ranging from approximately 4.84 pb to 0.0028 pb. For the visible mediator decay process, the production mechanism is $pp\to\mu^+\mu^-\text{MED}$ (where $\text{MED}\to\mu^+\mu^-$); for the invisible mediator decay process, the production mechanism is $pp\to\mu^+\mu^-\text{MED}$ (where $\text{MED}\to\text{DM}+\text{DM}$). For these two decay processes, 10  mediator mass BPs are selected respectively to generate corresponding signal cross-sections, and the orders of magnitude of them is quantitatively evaluated to determine whether it reaches the level required for LHC constraints with $\mathcal{L} = 35.9\ \mathrm{fb}^{-1}$. 

The recasting results show that for the 2$\ell$ channel, the signal production cross-sections of all models decrease with the increase of mediator mass, which is consistent with the variation trend of the exclusion lines in Ref.~\cite{Drees:2018hhs}. Under the assumptions that the coupling constants are within a reasonable range and the mediator decay branching ratios are greater than 99\%, taking the $\mathcal{L}_3$ model as an example: for the 2$\ell$ channel, within the mediator mass range of 10 to 250 GeV, the signal production cross-section decreases from $56.76\ \mathrm{fb}$ to $2.57\times10^{-3}\ \mathrm{fb}$; for the 4$\ell$ channel, the signal production cross-section is only $10^{-7}\ \mathrm{fb}$ at a mediator mass of 1 GeV and $8.49\times10^{-2}\ \mathrm{fb}$ at 350 GeV, and the signal production cross-sections of all intermediate mediator masses fall within this range. The above cross-section values are far lower than the reference thresholds given in Ref.~\cite{Drees:2018hhs}, failing to form effective exclusion constraints. Finally, the signal production cross-sections of the remaining six models across the entire mass range also do not reach the excluded levels, and none of them can be constrained by the LHC at 13 TeV with $\mathcal{L} = 35.9\ \mathrm{fb}^{-1}$. Furthermore, increasing the integrated luminosity of the 13 TeV LHC is expected to yield only weak exclusion limits for the seven muonphilic DM models considered in this work. 

\section*{Acknowledgments}
The authors gratefully acknowledge the valuable discussions and insights provided by the members of the China Collaboration of Precision Testing and New Physics. C.T.L., W.Y.C. and H.Q.L. are supported by the National Natural Science Foundation of China (NNSFC) under grants No.~12335005, No.~12575118, and the Special funds for postdoctoral overseas recruitment, Ministry of Education of China. 


\end{document}